\documentclass[aps,prc,superscriptaddress,showpacs,nofootinbib,floatfix,twocolumn]{revtex4}

\usepackage{graphicx}
\usepackage{epsfig}
\usepackage{amssymb}
\usepackage{pslatex}

\usepackage{dcolumn}
\usepackage{bm}

\newcommand{\sqrtsnn}{\sqrt{s_{_{NN}}}}

\newcommand{\pt}{\mbox{$p_{\ensuremath{\it T}}$}}
\newcommand{\mt} {\mbox{$m_\mathrm{T}$}}

\newcommand{\gaga}{\gamma\,\gamma}

\def\mean#1{\ensuremath{\left<#1\right>}}

\begin{document}
\title{High transverse momentum $\eta$ meson production\\ 
in p+p, d+Au and Au+Au collisions at $\sqrtsnn$ = 200 GeV}

\newcommand{\abilene}{Abilene Christian University, Abilene, TX 79699, U.S.}
\newcommand{\acadsin}{Institute of Physics, Academia Sinica, Taipei 11529, Taiwan}
\newcommand{\banaras}{Department of Physics, Banaras Hindu University, Varanasi 221005, India}
\newcommand{\barc}{Bhabha Atomic Research Centre, Bombay 400 085, India}
\newcommand{\bnl}{Brookhaven National Laboratory, Upton, NY 11973-5000, U.S.}
\newcommand{\caucr}{University of California - Riverside, Riverside, CA 92521, U.S.}
\newcommand{\ciae}{China Institute of Atomic Energy (CIAE), Beijing, People's Republic of China}
\newcommand{\cns}{Center for Nuclear Study, Graduate School of Science, University of Tokyo, 7-3-1 Hongo, Bunkyo, Tokyo 113-0033, Japan}
\newcommand{\columbia}{Columbia University, New York, NY 10027 and Nevis Laboratories, Irvington, NY 10533, U.S.}
\newcommand{\dapnia}{Dapnia, CEA Saclay, F-91191, Gif-sur-Yvette, France}
\newcommand{\debrecen}{Debrecen University, H-4010 Debrecen, Egyetem t{\'e}r 1, Hungary}
\newcommand{\fsu}{Florida State University, Tallahassee, FL 32306, U.S.}
\newcommand{\gsu}{Georgia State University, Atlanta, GA 30303, U.S.}
\newcommand{\hiroshima}{Hiroshima University, Kagamiyama, Higashi-Hiroshima 739-8526, Japan}
\newcommand{\ihepprot}{IHEP Protvino, State Research Center of Russian Federation, Institute for High Energy Physics, Protvino, 142281, Russia}
\newcommand{\illuiuc}{University of Illinois at Urbana-Champaign, Urbana, IL 61801}
\newcommand{\isu}{Iowa State University, Ames, IA 50011, U.S.}
\newcommand{\jinrdubna}{Joint Institute for Nuclear Research, 141980 Dubna, Moscow Region, Russia}
\newcommand{\kaeri}{KAERI, Cyclotron Application Laboratory, Seoul, South Korea}
\newcommand{\kangnung}{Kangnung National University, Kangnung 210-702, South Korea}
\newcommand{\kek}{KEK, High Energy Accelerator Research Organization, Tsukuba, Ibaraki 305-0801, Japan}
\newcommand{\kfki}{KFKI Research Institute for Particle and Nuclear Physics of the Hungarian Academy of Sciences (MTA KFKI RMKI), H-1525 Budapest 114, POBox 49, Budapest, Hungary}
\newcommand{\korea}{Korea University, Seoul, 136-701, Korea}
\newcommand{\kurchatov}{Russian Research Center ``Kurchatov Institute", Moscow, Russia}
\newcommand{\kyoto}{Kyoto University, Kyoto 606-8502, Japan}
\newcommand{\labllr}{Laboratoire Leprince-Ringuet, Ecole Polytechnique, CNRS-IN2P3, Route de Saclay, F-91128, Palaiseau, France}
\newcommand{\lawllnl}{Lawrence Livermore National Laboratory, Livermore, CA 94550, U.S.}
\newcommand{\losalamos}{Los Alamos National Laboratory, Los Alamos, NM 87545, U.S.}
\newcommand{\lpc}{LPC, Universit{\'e} Blaise Pascal, CNRS-IN2P3, Clermont-Fd, 63177 Aubiere Cedex, France}
\newcommand{\lund}{Department of Physics, Lund University, Box 118, SE-221 00 Lund, Sweden}
\newcommand{\muenster}{Institut f\"ur Kernphysik, University of Muenster, D-48149 Muenster, Germany}
\newcommand{\myongji}{Myongji University, Yongin, Kyonggido 449-728, Korea}
\newcommand{\nagasaki}{Nagasaki Institute of Applied Science, Nagasaki-shi, Nagasaki 851-0193, Japan}
\newcommand{\newmex}{University of New Mexico, Albuquerque, NM 87131, U.S.}
\newcommand{\nmsu}{New Mexico State University, Las Cruces, NM 88003, U.S.}
\newcommand{\ornl}{Oak Ridge National Laboratory, Oak Ridge, TN 37831, U.S.}
\newcommand{\orsay}{IPN-Orsay, Universite Paris Sud, CNRS-IN2P3, BP1, F-91406, Orsay, France}
\newcommand{\pnpi}{PNPI, Petersburg Nuclear Physics Institute, Gatchina,  Leningrad region, 188300, Russia}
\newcommand{\riken}{RIKEN, The Institute of Physical and Chemical Research, Wako, Saitama 351-0198, Japan}
\newcommand{\rikjrbrc}{RIKEN BNL Research Center, Brookhaven National Laboratory, Upton, NY 11973-5000, U.S.}
\newcommand{\saispbstu}{Saint Petersburg State Polytechnic University, St. Petersburg, Russia}
\newcommand{\saopaulo}{Universidade de S{\~a}o Paulo, Instituto de F\'{\i}sica, Caixa Postal 66318, S{\~a}o Paulo CEP05315-970, Brazil}
\newcommand{\seoulnat}{System Electronics Laboratory, Seoul National University, Seoul, South Korea}
\newcommand{\stonybrkc}{Chemistry Department, Stony Brook University, SUNY, Stony Brook, NY 11794-3400, U.S.}
\newcommand{\stonycrkp}{Department of Physics and Astronomy, Stony Brook University, SUNY, Stony Brook, NY 11794, U.S.}
\newcommand{\subatech}{SUBATECH (Ecole des Mines de Nantes, CNRS-IN2P3, Universit{\'e} de Nantes) BP 20722 - 44307, Nantes, France}
\newcommand{\tenn}{University of Tennessee, Knoxville, TN 37996, U.S.}
\newcommand{\titech}{Department of Physics, Tokyo Institute of Technology, Oh-okayama, Meguro, Tokyo, 152-8551, Japan}
\newcommand{\tsukuba}{Institute of Physics, University of Tsukuba, Tsukuba, Ibaraki 305, Japan}
\newcommand{\vandy}{Vanderbilt University, Nashville, TN 37235, U.S.}
\newcommand{\waseda}{Waseda University, Advanced Research Institute for Science and Engineering, 17 Kikui-cho, Shinjuku-ku, Tokyo 162-0044, Japan}
\newcommand{\weizmann}{Weizmann Institute, Rehovot 76100, Israel}
\newcommand{\yonsei}{Yonsei University, IPAP, Seoul 120-749, Korea}
\affiliation{\abilene}
\affiliation{\acadsin}
\affiliation{\banaras}
\affiliation{\barc}
\affiliation{\bnl}
\affiliation{\caucr}
\affiliation{\ciae}
\affiliation{\cns}
\affiliation{\columbia}
\affiliation{\dapnia}
\affiliation{\debrecen}
\affiliation{\fsu}
\affiliation{\gsu}
\affiliation{\hiroshima}
\affiliation{\ihepprot}
\affiliation{\illuiuc}
\affiliation{\isu}
\affiliation{\jinrdubna}
\affiliation{\kaeri}
\affiliation{\kangnung}
\affiliation{\kek}
\affiliation{\kfki}
\affiliation{\korea}
\affiliation{\kurchatov}
\affiliation{\kyoto}
\affiliation{\labllr}
\affiliation{\lawllnl}
\affiliation{\losalamos}
\affiliation{\lpc}
\affiliation{\lund}
\affiliation{\muenster}
\affiliation{\myongji}
\affiliation{\nagasaki}
\affiliation{\newmex}
\affiliation{\nmsu}
\affiliation{\ornl}
\affiliation{\orsay}
\affiliation{\pnpi}
\affiliation{\riken}
\affiliation{\rikjrbrc}
\affiliation{\saispbstu}
\affiliation{\saopaulo}
\affiliation{\seoulnat}
\affiliation{\stonybrkc}
\affiliation{\stonycrkp}
\affiliation{\subatech}
\affiliation{\tenn}
\affiliation{\titech}
\affiliation{\tsukuba}
\affiliation{\vandy}
\affiliation{\waseda}
\affiliation{\weizmann}
\affiliation{\yonsei}
\author{S.S.~Adler}	\affiliation{\bnl}
\author{S.~Afanasiev}	\affiliation{\jinrdubna}
\author{C.~Aidala}	\affiliation{\bnl}
\author{N.N.~Ajitanand}	\affiliation{\stonybrkc}
\author{Y.~Akiba}	\affiliation{\kek} \affiliation{\riken}
\author{J.~Alexander}	\affiliation{\stonybrkc}
\author{R.~Amirikas}	\affiliation{\fsu}
\author{L.~Aphecetche}	\affiliation{\subatech}
\author{S.H.~Aronson}	\affiliation{\bnl}
\author{R.~Averbeck}	\affiliation{\stonycrkp}
% ...
\collaboration{PHENIX Collaboration} \noaffiliation

\date{\today}        
\begin{abstract}
Inclusive transverse momentum spectra of $\eta$ mesons in the range 
$\pt \approx$ 2 -- 12~GeV$/c$ have been measured at mid-rapidity ($|\eta|<0.35$) by 
the PHENIX experiment at RHIC in p+p, d+Au and Au+Au collisions at 
$\sqrtsnn$ = 200~GeV. The $\eta$ mesons are reconstructed through their
$\eta\rightarrow\gaga$ channel for the three colliding systems as well as 
through the $\eta\rightarrow\pi^0\pi^+\pi^-$ decay mode in p+p and d+Au
collisions. The nuclear modification factor in d+Au collisions, 
$R_{dAu}(\pt)\approx$ 1.0 -- 1.1, suggests at most only modest $\pt$ broadening 
(``Cronin enhancement''). 
In central Au+Au reactions, the $\eta$ yields are significantly 
suppressed, with $R_{AuAu}(\pt)\approx$ 0.2. The ratio of $\eta$ to $\pi^0$ 
yields is approximately constant as a function of $\pt$ for the three colliding systems 
in agreement with the high-$\pt$ world average of $R_{\eta/\pi^{0}}\approx$ 0.5 
in hadron-hadron, hadron-nucleus, and nucleus-nucleus collisions 
for a wide range of center-of-mass energies ($\sqrtsnn\approx$ 3 -- 1800 GeV) 
as well as, for high scaled momentum $x_p$, in $e^+e^-$ annihilations at 
$\sqrt{s}=$ 91.2 GeV.
These results are consistent with a scenario where high-$\pt$ $\eta$ 
production in nuclear collisions at RHIC is largely unaffected by initial-state 
effects, but where light-quark mesons ($\pi^{0},\eta$) are equally suppressed 
due to final-state interactions of the parent partons in the dense medium 
produced in Au+Au reactions.
\end{abstract}
\maketitle

\tableofcontents


\section{INTRODUCTION}
\label{sec:intro}



Single-hadron production at large transverse momenta
($p_\mathrm{T}\gtrsim$ 2 GeV/$c$) in high-energy hadronic and nuclear
collisions results from the fragmentation of quarks and gluons
issuing from parton-parton scatterings with large momentum transfer $Q^2$.
Since the cross-sections for such hard processes can be calculated
perturbatively within Quantum Chromodynamics (QCD)~\cite{pQCD}, inclusive
high-$\pt$ hadrons (as well as jets, real and virtual direct photons,
and heavy quarks) have long been considered sensitive, well calibrated probes of the
small-distance QCD processes.
The study of inclusive hadron production at large $\pt$ in 
proton-proton interactions provides valuable information about 
perturbative QCD (pQCD), parton distribution functions (PDF) in the
proton, and fragmentation functions (FF) of the partons~\cite{geist90}. 
Furthermore, the use of polarized beams ($\vec{p}+\vec{p}$) allows one
to investigate the spin structure of the
proton~\cite{bunce00}. High-energy collisions of protons or deuterons
on nuclear targets (p,d+A) also provide interesting insights
on initial- and final-state QCD effects such as modifications of the
nuclear PDFs~\cite{arneodo,armesto06} and parton rescattering in the cold nuclear
medium~\cite{cronin}. Both effects are sensitive to
physics such as parton structure and evolution at small values 
of fractional momentum $x$ in the hadronic wave functions~\cite{iancu03}, 
and the dynamics of hadronization in cold nuclei~\cite{kopelio03,hwa04}.
Lastly, high-$\pt$ hadron production in nucleus-nucleus (A+A) reactions
is a sensitive probe of the properties of the strongly interacting QCD matter 
produced in the collision. Indeed, since perturbative processes happen at time scales 
$\tau\approx 1/\pt\lesssim$ 0.1 fm/$c$, the hard-scattered partons traverse and are 
potentially modified by the bulk matter formed shortly after the collision. 
In this context, the suppression of leading hadrons has been
postulated~\cite{gyulassy90} as a signal of ``jet quenching'' in a Quark-Gluon-Plasma 
(QGP) due to medium-induced energy-loss of the parent parton~\cite{bdmps,glv,wiedemann}.\\

All the aforementioned research topics have been addressed in detail by the rich physics program 
carried out at the Relativistic Heavy-Ion Collider (RHIC) at Brookhaven National Laboratory (BNL) 
during its first six years of operation (2000 -- 2006).  For example, the study of 
inclusive high-$\pt$ neutral pion production at mid-rapidity in p+p~\cite{ppg024}, 
$\vec{p}+\vec{p}$~\cite{phnx_spin_pi0}, d+Au~\cite{ppg028,ppg044} and Au+Au~\cite{ppg003,ppg014,ppg051,ppg054} 
collisions at $\sqrtsnn$ = 200 GeV, has provided valuable information respectively on:
\begin{description}
\item (i) the gluon-to-pion FF~\cite{ppg024,albino05}, 
\item (ii) the gluon contribution to the proton spin~\cite{vogelsang_spin,bourrely03,anselmino04}, 
\item (iii) initial-state effects in cold nuclear matter such as shadowing of the nuclear 
PDFs~\cite{eks98,deFlorian03,Kopeliovich03,vogt04}, 
Cronin broadening~\cite{accardi04,cattaruzza04,barna04}, and 
gluon saturation~\cite{albacete03,Iancu04,jamal05}, 
\item (iv) the thermodynamical and microscopic properties of hot and dense QCD matter~\cite{dde_qcdatwork}, 
such as the initial gluon rapidity density $dN^g/dy$~\cite{vitev_gyulassy} 
and the transport coefficient $\mean{\hat{q}}$~\cite{eskola04,dainese04} of the 
produced medium; and the mechanism of hadronization in a dense parton medium~\cite{recomb}.\\
\end{description}

In this paper, we extend previous PHENIX analyses of high-$\pt$ hadron production
in p+p~\cite{ppg024,phnx_spin_pi0}, d+Au~\cite{ppg028} and Au+Au~\cite{ppg014,ppg023,ppg051,ppg054} 
collisions at $\sqrt{s_{NN}}$ = 200 GeV, to include an additional identified particle, 
the $\eta$ meson, measured in the range $\pt$ = 2 -- 12 GeV/$c$. 
The spectra presented here are the hardest (i.e. have the
highest $\pt$)
ever measured for the $\eta$ meson\footnote{Before this measurement, 
only the ISR AFS Collaboration p+p$\rightarrow\eta$+X measurement for $\pt$ = 3 -- 11 GeV/$c$ at 
$\sqrt{s}$ = 62.4 GeV~\cite{Kourkoumelis79} and the single CDF $\eta/\pi^0$ point 
measured at $\pt$ = 12 GeV/$c$ in $\bar{p}+p$ collisions at $\sqrt{s}$ = 1800 GeV~\cite{Abe93} 
had comparable maximum $\pt$ values.} in p+p, p+A and A+A collisions.
The high $\pt$ reach of the $\eta$ helps to characterize the mechanisms of truly perturbative 
parton-parton scatterings and parton fragmentation in different QCD environments 
(p+p, d+A and Au+Au).
The $\eta$ data from p+p collisions are presented here as a baseline for 
medium effects in d+Au and Au+Au. Once a parametrization of the $\eta$ FF in $e^+e^-$ 
is performed (see Section~\ref{sec:eta_pi0_ee}), the observed p+p cross section will additionally 
allow a test of pQCD predictions. Such a FF parametrization would be useful in particular 
in the light of upcoming high-$\pt$ $\eta$ asymmetry data obtained with polarized beams 
of relevance for the proton spin program~\cite{phenix_eta_spin}.

For d+Au and Au+Au reactions, we present the single $\eta$ spectra, the $\eta$ 
nuclear modification factors, and the $\eta$-to-$\pi^0$ ratio measured as a function of 
$\pt$ in different centralities. Within uncertainties, the d+Au spectra for all centralities 
are consistent with the p+p yields scaled by the corresponding number of incoherent 
nucleon-nucleon ($NN$) collisions. The maximum amount of $\pt$ broadening 
seen in the $\eta$ data is 10\%, $R_{dAu}(\pt)\approx$ 1.0 -- 1.1. Such a result confirms
the limited influence of cold nuclear matter effects, such as shadowing, Cronin
broadening or recombination, on high-$\pt$ meson production at mid-rapidity at RHIC~\cite{ppg028,ppg044}.
On the other hand, the factor of $\sim$5 deficit of inclusive $\eta$ yields observed 
above $\pt\approx$ 4 GeV/$c$ in central Au+Au compared to binary-scaled p+p collisions, 
$R_{AuAu}(\pt)\approx$ 0.2, is the same as that found for high-$\pt$ $\pi^0$~\cite{ppg014,ppg054} 
and for inclusive charged hadrons~\cite{ppg023,star_hipt_200}.
Such a common suppression pattern for $\pi^0$, $\eta$ and $h^\pm$ is expected if the 
energy loss takes place at the parton level in the dense medium formed in the reaction
{\it prior} to its fragmentation into a given hadron in the vacuum. Indeed, in this case 
the high-$\pt$ deficit will just depend on the energy lost by the parent light-quark 
or gluon (i.e. on the initial density of scatterers in the produced medium) and not on the nature 
of the final leading hadron whose production will be described by the same {\it universal} 
probabilities (fragmentation functions) which govern vacuum hadron production 
in more elementary systems. Such an interpretation is supported by the fact that the 
ratio of $\pt$-differential cross-sections of $\eta$ mesons with respect to $\pi^0$ 
in Au+Au, d+Au and p+p collisions is approximately constant, 
$R_{\eta/\pi^0}\approx$ 0.40 -- 0.48, which is consistent with the world average 
measured: (i) in hadron-hadron, hadron-nucleus and nucleus-nucleus collisions above 
$\pt\approx$ 3 GeV/$c$; as well as (ii) in electron-positron annihilations at the $Z$ pole 
($\sqrt{s}=$ 91.2 GeV) for {\it energetic} $\eta$ and $\pi^0$ with scaled momenta 
$x_{p} = p_{\ensuremath{\it hadron}}/p_{\ensuremath{\it beam}}\gtrsim$ 0.35.
Comparison of our data to a world compilation of $\eta/\pi^0$ ratios is done 
in the last section of the paper.\\

In addition to their interest as a {\it signal} in their own right, 
reliable knowledge of the production of $\eta$ mesons in p+p, d+Au and Au+Au reactions 
is also required 
in order to determine and statistically remove the {\it background} of secondary 
$e^\pm$ and $\gamma$ for other measurements such as single electrons (from heavy-quark 
decays)~\cite{phenix_charm,phenix_dielec}, dielectrons~\cite{phenix_dielec} 
and direct photons~\cite{ppg042,pp_gamma}. 
Indeed, $\eta$ mesons constitute the second most important source after the $\pi^0$ 
of decay electrons (Dalitz and conversion) and photons contributing to these backgrounds.\\

The paper is organized as follows. Section~\ref{sec:exp} presents a 
description of the experimental setup and detector systems used in 
this work. Section~\ref{sec:analysis} provides an explanation of the 
analysis methods employed to obtain the $\eta$ data. 
Section~\ref{sec:results} presents and compares the $\eta$ results ($\pt$ spectra, 
nuclear modification factors, and $\eta/\pi^0$ ratios) measured in p+p, d+Au and 
Au+Au at $\sqrt{s_{_{NN}}}$ = 200~GeV, and discusses the relative role of cold 
nuclear (d+Au) and hot and dense medium (Au+Au) effects on high-$\pt$ meson production.
In particular, Section~\ref{sec:eta_pi0_syst} discusses the measured $\eta$-to-$\pi^0$
ratios in the context of different phenomenological models of high-$\pt$ hadron 
production as well as in comparison to other experimental results measured in 
high-energy particle collisions at different center-of-mass energies. 
A less detailed presentation of a subset of these $\eta$ results has already been 
published in ~\cite{ppg051}.

\section{EXPERIMENTAL SETUP}
\label{sec:exp}

\begin{figure}[htb] 
\includegraphics[width=1.0\linewidth]{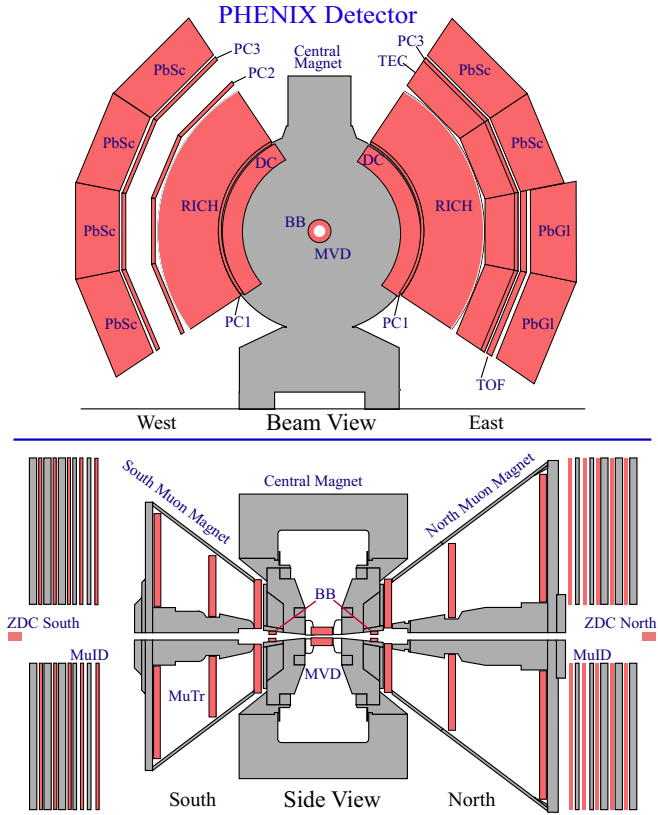} 
\caption{The PHENIX experimental setup during Run-2 and Run-3 at RHIC. The detectors used in the present analysis 
are the 8 EMCal (PbSc, PbGl) sectors for photon detection ($\eta\rightarrow\gaga$), the drift chamber (DC) and two 
layers of multiwire proportional chambers with pad readout (PC) for charged pion detection 
($\eta\rightarrow\pi^0\pi^+\pi^-$); as well as the two Beam-Beam Counters (BBC) and the two Zero-Degree 
Calorimeters (ZDC) for global event characterization.} 
\label{fig:phenix} 
\end{figure}

The PHENIX experiment at the RHIC
facility~\cite{nim_rhic} at BNL is specifically 
designed to measure hard QCD probes such as high-$\pt$ hadrons, direct photons, 
leptons, and heavy flavor production. PHENIX achieves good mass and particle 
identification (PID) resolutions as well as small granularity by combining 13 detector 
subsystems ($\sim$350,000 channels) divided into: 
(i) two central arm spectrometers for electron, photon and hadron measurements at 
mid-rapidity ($|\eta|<0.35$, $\Delta\phi=\pi$); 
(ii) two forward-backward ($|\eta|$ = 1.2 - 2.2, $\Delta\phi = 2\pi$) spectrometers 
for muon detection; and 
(iii) two global (inner) detectors for trigger and centrality selection.
A detailed description of the complete detector can be found elsewhere~\cite{nim_phenix}.
The data presented in this paper were obtained during the Run-2 (2001--2002, Au+Au) 
and Run-3 (2003, d+Au, p+p) operations at RHIC. The layout of the PHENIX detector 
during these run periods is shown in Fig.~\ref{fig:phenix}. The primary detectors 
used to obtain the present results are the PHENIX central arm

\clearpage

\noindent spectrometers, 
particularly the electromagnetic calorimeters (EMCal)~\cite{nim_emcal} and the 
charged particle tracking devices (the Drift Chamber (DC)~\cite{nim_dc} and Pad Chambers (PC)~\cite{nim_pc}). 
In addition, the Beam-Beam Counters (BBC)~\cite{nim_bbc} and the 
Zero-Degree Calorimeters (ZDC)~\cite{nim_zdc} are used for triggering, 
event characterization and (Au+Au and d+Au) centrality determination.\\


\begin{table*}[t]
\caption{\label{tab:evts_sampled}
Events sampled and integrated luminosity (after vertex cuts) in the $\eta$ analyses
for p+p, d+Au and Au+Au collisions. The equivalent p+p luminosities in d+Au (Au+Au) have
been obtained normalizing their corresponding luminosities by $2\,A$ ($A^2$) factors as
expected for hard cross-section scaling.}
\begin{ruledtabular}\begin{tabular}{lcccccc}
collision\hspace{5mm} & \multicolumn{2}{c}{Events sampled} & \multicolumn{2}{c}{\hspace{5mm}Total integrated luminosities\hspace{5mm}} & \multicolumn{2}{c}{\hspace{5mm}BBC attributes\hspace{5mm}} \\
system    &  \hspace{5mm} MB (LVL1) trigger \hspace{5mm}& High-$\pt$ (Gamma) trigger \hspace{5mm}& absolute & equivalent p+p & cross section & efficiency \\ \hline
p+p       &  $25.2\cdot 10^{6}$  &  $49.3\cdot 10^{8}$ &  216 nb$^{-1}$ & 216 nb$^{-1}$ & 23.0 mb $\pm$ 9.7\% & (55$\pm$5)\%\\
d+Au      &  $58.3\cdot 10^{6}$  &  $29.2\cdot 10^{8}$ & 1.5 nb$^{-1}$ & 590 nb$^{-1}$ & 1.99 b $\pm$ 5.2\% & (88$\pm$4)\%\\
Au+Au     &  $34\cdot 10^{6}$    &  $30\cdot 10^{6}$  &  9 $\mu$b$^{-1}$ & 230 nb$^{-1}$ & 6.315 b $\pm$ 8.4\%& (92$\pm$3)\%\\
\end{tabular}\end{ruledtabular}
\end{table*}


\subsection{Electromagnetic Calorimeter (EMCal)}

The $\eta$ mesons are detected in PHENIX via their $\gaga$ (branching ratio BR = 39.43\%)
and $\pi^0\pi^+\pi^-$ (BR = 22.6\%) decays~\cite{pdg}. Photons from the direct $\gaga$ 
channel as well as from the secondary (daughter) $\pi^0$ decays are measured in the 
PHENIX EMCal, which has a pseudo-rapidity acceptance of $-0.35 < \eta < 0.35$ 
and covers $\pi$ radians in azimuth. The electromagnetic calorimeter is divided into 
eight sectors with two distinct detection technologies (see Fig.~\ref{fig:phenix}).
A lead-scintillator 
calorimeter (PbSc) consists of 15552 individual lead-scintillator sandwich modules 
(5.54 cm $\times$ 5.54 cm $\times$ 37.5 cm, 18$X_0$), grouped in six sectors
located at a radial distance of 5.1\,m from the beam line, covering a total 
solid angle of $\Delta\eta \approx 0.7$ and $\Delta \phi \approx 3\pi/4$. 
A lead-glass \v{C}erenkov calorimeter (PbGl) comprising two sectors, with a total
of 9216 modules (4 cm $\times$ 4 cm $\times$ 40 cm, 14.4$X_0$), is located at a 
radial distance of $\sim$5\,m from the beam pipe and covers a total solid angle 
at mid-rapidity of $\Delta\eta \approx 0.7$ and $\Delta \phi = \pi/4$.
The corresponding $\Delta\eta \times \Delta \phi$ acceptance of a single tower at 
$\eta = 0$ is $0.011\times 0.011$ and $0.0075\times 0.0075$ for the PbSc and PbGl 
calorimeters, respectively. The chosen transverse size of the towers is not much
larger than their corresponding Moli\`ere radius ($\rho_M$ = 3.0 cm and 3.7 cm for 
PbSc and PbGl, respectively) so that most of the electromagnetic showers extend 
over several modules resulting in an improved position resolution based on a 
``center of gravity'' reconstruction of the impact point of the photon clusters.\\


The energy calibration of the PbSc modules was obtained from the original 
beam-test values and redundantly confirmed with (i) the position of the reconstructed
$\pi^0$ mass peak, (ii) the energy deposit from minimum-ionizing charged particles 
traversing the calorimeter, as well as with (iii) the expected 
$E_{\ensuremath{PbSc}}/p_{\ensuremath{tracking}}\sim$1 value measured for 
electrons and positrons identified in the Ring-Imaging \v{C}erenkov (RICH)
detector and whose momentum was measured in the tracking detectors.
In the PbGl modules, the reference energy calibration from the 
original beam-test values is corrected with the time-dependent gain corrections
obtained with a light-emission-diode (LED) system for the lead-glass calorimeter. 
The LEDs emit light with known intensity, so gain fluctuations can be detected.
The final PbGl calibration is obtained by comparing the measured $\pi^0$ 
peak position to its nominal value.


\subsection{Central Arm Tracking}

Charged pions are measured with the PHENIX central tracking system combining
information from the drift and pad chambers.
The momenta of the $\pi^\pm$ are measured at a radius of 2.0 m from the event vertex by 
the Drift Chamber (DC). The DC, located outside the field of PHENIX 
central magnets, uses several layers of wires to reconstruct the angle of the track, which
is inversely proportional to its momentum. The DC momentum resolution is determined to 
be 0.7$\oplus$1.1\%$\pt$ (GeV/$c$). The polar angle of the track is measured by 
Pad Chamber 1 (PC1), a multi-wire proportional chamber located just beyond 
the DC. The last pad-chamber layer, PC3, at a radius of 5.0 m and directly in front of the EMCal, 
is used in this analysis for two purposes: to confirm the track by matching to a PC3 hit, 
as well as to veto an EMCal cluster produced by a charged particle track.\\

The DC momentum scale is checked by the reconstruction of the correct mass of
(i) $\pi^\pm$, $K^\pm$, $p,\bar{p}$ identified with the time-of-flight system~\cite{nim_pid}, 
and (ii) $\omega$, $\phi$, $J/\Psi$ mesons decaying into the $e^+e^-$ channel identified with the 
RICH and EMCal. The momentum scale is thus known with an accuracy better than 0.2\%. 
Since at low $\pt$, the momentum resolution is better when measured with the tracking system 
than that using the energy measured via calorimetry, and given that the momentum range of the 
three $\eta$ decay products has a relatively low $\pt$, the uncertainties in the 
tracking system calibration are less important in the $\pi^0\pi^+\pi^-$ measurement than 
in the $\gamma\gamma$ decay channel. As a result, the tracking devices provide 
better accuracy for the $\eta$ mass reconstruction than the EMCal.


\subsection{Global Detectors}

Triggering and global event characterization is achieved using the Beam-Beam Counters 
(BBC) and the Zero-Degree Calorimeters (ZDC).
The two BBC are placed around the beam pipe 1.44~m in each direction from the nominal 
interaction point. Each BBC consists of 64 hexagonal quartz \v{C}erenkov
radiators closely packed in an approximately azimuthally symmetric configuration.
The BBC are used to count the charged particle multiplicity in the pseudo-rapidity 
range $3.0 < |\eta| < 3.9$, to provide the start time for 
time-of-flight measurement, and to give the collision vertex position along the interaction 
diamond with a typical resolution of 0.6 (2)~cm in Au+Au (p+p) collisions~\cite{nim_bbc}. 
In d+Au collisions, the centrality of the collision is determined by measuring the 
charge deposited in the BBC in the $Au$ beam direction~\cite{ppg036}; whereas 
in Au+Au reactions, the correlation between the BBC charge sum and the ZDC total 
energy is used for centrality determination~\cite{ppg001} (see next Section).
The ZDC are small hadronic calorimeters that measure the energy carried by spectator 
neutrons at very forward angles. They are placed 18~m up- and downstream of the 
interaction point along the beam line ($|\theta|<$ 2 mrad). Each ZDC consists of 
three modules with a depth of 2 hadronic interaction lengths read out by a 
single photo-multiplier tube (PMT). Both time and amplitude are digitized for each 
PMT along with the analog sum of the three PMT signals for each ZDC.\\


\section{DATA ANALYSIS}
\label{sec:analysis}

In this section, we describe the event selection criteria, the
reaction centrality determination in d+Au and Au+Au collisions, the
$\eta$ identification and reconstruction procedures in the
$\eta\rightarrow\gaga$~ and $\eta\rightarrow\pi^0\pi^+\pi^-$ channels
and the various corrections (geometrical acceptance, reconstruction
efficiency, trigger, absolute cross-section normalization) applied to
the raw data. The systematic uncertainties of the measurements are
discussed at the end of the section.



\subsection{Event Selection}
\label{sec:event_selection}

The data presented in this paper were collected under
two general trigger conditions. The first sample, consisting of minimum-bias (MB)
events with vertex position along the beam axis $|z| <$~30 cm, was conditioned on a local-level-1 (LVL1) trigger 
generated by coincidences between the two BBC (in the case of p+p and d+Au) or by coincidences between
the BBC and ZDC detectors (in the case of Au+Au). The MB trigger cross
sections measured by the BBC in p+p and d+Au collisions are respectively
23.0 mb $\pm$ 9.7\% mb and 1.99 b $\pm$ 5.2\%~\cite{dau_totalxsec},
whereas the Run-2 Au+Au minimum bias trigger has some inefficiency for the most peripheral
interactions and records only $92.2^{+2.5}_{-3.0}$\% of $\sigma_{AuAu}$~\cite{ppg014}.
In other words, the LVL1 triggers accept respectively (55$\pm$5)\%, (88$\pm$4)\%,
and (92$\pm$3)\% of the total inelastic cross sections:
$\sigma_{pp}^{inel}$ = 42 $\pm$ 3 mb,
$\sigma_{dAu}^{inel}$ = 2260 $\pm$ 100 mb, and
$\sigma_{AuAu}^{inel}$ = 6850 $\pm$ 540 mb.
A second ``photon-triggered'' sample, requiring electromagnetic showers
above a given threshold in the EMCal (with or without the MB BBC
requirement), has been used to extend the $\eta$ measurements to higher $\pt$.
The details of this level-2 (LVL2) software trigger are described in~\cite{ppg054}.
The total number of events collected in the MB and photon-triggered samples (after
vertex cuts) as well as the integrated luminosities for each collision system are listed
in Table~\ref{tab:evts_sampled}.

\subsection{Centrality Determination (d+Au, Au+Au)}
\label{sec:cent_determin}

\begin{figure}
\includegraphics[width=1.0\linewidth]{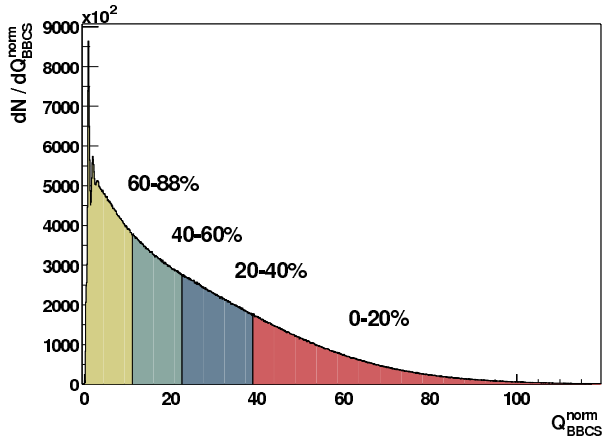}
\caption{Distribution of the normalized charge in the south Beam-Beam 
Counter (BBCS) in d+Au collisions at $\sqrtsnn$ = 200 GeV.
  The normalization is done such that the normalized charge
  corresponds to the number of hits.}
\label{fig:bbc}
\end{figure}

The events in d+Au collisions are classified into four different centrality
classes given in percentiles of the total cross-section: 0--20\%, 20--40\%, 40--60\%
and 60--88\%, with the latter being the most peripheral.
The reaction centrality is related to the number of hits in the south
Beam-Beam Counter (BBCS), which
is proportional to the number of participant nucleons in the gold nucleus~\cite{ppg036}.
The distribution of the normalized charge in the BBCS and the classification into
different centrality classes is shown in Fig. \ref{fig:bbc}.
In order to obtain reasonably large statistics in each Au+Au centrality class,
three centralities are used in the current Au+Au analysis: 0--20\% (central), 20--60\% (semicentral)
and 60--92\% (peripheral), determined by cuts in the correlated distribution of the
charge detected in the BBC and the energy measured in the ZDC~\cite{ppg001}.
A Glauber Monte Carlo model combined with a simulation of the BBC (plus ZDC)
response allows determination of the mean value of the associated
nuclear overlap function $\mean{T_{dA}}$ ($\mean{T_{AA}}$) for 
each d+Au (Au+Au) centrality bin.
Table~\ref{tab:T_AA} lists the mean value of the  nuclear overlap function for different
centralities in both systems.

\begin{table}[ht]
\caption{\label{tab:T_AA}
Values of the average nuclear overlap function $\mean{T_{dA}}$ and
$\mean{T_{AA}}$ for the different centralities considered in d+Au and Au+Au
reactions respectively.}
\begin{ruledtabular}\begin{tabular}{ccc}
centrality bin \hspace{5mm} & $\mean{T_{dA}}$ (mb$^{-1}$) & $\mean{T_{AA}}$ (mb$^{-1}$)  \\ \hline
 min bias & 0.20 $\pm$ 0.01  &  6.14 $\pm$ 0.45 \\
 0-20\%  & 0.36 $\pm$ 0.02   &  18.5 $\pm$ 1.3 \\
20-40\%  & 0.25 $\pm$ 0.017  &  - \\
20-60\%  &      -            &  4.6 $\pm$ 0.4 \\
40-60\%  & 0.17 $\pm$ 0.014  &  - \\
60-88\%  & 0.073 $\pm$ 0.007 &  - \\
60-92\%  &      -            &  0.3 $\pm$ 0.1  \\ 
\end{tabular}\end{ruledtabular}
\end{table}



\subsection{$\eta\rightarrow\gaga$~ Reconstruction}
\label{sec:eta_gaga_reco}

The main mode of $\eta$-meson reconstruction in PHENIX is via the electromagnetic
channel $\eta\rightarrow\gaga$. PHENIX has published the results of a number of
$\pi^0\rightarrow\gaga$ measurements in the EMCal for different colliding
systems~\cite{ppg003,ppg014,ppg024,phnx_spin_pi0,ppg028,ppg044,ppg054,ppg051}.
The technique for identifying the photons and reconstructing the $\pi^0$ yields as a function
of $\pt$ and centrality is now well established and is exactly the same one
used here to obtain the corresponding $\eta$ yields. Although the reconstruction
methods are identical, the p+p and d+Au analyses do not suffer from the
large particle multiplicity background that the Au+Au $\eta$ reconstruction faces,
and there are a few differences between Au+Au and the other studies. In the Au+Au case, the
applied analysis cuts (photon identification, invariant mass reconstruction,
and other cuts) are tighter than in the p+p and d+Au cases.  Additionally, in
order to deal with cluster overlap effects appropriately, the Au+Au analysis uses
a full GEANT~\cite{geant} Monte Carlo (MC) simulation, in which the simulated single $\eta$
are embedded into real events, rather than using a tuned fast MC simulation without embedding.
These differences are explained in separate subsections below.


\subsubsection{Photon reconstruction in EMCal}

Electromagnetic clusters are reconstructed in the EMCal sectors by finding
contiguous calorimeter towers with pulse heights above the ADC pedestal value.
The energy of each EMCal cluster is corrected for angular dependence
and non-linearity based on test-beam results and simulation. The
linearity correction for the two detector types is different, with the
PbGl having a stronger dependence on the energy. The correction
factors for a photon with a detected energy of 1 GeV (10 GeV) are 1.00
(0.95) for the PbSc and 1.05 (0.975) for the PbGl, respectively. The
PbGl calorimeter also shows a stronger variation of the measured
photon energy with the angle of incidence on the detector surface:
at $20^{\circ}$ the measured energy is reduced by 5\% compared to
perpendicular incidence ($0^{\circ}$), while in the PbSc this reduction is
only of the order of 2\%.\\

Since we are interested in high-$\pt$ $\eta$ production, only EMCal
clusters with energy above 1 GeV are selected for further analysis.
In addition, fiducial cuts excluding the edges of each EMCal sector
as well as an area of $3\times 3$ towers
around the towers that have been determined to be hot or dead,
were applied in order to exclude clusters with incorrectly reconstructed
energies.
Among the clusters passing the cuts, photon candidates are identified
by applying standard particle identification (PID) cuts based on
time-of-flight (TOF) and shower profile. Both cuts are applied to
reject slower and broader showers which are mostly of hadronic origin.
For the PbSc we require the measured cluster TOF to be $t_{\rm clust}
< L/c \pm 1.2$~ns where $L$ is the straight-line path from the
collision vertex to the reconstructed cluster centroid. For the
PbGl we require reconstructed clusters to have times, $t_{\rm
clust} < L/c \pm 2$~ns; the difference is due to the difference in
intrinsic timing resolution of the two calorimeter technologies.
Shower profile cuts are based on rejecting those clusters whose energy
deposition among the modules, and in particular in the most central
tower of the cluster, are not consistent, within a given $\chi^2$-test
limit, with the shower shape expected for electromagnetic showers
as parametrized from test-beam data~\cite{nim_emcal}.\\

In the most central Au+Au events, the EMCal typically detects
$\gtrsim$300 clusters, corresponding to a detector occupancy of $\sim$10\% in
terms of hit towers, resulting in a non-negligible
probability of particles making clusters which overlap. In order to
minimize the effects of cluster overlaps due to high multiplicity,
two methods are used to determine the cluster energy. First the
energy of each cluster in the PbSc calorimeter is determined from the
sum of all contiguous towers with deposited energy above a given
threshold ($E_{tower}$ = 15 MeV, typically). Alternatively, an
extrapolation (using a standard electromagnetic shower profile for an
event with zero background) from the measured core energy ({\it
ecore}) in the four central towers to the full cluster energy is used.
For the latter case, the {\it ecore} energy was computed from the
experimentally measured center of gravity, central shower energy, and
impact angle in the calorimeter using a parametrized shower profile
function obtained from electromagnetic showers measured in the beam
tests. Such an {\it ecore} energy represents an estimate of the true
energy of a photon impinging on the PbSc unbiased by background
contributions from other particles produced in the same event and
depositing energy in the neighborhood of a given cluster. The use of
{\it ecore} instead of the total cluster energy for photon
reconstruction helped considerably to minimize the effects of
cluster overlaps in central Au+Au collisions.


\subsubsection{Raw $\eta$ yield extraction (p+p and d+Au)}
\label{sec:reco_pp_dAu}

\begin{figure}[th]
\includegraphics[width=1.0\linewidth]{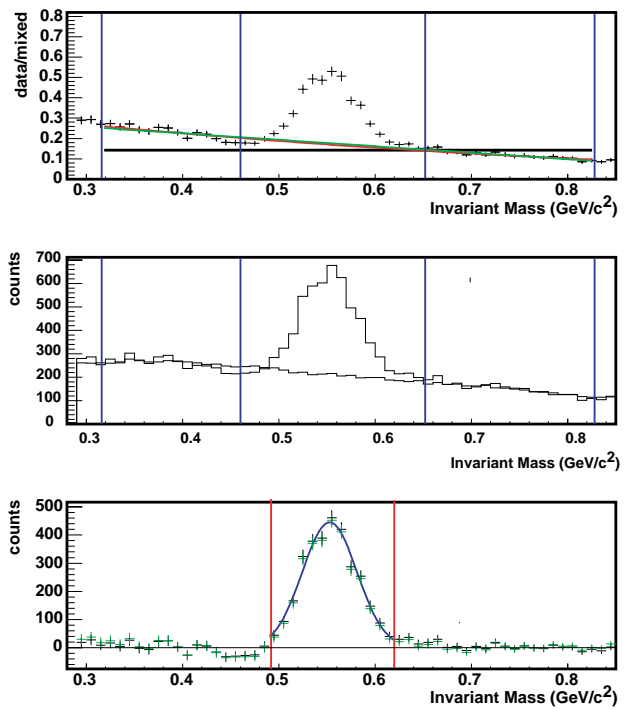}
\caption{(color online) Invariant mass distribution of photon candidate pairs measured in
p+p collisions for the default PID cuts with pair transverse momenta
4.0 GeV/$c < \pt <$ 4.5 GeV/$c$.
Top: Ratio of real and mixed event distributions, and background fits.
The red fit is used for the
background parametrization; the green fit for estimating the systematic
uncertainty. Middle: Real invariant mass spectrum and scaled background.
Bottom: Final distribution with the scaled background subtracted from
the real event distribution (black entries); the green entries result
from the background fit for estimating the systematic
error. Additionally, the peak is fitted with a Gaussian to get its mean and
sigma.}
\label{fig:invmass_pp}
\end{figure}

\begin{figure}[th]
\includegraphics[width=1.0\linewidth]{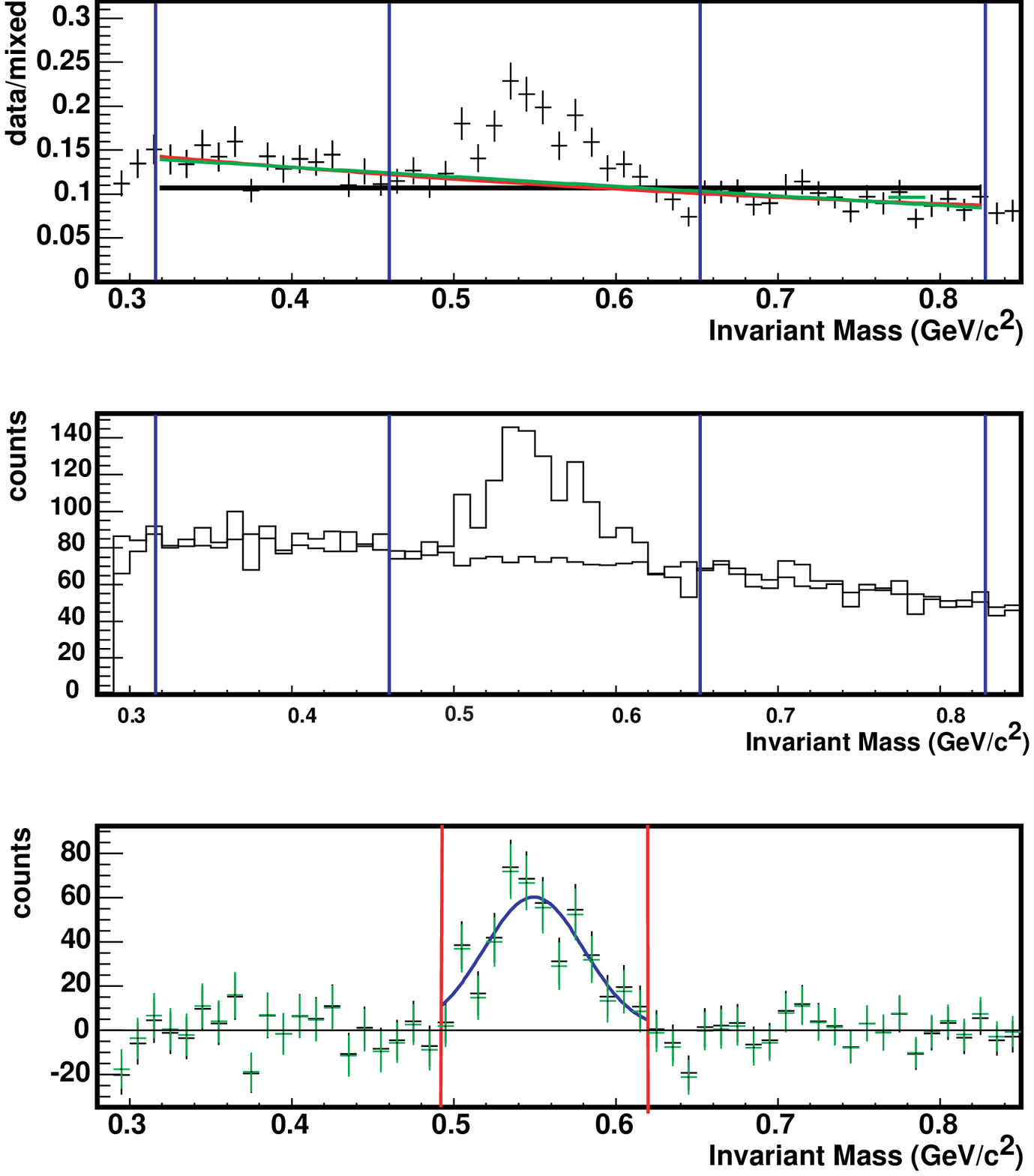}
\caption{(color online) Invariant mass distribution of photon candidate pairs measured in
minimum bias d+Au collisions for the default PID cuts with
pair transverse momenta 3.5 GeV/$c < \pt <$ 4.0 GeV/$c$.
Top: Ratio of real and mixed event distribution, and background fits
(the red fit is used for the
background parametrization, the green fit for estimating the systematic
uncertainty). Middle: Real invariant mass spectrum and scaled background.
Bottom: Final distribution with the scaled background subtracted from
the real event distribution (black entries). The green entries result
from the background fit for estimating the systematic
error. Additionally, the peak is fitted with a Gaussian to get its mean and
sigma.}
\label{fig:invmass_dau}
\end{figure}

The $\eta$ yields are obtained by an invariant mass analysis of photon
candidate pairs having asymmetries
$\alpha = |E_{\gamma_1}-E_{\gamma_2}|/(E_{\gamma_1}+E_{\gamma_2})<$ 0.7.
The cut on the asymmetry $\alpha$ reduces the background since high-$\pt$ combinatorial
pairs are strongly peaked near $\alpha = 1$ due to the steeply falling spectrum
of single photon candidates. The resulting invariant mass spectra obtained
for proton-proton and deuteron-gold collisions are shown in Figs.~\ref{fig:invmass_pp}
and~\ref{fig:invmass_dau} respectively for two typical $\pt$ bins.
A peak is seen at about $550$ MeV/$c^2$, the expected mass of the $\eta$ meson.
The measured peak position is modified by detector effects which lead to energy
smearing. The combined effects of the energy resolution of the detector, the steeply falling single photon
spectrum, and the finite size of the energy bins lead to a
smearing of the measured photon energies which widens the $\eta$ signal.
As a result the average peak position in the invariant mass spectra is about
9 MeV/$c^2$ larger than the nominal mass of the $\eta$ meson, an effect which
is well reproduced by the simulation.\\

The combinatorial background below the peak signal is estimated with the
event mixing method in which clusters from different events with similar event vertex
(and centrality class in d+Au) are combined to produce a ``background'' invariant
mass distribution. This background is normalized to the real invariant mass spectrum
and then subtracted from the invariant mass spectrum of the real events.
To estimate the normalization of the background, the distribution of the real events is
first divided by the mixed event distribution. This ratio is shown in the upper
panel of Figs.~\ref{fig:invmass_pp} and \ref{fig:invmass_dau}. The normalization function
is estimated by a fit in the region outside the peak. The spectrum, fitted to a
second-degree polynomial, is shown in the region denoted by the vertical lines
in the upper and the middle panels of Figs.~\ref{fig:invmass_pp}
and~\ref{fig:invmass_dau}. The final real event mass distribution after the
background subtraction is shown in the lower panel of
Figs.~\ref{fig:invmass_pp} and~\ref{fig:invmass_dau}.\\

The interval over which the background is adjusted is limited by two considerations:
 the expected $\eta$ peak position $m$ and width $\sigma$.
Both were estimated in a first analysis of the spectra and set to
$m = 556\;\mathrm{MeV}/c^2$ and $\sigma = 32\; \mathrm{MeV}/c^2$.
The background interval includes the region between $m-7.5\sigma$ and $m+8.5\sigma$
($320\; \mathrm{MeV}/c^2$ and $830\; \mathrm{MeV}/c^2$) excluding
the peak region $m \pm 3\sigma$
($460\; \mathrm{MeV}/c^2 < m_{inv} < 650\; \mathrm{MeV}/c^2$).
For higher transverse momenta, the background almost vanishes and
thus the estimation of the normalization by a fit leads to large
errors. Hence an alternative method is used where the fit function is
replaced by the ratio of the number of photon pairs in the normalization
region in the real and the mixed event distributions.



Finally, the total number of $\eta$ in a given $\pt$ bin is obtained
by integration of the invariant mass distribution within 3$\sigma$
around the $\eta$ peak position.
The statistical error of the peak extraction is estimated as done for
the $\pi^0$ and described in~\cite{ppg054}. The uncertainty of the background
parametrization is estimated by calculating the error on the ratio of the
integrals of the real and the mixed event distributions in the region of
the background fit. Above $\pt$ = 10 GeV/$c$ in p+p collisions, the mixed
event background does not work as expected as there are some entries in
the mixed event background but not in the $\eta$ region. For these cases,
the background was estimated by integrating the real event distribution
outside the peak (in the fit region mentioned above) and scaling this
to the $\eta$ integration region. This background is then subtracted
from the real event distribution to get the number of $\eta$.
The error in this case is estimated with $\sigma_{Sig}^2 = S + 2B$,
$S$ being the $\eta$ signal and $B$ being the background.
The integral of the invariant mass distribution after the background
subtraction is calculated in a reduced interval $m-2\sigma;m+2\sigma$
($492$ MeV/$c^2$ -- $620$ MeV/$c^2$).\\


\begin{figure}[th]
\includegraphics[width=1.0\linewidth]{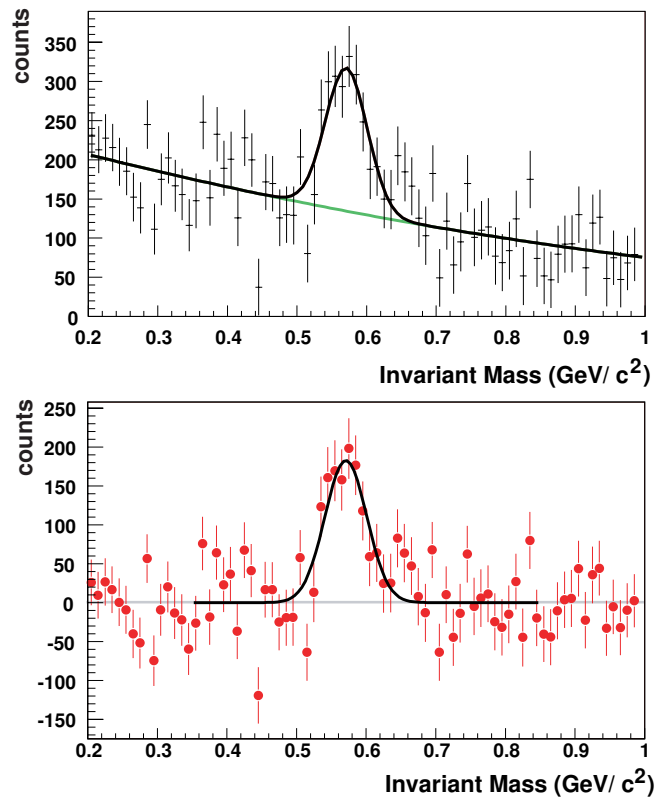}
\caption{Top: Invariant mass distribution of pairs of photon candidates measured
in minimum bias Au+Au with pair momenta $\pt$ = 3.5 - 4.5 GeV/$c$ around the $\eta$
mass fitted to a Gaussian plus exponential. Bottom: Final $\eta$ signal after 
mixed-event (and residual) background subtraction.}
\label{fig:invmass_auau}
\end{figure}

\subsubsection{Raw $\eta$ yield extraction (Au+Au)}
\label{sec:reco_auau}

The $\eta$ yields for Au+Au are determined by calculating the
invariant mass of photon pairs with asymmetries $\alpha = |E_{\gamma
1}-E_{\gamma 2}|/(E_{\gamma 1}+E_{\gamma 2})<$ 0.5, a value tighter
than that used for the p+p and d+Au cases in order to reduce the
larger uncorrelated background in Au+Au collisions, and binned in
$\pt$. The $\eta$ yield in each $p_{T}$ bin is determined by
integrating the background-subtracted $\gaga$ invariant mass
distribution around the $\eta$ peak. The combinatorial background is
obtained by combining uncorrelated photon pairs from different events
with similar centrality and vertex, and normalizing the distribution
in a region below ($m_{inv}$ = 400 -- 450 MeV/$c^2$) and above
($m_{inv}$ = 750 -- 1000 MeV/$c^2$) the $\eta$ mass peak (Fig.~\ref{fig:invmass_auau} top).
After the mixed background subtraction, the resulting distribution is
fitted to a Gaussian plus an exponential (or linear, see below) function to account
for the residual background -- more important at low $\pt$ -- not completely
removed by the event-mixing technique. 
The bottom plot in Fig.~\ref{fig:invmass_auau} depicts the $\eta$ signal after
mixed (and residual) background subtraction. To estimate the uncertainty in the
subtraction of the residual background, different pair asymmetries and an alternative
linear parametrization were used (see Section~\ref{sec:sys_error}).
The signal-to-background (S/B) ratio in peripheral (central) Au+Au collisions
is approximately 1.3 (1.5) and 0.05 (0.002) for the highest and lowest $\pt$,
respectively.  The signal-to-background ratio is comparable for central and
peripheral collisions at the highest $\pt$ because the spectrum in the central
data extends to higher $\pt$ than that in the peripheral.


The scaled mixed-event distribution is subtracted from the real-event distribution
to produce a statistical measure of the true  $\eta$ yield. The result
of such a subtraction procedure is shown in the bottom plot of
Fig.~\ref{fig:invmass_auau}. The raw  $\eta$ yield is obtained by
integrating the subtracted invariant mass distribution in a range
determined by the mean and the width of the $\eta$ peak and given by
$m_{\rm inv} \in \left[m_{\eta} - 2\sigma_{\eta},m_{\eta} +
2\sigma_{\eta}\right]$. Th analysis described above is applied
in bins of $\Delta \pt = 1$~GeV/$c$ for $\pt$ = 2 -- 4~GeV/$c$ and
$\Delta \pt = 2$~GeV/$c$ above. We cease attempting to extract $\eta$
yields at high $\pt$ when the number of pairs within the selected
(background-subtracted) $\eta$ mass window falls below 4 counts.




\subsubsection{Acceptance correction}

\begin{figure}
\includegraphics[width=1.0\linewidth]{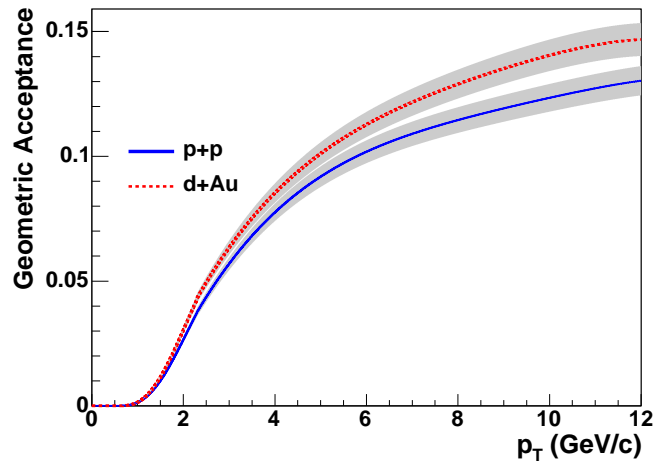}
\caption{Geometric acceptance (including dead channels) for the $\eta$ meson
as a function of $\pt$ measured in the EMCal in both d+Au (dashed curve) and  p+p (solid curve)
collisions at PHENIX in Run-3.}
\label{fig:acceptance}
\end{figure}

The geometric acceptance is evaluated using a fast Monte-Carlo (fastMC)
program based on routines from the JETSET library \cite{Sjo94}
to simulate the $\eta\rightarrow\gaga$~ 
decays and determine the geometric acceptance of the calorimeter
for the decaying photons. 
The acceptance correction accounts for the fraction of produced
$\eta$ mesons whose decay photons will not actually hit the detector due
to the finite solid angle covered by the detector. A decay photon will be accepted by the EMCal
in the fastMC when it hits the active surface of the detector covering the
pseudorapidity range $-0.35 < \eta < 0.35$ (computed using a realistic distribution
of event vertices within $|z|<$ 30 cm) and 2$\times$90$^\circ$ in azimuth.
The acceptance shows a strong dependence on the transverse momentum because
the opening angle of the decay photons decreases with increasing $\pt$.
Thus, the probability that both decay photons hit the detector decreases
for small values of $\pt$.  The acceptance for p+p and d+Au
collisions is shown in Fig.~\ref{fig:acceptance}. The acceptance is
influenced by the geometry of the whole detector as well as by dead and hot modules in the p+p
and d+Au cases (for Au+Au the efficiency losses
due to dead and hot modules are computed from the full GEANT3 simulation
plus ``embedding'' and are accounted for in the {\it efficiency} loss
correction). Due to a different number of masked out modules, the
acceptance is not exactly the same in p+p and d+Au collisions.

\subsubsection{Efficiency corrections of the raw $\eta$ yields (p+p, d+Au)}
\label{sec:MC}

\paragraph{Reconstruction efficiency correction}

The reconstruction efficiency takes into account that the measured
$\eta$ spectrum in the detector is different from the real physical spectrum,
i.e.~the reconstruction efficiency is defined as the ratio between the
output and the input $\eta$ spectra:
\begin{eqnarray}
\varepsilon (\pt) =
\frac{dN_{\eta}/d\pt|_{output}}{dN_{\eta}/d\pt|_{input}}
\label{eq:efficiency_def}
\end{eqnarray}
as obtained using the fastMC, which parametrizes all
the detector effects on the input spectrum (EMCal energy and
position resolution, efficiency losses due to $\gamma$ identification
cuts and $\gaga$~ reconstruction procedure, etc.). A realistic input
$\eta$ spectrum $dN/d\pt|_{input}$ is used as an initial spectrum for
the efficiency calculation and an iterative procedure is performed, in
which the corrected output spectrum is used as the input spectrum of
the next iteration. To simulate detector effects, the smeared
energies and hit positions of the decay photons are parametrized in
the fastMC. The energy smearing has a constant and an energy-dependent 
term and follows the functional form
\begin{eqnarray}
\sigma_E/E = \frac{A}{\sqrt{E/\mbox{GeV}}}\oplus B\mbox{ .}
\label{eq:energysmearing}
\end{eqnarray}

The parameters for Eq.~(\ref{eq:energysmearing}) are given in
Table~\ref{tab:energysmearing} for the different collision systems and
the two EMCal detector types. The initial values have been
taken from the detector response obtained in the beam
tests~\cite{nim_emcal} and re-tuned for real run conditions
in previous $\pi^0$ analyses~\cite{ppg024,phnx_spin_pi0}.
During the cross-checks between simulated and real data it was found that
the energy scale of the EMCal was slightly shifted compared to the parametrized results.
Since the energy scale is estimated experimentally by fitting the location of the
$\pi^0$ peak, and the position of the peak is also affected by
reconstructed secondary neutral pions from $K^0_s$ decays which themselves
decay off vertex, an additional correction is applied in the fastMC
shifting the energy scale by 0.7\%.
After this correction, the position and the width of the simulated $\eta$ peaks
are confirmed to be consistent with the position and the width measured
in the data for all $\pt$ bins.

\begin{table}[ht]
\caption{\label{tab:energysmearing}
Parameters for energy smearing, Eq.~(\protect\ref{eq:energysmearing}),
as used in the fastMC for the different EMCal detector types and the different
collision systems.}
\begin{ruledtabular}\begin{tabular}{llcc}
collision & detector & energy-dependent & constant \\
system    & type     & term (A)         & term (B) \\ \hline
p+p       & PbGl     & 0.085                     & 0.059\\
p+p       & PbSc     & 0.082                     & 0.050\\ 
d+Au      & PbGl     & 0.085                     & 0.059\\
d+Au      & PbSc     & 0.082                     & 0.050\\ 
\end{tabular}\end{ruledtabular}
\end{table}

The efficiency correction also takes into account the different cuts
used for particle identification. The simulation must consider the
loss of photons and thus of $\eta$ due to the applied shower shape (or ``dispersion'')
and energy threshold cuts. The effect of the dispersion cut is estimated by a
comparison of uncorrected spectra without a PID cut with the spectra obtained
with the different PID cuts. The spectra are obtained with a sharp asymmetry ($\alpha$)
cut and as a function of $(E_1+E_2)/2$. The resulting loss of $\eta$ is translated
into a photon loss probability, which is then used in the simulation. The energy
cut is reproduced by rejecting photon hits according to an energy-dependent survival
probability in the simulation. Finally, the simulation reconstructs the
invariant mass and the transverse momentum of the $\eta$ from the reconstructed
(smeared) information. Only particles inside the interval used for the integration
of the real peak are accepted. The overall $\eta$ efficiency losses
obtained by this method are of the order of $\varepsilon_{\eta\rightarrow\gaga}$ =
76\% $\pm$ 3\% (dominated by the asymmetry cut and the invariant mass yield
extraction procedure) and are flat within 1-2\% in the whole $\pt$ range measured
for both (p+p and d+Au) colliding systems.\\

\paragraph{Photon conversion correction}

Some of the produced $\eta$ are not reconstructed due to conversions of
one or both decay photons in the inner regions of the PHENIX detector.
Such an effect is not included in the fastMC and is computed independently
using a full simulation of the detector including a realistic
description of the material in front of the EMCal.
The correction factors obtained from this analysis are $1.067 \pm 0.003$ for PbSc
West, $1.052 \pm 0.004$ for PbSc East and $1.076 \pm 0.005$ for PbGl,
as the material between the collision vertex and the EMCal is different in
the east and the west arm and between PbGl and PbSc.\\



\subsubsection{Efficiency corrections of the raw $\eta$ yields (Au+Au)}

In the Au+Au case, the detection efficiency is determined using a full
PISA (PHENIX Integrated Simulation Application) GEANT3-based
Monte Carlo (MC) program of the PHENIX detector in order to simulate the
complete response of the calorimeter to single $\eta$ decays.
The nominal energy resolution was adjusted in the simulation by adding an
additional $\pt$-independent energy smearing of $\sim3\%$ for each PbSc tower.
The shape, position, and width of the $\eta$ peak measured for all $\pt$ and
centralities were thus well reproduced by the simulated data.
The data from each simulated $\eta$ is embedded into real Au+Au events
and the efficiency for detecting the embedded $\eta$ is evaluated analyzing
the merged events with the same analysis cuts used to obtain the real
yields. Using this technique we determine efficiency corrections that account
not only for the energy resolution and position resolution of the calorimeter,
but also for the losses due to overlapping clusters in a real Au+Au
event environment.  The embedding also permits a precise determination
of the effect of edge cuts and bad modules. Though these effects can
in principle be considered as geometric acceptance corrections (as done in the
p+p and d+Au analyses), they depend not only on the geometry but also on the
energy deposition of an electromagnetic shower in the different calorimeter towers.
Lastly, in the full-simulation plus embedding procedure we additionally
have control over the effects of photon conversions, as the GEANT simulation
considers the material in front of the EMCal and the information
whether a decay-photon converts is kept for evaluation in the efficiency
determination.\\

The input $\eta$ spectrum embedded in real events is weighted to match a functional
form fit to the measured $\eta$ spectrum so that the correct folding of the
$\eta$ spectrum with the resolution is obtained.  This procedure is iterated,
with the fit of the $\pt$ dependence of the input weights adjusted as the
estimate of the efficiency correction improves, until the procedure converges
within the nearly \pt-independent statistical error of the embedded sample,
approximately 3\%. The final overall $\eta$ yield reconstruction efficiency
correction factor was $\sim$3 with a centrality dependence of $\lesssim$20\%.
The losses were dominated by fiducial and asymmetry cuts.




\subsection{$\eta\rightarrow\pi^0\pi^+\pi^-$ Reconstruction}
\label{sec:eta_3pi_reco}

\subsubsection{Raw $\eta$ yield extraction}

The second mode of $\eta$-meson reconstruction in PHENIX is via the
three-body decay channel $\eta\rightarrow\pi^0\pi^+\pi^-$ with
branching ratio BR = 22.6\% $\pm$ 0.4\%. This mode has been used for
the p+p and d+Au data, but not for the Au+Au where the large detector
occupancy makes the signal very difficult to extract. Reconstruction
starts with identifying the $\pi^0$ candidates among the pairs of
EMCal $\gamma$-clusters with energy $E_\gamma >$ 0.2 GeV in the same
way described in the previous section for the direct
$\eta\rightarrow\gaga$~ channel. The mass of a candidate is required
to be within two standard deviations from the peak position of $\pi^0$.
The peak position and its width are determined by the $\pi^0$ decay
kinematics and EMCal resolution for each of the clusters and its
position. These parameters were found to be consistent with the
expected values. Selected $\pi^0$\ candidates with transverse
momentum $\pt >$ 1.0 GeV/$c$ are assigned the exact mass of the meson
and measured $\pt$ of the pair. These candidates are further combined
into triplets with positive and negative particle tracks measured by
DC and PC1 to have momentum in the range 0.2 GeV/$c < \pt <$ 4.0
GeV/$c$. No particle identification was used
on the charged tracks.\\

\begin{figure}
\includegraphics[width=1.0\linewidth]{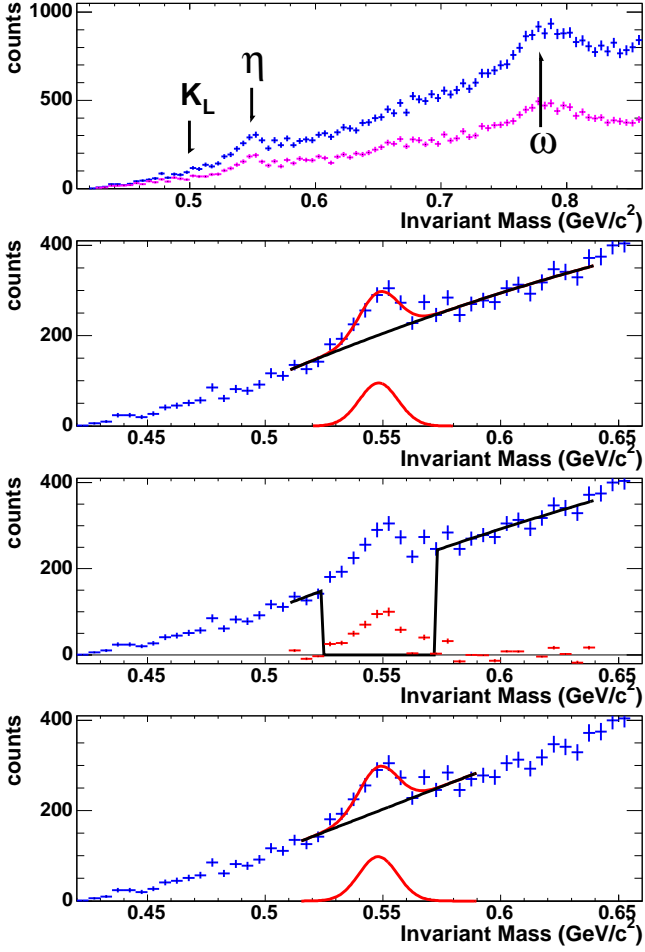}
\caption{Invariant mass distribution of pion triplets ($\pi^{0}\pi^{+}\pi^{-}$)
measured in p+p collisions at $\sqrt{s}$ = 200 GeV in the range $m_{inv}$ = 0 - 0.9
GeV/$c^2$ (top panel) showing $\eta$- and $\omega$-meson peaks.  Lower panels showing
$\eta$-mass region $m_{inv}$ = 0.42 - 0.66 GeV/$c^2$ demonstrate three different methods
of the extraction of the raw yields.  See text for a detailed explanation of each
method.}
\label{fig:invmass_three}
\end{figure}

In order to extract the raw $\eta$ yields the mixed-event subtraction
technique was not used in this case because it does not adequately
reproduce the shape of the background in the real events. The most
important physical reason for this is that there are a significant
number of correlated tracks among the $\pi^+\pi^-$ pairs coming from
various heavier particle decays. The yield extraction was done by
simultaneous fitting of the peak and the background in the adjacent
region. The characteristic peak in the three-particle mass
distribution is shown in the top panel of
Fig.~\ref{fig:invmass_three}. The position of the peak is consistent
with the nominal mass of the $\eta$-meson within the statistical
error of the fit shown in the figure. The measured 8~MeV/$c^2$ width
of the peak is narrower than in the $\eta\rightarrow\gaga$~ decay
channel. Unlike the $\gaga$~ channel where the full width of the peak
is defined by the EMCal resolution alone, in $\pi^{0}\pi^{+}\pi^{-}$
only 1/4 of the measured mass is derived from an EMCal-based measurement. Given the relatively low $\pt$ of the
decay products, tracking has better resolution than the
calorimeter. These two effects result in higher accuracy of the mass
measurement and smaller width of the peak compared to
the $\eta\rightarrow\gaga$~ analyses.\\

The raw yield numbers were extracted by simultaneously fitting the
signal with a Gaussian function and the background to a quadratic function.
The fit was limited to the mass window of 510~MeV/$c^{2}
< m_{3\pi} <$~640~MeV/$c^{2}$. The lower limit is chosen to avoid the
region where the $K_{L}^{0}\rightarrow\pi^{0}\pi^{+}\pi^{-}$ decay (branching
ratio BR = 13\%) yields an additional signal at and above
$m_{K_{L}^{0}}$ = 498~MeV/$c^{2}$. The upper limit is chosen at a safe distance
from the $\omega\rightarrow\pi^{0}\pi^{+}\pi^{-}$ (BR = 89.1\%) peak at
782~MeV/$c^{2}$ with a width of 20-25~MeV/$c^{2}$.
An example of the fit is shown in the second panel in Fig.~\ref{fig:invmass_three}.
We also compared the result of such a combined signal+background fit
with separate fitting of the background. For that, the region under the peak,
530~MeV/$c^{2} < m_{3\pi} <$~570~MeV/$c^{2}$, was rejected from the fit and
the background was approximated by the quadratic function. The function was
interpolated and subtracted from the histogram in Fig.~\ref{fig:invmass_three}
(third panel). The histogram counts in the region initially rejected were summed
up to calculate the yield.\\

In addition, simultaneous fitting was done in the restricted
window below $m_{3\pi} <$580~MeV/$c^{2}$, with the background
approximated by a linear function. The same three fits were repeated applying
an additional condition in the analysis. Each charged track was required to
match a hit in PC3 or in the EMCal in case a track missed the active area of PC3.
The resultant invariant mass spectrum is shown by the lower curve in the top panel of Fig.~\ref{fig:invmass_three}.
The amplitude of the signal is reduced by about a factor of two because many
tracks fall outside the acceptance of these two detectors, but the background
is also reduced and, more importantly, modified in its shape. The overall
significance of the results with and without matching is approximately the same.
Signal loss due to matching can be corrected with the simulation with small systematic
uncertainty and the results can be compared to deduce the accuracy of the yield
extraction procedure. Thus, for each $\pt$ point we obtain six statistically
correlated measurements of the raw yields. The first measurement with its
statistical error is used in further analysis and the variance of the six
measurements provides the estimate of the systematic errors of the
yield extraction.


\subsubsection{Acceptance and efficiency corrections of the raw $\eta$ yields}

Similar corrections as described for the $\gaga$~ decay channel need
to be applied to the $\eta \rightarrow \pi^{0}\pi^{+}\pi^{-}$ raw
yields. However, for the 3-pion analysis, we use the full
detector simulation and both corrections, namely the acceptance and
the efficiency corrections, are computed at the same time. A MC hadron decay generator was used to produce
initial $\eta$-mesons with a $\pt$ distribution providing
satisfactory statistical significance in all bins after acceptance
and trigger losses. The full GEANT-based PISA simulation was updated
with the three-body decay of the $\eta$-meson and used to decay
$\eta$-mesons. PISA also performs the full simulation of the PHENIX
detector and generates the response of all its subsystems up to the
electronics-signal level, which was then processed by standard PHENIX
reconstruction software. Special attention was paid to verify that
the simulation code represented the real configuration of the
detector, and that the $\pi^{0}$ peak parameters in the real data and
simulation were consistent with each other. The reconstruction of the
simulated data was carried out using the
same steps and tools as the real data.



\begin{figure}
\includegraphics[width=1.0\linewidth]{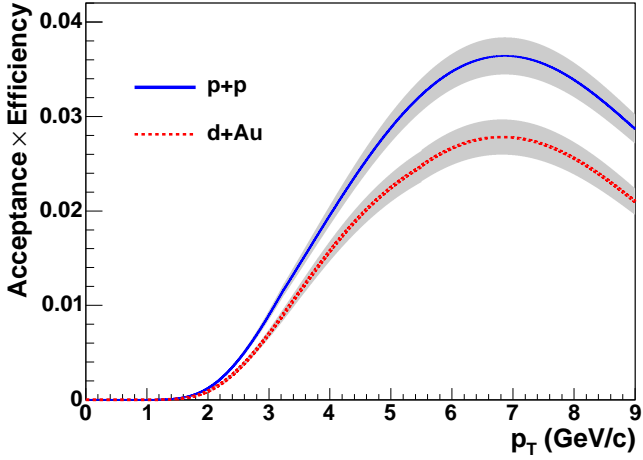}
\caption{Acceptance$\times$efficiency for the $\eta \rightarrow \pi^{0}\pi^{+}\pi^{-}$
as a function of $\pt$ in both d+Au (dashed curve) and p+p (solid curve) collisions at PHENIX in Run-3.}
\label{fig:3pi-acceptance}
\end{figure}

Figure~\ref{fig:3pi-acceptance} shows the combined
efficiency$\times$acceptance as a function of $\pt$ for the
three-pion decay analysis. In order to compare this with the $\gaga$~
decay channel reconstruction efficiency one needs to multiply the
acceptance curve shown in Fig.~\ref{fig:acceptance} with the obtained
$\varepsilon_{\eta\rightarrow\gaga}$ = 76\% $\pm$ 3\% overall
efficiency loss. The three-body decay combined acceptance is
significantly lower than the acceptance of the $\gaga$~ decay
channel. With comparable branching ratios of the two modes the
resulting statistics in the three-body decay mode is expected to be
smaller. The decrease of the efficiency at high $\pt$ is due to 
the momentum cut on the $\pi^{\pm}$ to be below 4.0~GeV/$c$. Above
that threshold the track sample is contaminated by products of
in-flight decays of long-lived particles with mismeasured momentum.


\subsubsection{Phase-space density correction}

\begin{figure}[hbt]
\includegraphics[width=1.0\linewidth]{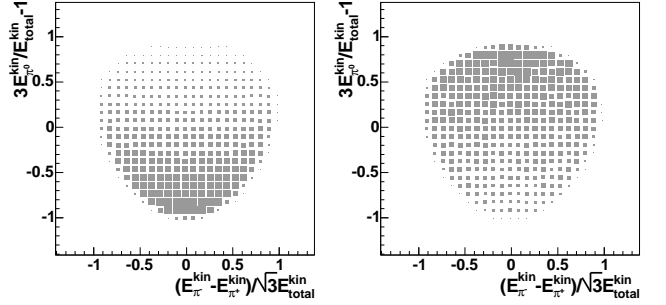}
\caption{The phase-space density of the $\eta \rightarrow
\pi^{0}\pi^{+}\pi^{-}$ decay~\protect\cite{eta-phase1,eta-phase2,eta-phase3}
(left panel). PHENIX reconstruction and trigger efficiencies in p+p
for the  $\eta \rightarrow \pi^{0}\pi^{+}\pi^{-}$ decay (right panel).}
\label{fig:dalitz}
\end{figure}

\begin{figure}[hbt]
\includegraphics[width=1.0\linewidth]{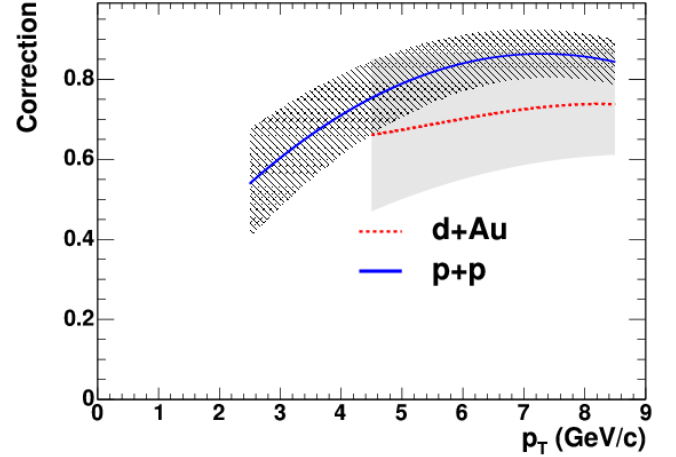}
\caption{Phase-space density correction for p+p (solid curve) and d+Au (dashed curve) event
samples as a function of $p_{T}$.}
\label{fig:dalitz_correction}
\end{figure}

The $\eta \rightarrow \pi^{0}\pi^{+}\pi^{-}$ decay channel required an
additional correction to take into account the uneven distribution of the momenta
of the three pions within the kinematically allowed region. Such a distribution,
taken from~\cite{eta-phase1,eta-phase2,eta-phase3}, is shown in the left panel
of Fig.~\ref{fig:dalitz}. The vertical axis is the fraction of kinetic energy
carried by the $\pi^0$ in the $\eta$-meson rest frame. The horizontal axis shows
the difference between kinetic energies of $\pi^{-}$ and $\pi^{+}$
divided by the total in the same system.
The left plot of Fig.~\ref{fig:dalitz} shows that on average the $\pi^0$-meson
carries less kinetic energy, and thus momentum, than the two charged
$\pi$-mesons. The right panel shows the PHENIX reconstruction efficiency
including geometrical acceptance, high-$\pt$ trigger efficiency (see next Section),
and analysis cuts. The latter two effectively select higher momentum
$\pi^0$ and lower momentum $\pi^\pm$ in the lab frame.
In the $\eta$-meson rest frame these translate into the effect opposite of
what is shown in the left panel. In order to correct for that we used the
following approach.
The uniform distribution of the phase-space density produced by the simulated 
event generator was weighted according to the known probabilities of the $\pi$-meson
momenta to be observed in the $\eta$-meson decay. The corresponding
correction was deduced by comparing the reconstruction efficiencies with and
without applying weights. The systematic uncertainties associated with the
measurement of the phase-space density accuracy were thus obtained.
This correction is shown in Fig.~\ref{fig:dalitz_correction}.
The correction factor is calculated in the range where data is available.

\subsection{Trigger Corrections and Absolute Cross-Section Normalization}

\subsubsection{Minimum-bias trigger efficiency}

The minimum bias trigger does not detect every collision; only a
certain fraction $\varepsilon_{trig}$ of the inelastic collisions and
a fraction $\varepsilon_{\eta}$ of the $\eta$ mesons can be observed.
The spectra have to be corrected for both of these effects. The
correction factors $\varepsilon_{trig}/\varepsilon_{\eta}$, determined
in~\cite{ppg036} for d+Au collisions, are shown in Table~\ref{tab:minbias_trig}.
In the case of p+p collisions, as well as in 

\newpage

\begin{table}[th]
\caption{\label{tab:minbias_trig}
Correction factors ($\varepsilon_{trig}/\varepsilon_{\eta}$),
  due to the efficiency of the MB
  trigger for different d+Au centralities~\protect\cite{ppg036}.}
\begin{ruledtabular}\begin{tabular}{ll}
collision  system         & correction factor \\ \hline
d+Au 0-20\% central         & 0.95 \\
d+Au 20-40\% semicentral    & 0.99\\
d+Au 40-60\% semiperipheral \hspace{5mm}& 1.03 \\
d+Au 60-88\% peripheral     & 1.04 \\ 
\end{tabular}\end{ruledtabular}
\end{table}

\noindent MB d+Au
collisions, one can directly determine the inelastic $\eta$ cross
section. Therefore, one does not apply the correction factors mentioned
above but rather multiplies the spectra by the total cross section observed 
by the BBC, found to be 23.0 mb $\pm$ 9.7\% in Run-3 p+p
collisions and 1.99 b $\pm$ 0.10 b in Run-3 d+Au
collisions~\cite{ppg038}. An additional correction has to be applied for
the bias of the BBC to high-$\pt$ $\eta$. It is found to be 0.79 for
p+p~\cite{ppg044} and 0.94 for d+Au~\cite{ppg038} collisions.\\

\begin{figure}[b]
\includegraphics[width=0.9\linewidth]{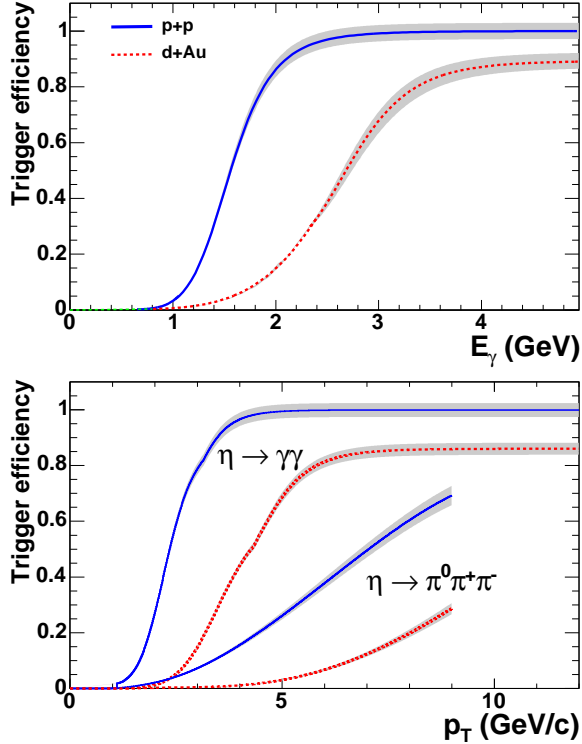}
\caption{Simulation result for the efficiency of the Gamma trigger in
d+Au collisions. The Gamma trigger efficiency for a single $\gamma$
is shown in the top panel for p+p (solid curve) and d+Au (dashed
curve). The gray band is the error of the measurement. The lower
panel shows the recalculated trigger efficiency for $\gaga$~ and
$\pi^{0}\pi^{+}\pi^{-}$ channels for both collision systems.}
\label{fig:trigger}
\end{figure}


\subsubsection{High-$\pt$ Gamma-trigger efficiency}
\label{sec:hipt_trigger}

The efficiency of the high-$\pt$ trigger has to be studied as well
to get $\eta$ spectra for the Gamma-triggered data at high
transverse momenta, as previously performed for PHENIX $\pi^0$
analyses~\cite{ppg024,ppg054}.  The Gamma triggers in
PHENIX are implemented by adding together amplitudes in $4\times4$
adjacent EMCal towers during data taking and comparing them to a
pre-set threshold.  In the case of p+p the threshold was set to
correspond to $E_\gamma$ = 1.5 GeV while for d+Au it was set at
$E_\gamma$ = 3.5 GeV. In the case of Au+Au, triggering was performed
by a LVL2 software algorithm run over the MB-triggered events 
during data taking, such that the number of rejected
minimum bias events were recorded.  This allowed two different threshold
triggers to be employed based on event centrality in Au+Au:
$E_\gamma$ = 1.5 GeV for the 60-92\% peripheral sample and $E_\gamma$ = 3.5 GeV
for the more central event selections.\\

\begin{figure}[b]
\centering
\includegraphics[width=1.0\linewidth]{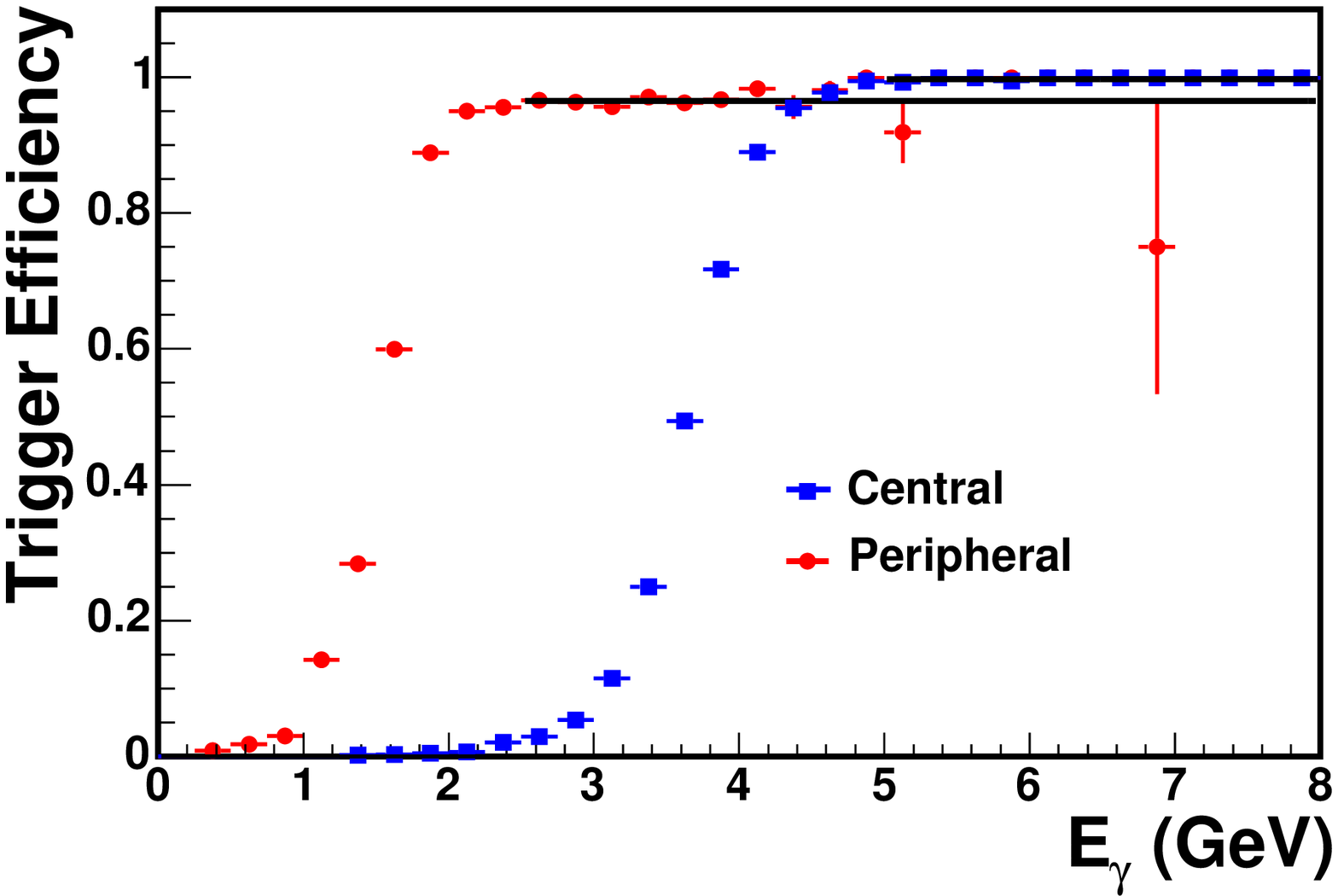}
\includegraphics[width=1.0\linewidth]{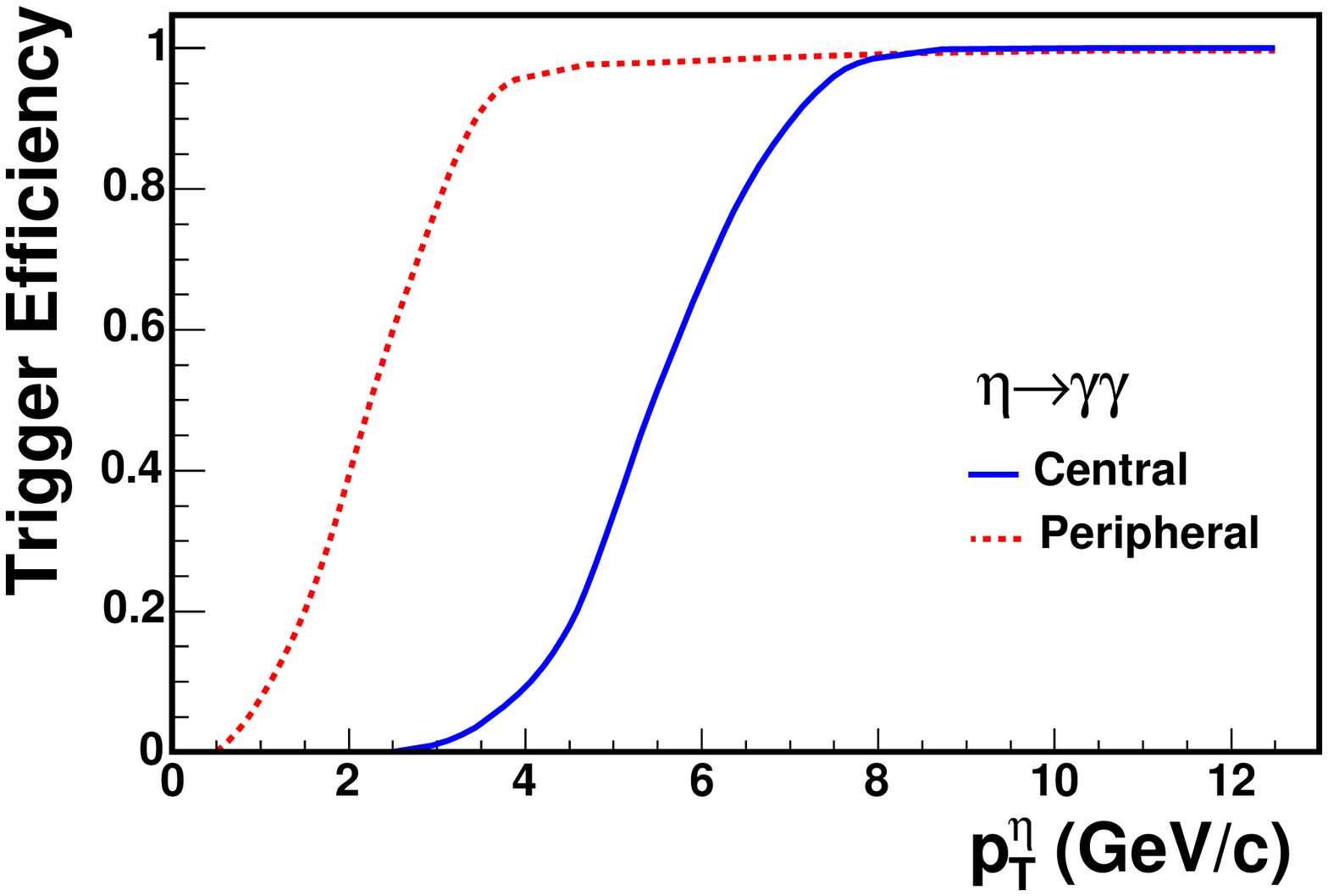}
\caption{(color online) Level-2 trigger photon (top) and $\eta$ (bottom)
efficiencies for central Au+Au ($E_\gamma$ = 3.5 GeV threshold) and
peripheral Au+Au ($E_\gamma$ = 1.5 GeV) collisions as in
Fig.~\ref{fig:trigger}. For deriving the $\eta$ efficiencies, the
histograms in the top panel were directly used, with the black lines
denoting constant fits to the above-threshold asymptotic value, at
$>$99.7\% for the central trigger.  The central (peripheral) LVL2
$\eta$ sample was used only above $p_T >$ 5 (2) GeV/c.}
\label{fig:trigger2}
\end{figure}

The trigger efficiency curves versus the energy of a single photon
for two different threshold settings used in p+p and d+Au
collisions are shown in the top panel of Fig.~\ref{fig:trigger}.
Based upon these curves, the 2-$\gamma$ efficiency is calculated as
for the previous $\pi^0$ analyses~\cite{ppg024,ppg054} using
the fastMC calculation.  For this calculation, the single-photon
trigger turn-on curve is represented by an integrated Gaussian for
the d+Au analysis and by the integrated sum of two Gaussians for
the p+p analysis.  In the case of the Au+Au LVL2 triggers, the
high-statistics measurement of the single-photon efficiency, which
for the central trigger reaches $\sim$100\% above threshold, is used
itself as shown in the top panel of Fig.~\ref{fig:trigger2}.  The
derived $\pi^0$ efficiency is checked by comparing the ratio of the
number of $\pi^0$ in MB events that carry the trigger
flag to the number of $\pi^0$ in all MB events.  In this
way the normalization of the LVL2 data sample relative to the MB
data sample is confirmed to be accurate to $2\%$.\\

In the same way we determine the $\eta \rightarrow \gamma\gamma$
trigger efficiency which is shown in the lower panel of
Figs.~\ref{fig:trigger} and \ref{fig:trigger2}. In the $\eta
\rightarrow \pi^{0}\pi^{+}\pi^{-}$ decay channel where the
statistics is very limited, we use the measured single-photon
trigger efficiency curves shown in the top panel and full detector
MC to determine the efficiency of the trigger. The derived curves
for p+p and d+Au are also shown in the lower panel of
Fig.~\ref{fig:trigger}. One can see that the trigger efficiency
plateaus at a $p_{T}$ of the $\eta$ about twice the energy of the
threshold in the case of the $\gamma\gamma$ decay channel, but in the
three-body decay mode where the trigger can only be fired by one of the
$\gamma$ from $\pi^{0} \rightarrow \gamma\gamma$, it requires the $p_{T}$
of the $\eta$ to be approximately 4 times the threshold. In
central Au+Au the $\eta$ efficiency reaches 50\% ($\sim$100\%)
for $\eta$ above $\pt = 5\,(7-8)\;\mbox{GeV}/c$, as shown
in Fig.~\ref{fig:trigger2}, bottom panel.  The LVL2 data were used
only for $p_T$ regions where the trigger had better than $\sim$ 50\%
$\eta$ efficiency: $p_T >$ 5 GeV/c for the central trigger and $p_T
>$ 2 GeV/c for peripheral.


\subsubsection{Cross-section normalization}


The invariant cross-sections for $\eta$ production as a function of $\pt$ in MB p+p and
d+Au collisions are obtained from the measured number of counts in each $\pt$ bin via
\begin{equation}
E\frac{d^3\sigma}{d^3p} \equiv \frac{1}{2\pi \pt N_{\rm evt}}
\frac{1}{\mathcal{L}}\, \frac{1}{BR} \,\frac{1}{Acc(\pt) \, \varepsilon(\pt) \, \varepsilon_{\rm trig}(\pt)}
\frac{N(\Delta \pt)}{\Delta \pt \, \Delta y},
\label{eq:sigma_eta}
\end{equation}
where $Acc$, $\varepsilon$, and $\varepsilon_{\rm trig}$ are the
acceptance, reconstruction efficiency, and trigger efficiency, respectively,
determined in the previous Section; $BR$ = 0.3943 $\pm$ 0.0026 is the
known $\gaga$~ decay branching ratio of the $\eta$ meson, and
$\mathcal{L}$ is the integrated luminosity obtained using the
absolute inelastic cross-section normalization (see
Section~\ref{sec:event_selection}). The invariant {\it yields} as a
function of $\pt$ for a given bin in collision centrality in d+Au and
Au+Au collisions are obtained via

\begin{widetext}

\begin{equation}
 \frac{1}{2\pi \pt} \frac{d^2N_{\rm cent}}{d\pt dy} \equiv \frac{1}{2\pi \pt N^{\rm event}_{\rm cent}}
\frac{1}{BR} \,\frac{1}{Acc(\pt) \, \varepsilon(\pt, {\rm cent}) \, \varepsilon_{\rm trig}(\pt)}
\frac{N(\Delta \pt, {\rm cent})}{\Delta \pt \, \Delta y}.
\label{eq:invyld}
\end{equation}

\end{widetext}

A final bin-shift correction is needed to take into account the fact that the data points
of the $\eta$ spectra are plotted at the center of each given $\pt$ interval (bins whose
width is as large as $\Delta \pt$ = 2 GeV/$c$), which, due to the exponentially falling spectrum,
does not represent the true physical value of the yield in the interval~\cite{lafferty94}.
Usually, either the correction is applied displacing the $x$-values horizontally (i.e.
the center of the $\pt$ bin is decreased) keeping their $y$-value, or the $y$-values are
moved vertically (i.e.~the yields are decreased) keeping the $\pt$-values at the center
of the bin. The second method (yield correction) is preferred here because it facilitates
taking bin-to-bin $\pt$ ratios of spectra (with slightly different shapes) from different
collisions systems. The net effect of this recipe is a small (few \%) shift downwards of
the invariant $\eta$ yields in each $\pt$ bin.



\subsection{Systematic Uncertainties}
\label{sec:sys_error}

\subsubsection{$\eta\rightarrow\gaga$~ analysis (p+p and d+Au)}

All systematic errors for the p+p and the d+Au analysis are summarized
in Table~\ref{tab:errtable}. Hereafter, the errors are categorized by type:

({\bf A}) point-to-point error uncorrelated between $\pt$ bins,

({\bf B}) $\pt$ correlated, all points move in the same direction
but not by the same factor,

({\bf C}) an overall normalization error in which all points
move by the same factor independent of $\pt$.

The cross section measurement of the MB trigger has a type-C uncertainty of
9.7\% in p+p and 5.2\% in d+Au. All other systematic errors are of
type B, i.e.~they are $p_{T}$ correlated.

\begin{table*}[th]
\caption{\label{tab:errtable}
Systematic errors of the $\eta$ measurement in p+p and
d+Au (Run-3) for different $\pt$ bins. The error of the peak extraction has a very
steep slope at low $p_{T}$.}
\begin{ruledtabular}\begin{tabular}{lccccc}
error source          & $\pt$-indep.& 3 GeV/$c$ & 5 GeV/$c$ & 10 GeV/$c$ & type \\ \hline
peak extraction      & & 14.5\% (p+p), 9.5\% (d+Au) & 6\% & 6\% & B \\
geometric acceptance & & 4.5\% & 4.5\% & 4.5\% & B \\
$\eta$ reconstruction efficiency & &  1.3\% & 2.3\% & 3.6\% & B \\
global energy scale & & 5.5\% & 7.0\% & 8.4\% & B \\
energy scale linearity & & 1.5\% & 0.4\% & 4.3\% & B \\
Gamma-trigger efficiency & &  9\% (p+p), - (d+Au) & 0\% (p+p), 2.5\% (d+Au) & 0\% & B \\
conversion correction & 2.0\%  & &  & & B \\ \hline
absolute cross-section normalization & 9.7\% (p+p), 5.2\% (d+Au) & &  & & C \\ 
\end{tabular}\end{ruledtabular}

\caption{\label{tab:sys_errors}
List of systematic uncertainties in the PbSc $\eta$
measurement in Au+Au collisions (Run-2). Ranges generally correspond
to uncertainties from the lower $p_T$ to the higher $p_T$ values of the measurement.}
\begin{ruledtabular}\begin{tabular}{lccc}
error source            & ~percent error &  type \\ \hline
raw yield (peak) extraction (point-to-point) \hspace{5mm}  & 0-31\%  & A \\
raw yield (peak) extraction ($\pt$ correlated)\hspace{5mm} & 10-20\%  & B \\
energy scale            & 3-8\% & B \\
PID cuts                & 8\%   & A \\
geometric acceptance    & 4-2\% & B \\
trigger efficiency      & 5-2\% & B \\
reconstruction efficiency   & 2\% & A
\\
\end{tabular}\end{ruledtabular}

\caption{\label{tab:errtable_3pi} Systematic errors of the $\eta\rightarrow\pi^{0}\pi^{+}\pi^{-}$
measurement in p+p and d+Au collisions (Run-3). The first number corresponds to the p+p data
and the number in parentheses to d+Au in cases where it is different from p+p.}
\begin{ruledtabular}\begin{tabular}{lcccccccc}
error source      & $\pt$ indep. & 3 GeV/$c$ & 4 GeV/$c$ & 5 GeV/$c$ & 6 GeV/$c$ & 7 GeV/$c$ & 8 GeV/$c$& type\\ \hline
EMCal geometrical acceptance     &  4\% (4\%) & & & & & & &                                                       C \\
DC-PC1 acceptance              &  2\% (2\%) & & & & & & &                                                       B \\
acceptance variation           &0.5\% (3\%) & & & & & & &                                                       B \\
PC3-EMCal matching               &  2\% (2\%) & & & & & & &                                                       B \\
$\pi^0$ selection              &            &  3\%&   3\% & 3\% (3\%) &  3\% (3.5\%) & 3\% (4\%) & 3\%   (4\%)& B \\
conversion uncertainty         &  3\% (3\%) & & & & & & &                                                       C \\
EMCal energy resolution          &            &  2\%& 2.5\% & 3\% (5\%) &  4\% (5\%)   & 5\% (5\%) & 5\%   (5\%)& B \\
EMCal energy scale               &            &  3\%&   3\% & 3\% (4\%) &3.5\% (4\%)   & 4\% (4\%) & 5\% (4.5\%)& B \\
Gamma-trigger efficiency         &            &  5\%&   5\% & 5\% (5\%) &  5\% (5\%)   & 5\% (5\%) & 4\% (4.5\%)& B \\
Gamma-trigger run-by-run variation \hspace{5mm}      &  4\% (4\%) & & & & & & &     B \\
peak extraction in data (fit)  &            & 10\%&  13\% &20\% (30\%) & 13\% (20\%) &23\% (20\%) &30\% (15\%)& A \\
peak extraction in data (width)&            & 10\%&  10\% &10\% (15\%) & 10\% (15\%) &10\% (10\%) &12\% (20\%)& A \\
peak extraction in simulation  &  3\% (5\%) & & & & & & &                                                       B \\
branching ratio uncertainty    &1.8\% (1.8\%)& & & & & & &                                                      C \\
phase space corrections        &            & 20\%&  15\% &11\% (27\%) &  8\% (24\%) & 7\% (20\%) & 7\% (19\%)& B \\
MB trigger                     &\hspace{2mm}9.7\% (5.2\%)\hspace{2mm} & & & & & & &    C \\
trigger bias                   &2.5\% (1\%) & & & & & & &                                                       C \\\hline
total                          &            & 28\%&  26\% &29\% (45\%) & 24\% (38\%) &30\% (33\%) &36\% (35\%)\\
\end{tabular}\end{ruledtabular}
\end{table*}

The error of the raw yield (peak) extraction was estimated, as
described in~\cite{ppg054}, calculating the error of the ratio of the
integrals of the real and the mixed event distributions in the region of
the background fit. The systematic error in peak extraction differs
from the systematic error estimated for neutral pions in~\cite{ppg054} because
the background in the $\eta$ region cannot be estimated as well as the
background in the $\pi^0$ region. This type-B error, estimated
to be 4\% higher than for pions, becomes dominant at very low transverse momenta
due to the small S/B ratio.
The error on the acceptance correction includes fiducial cuts on the edges of the
EMCal sector as well as cuts around towers that have been determined to be hot or dead.
The uncertainty in the MC (GEANT) description of the detector geometry is estimated
varying these cuts slightly in the fastMC and in the embedded events
(Au+Au). Those variations are found to result in differences in the yields of less than 5\%.
Different combinations of particle ID cuts were used in the analysis
to estimate the uncertainty related to the photon identification.
The differences among the various samples are less than 4\% for
all the different PID cuts for p+p as well as for d+Au reactions.
The error in the reconstruction efficiency contains this difference.
The most important source of uncertainty at high $\pt$ is related to the
energy scale. The $\eta$ peak positions and widths observed in the data
are not reproduced to better than 1.5\%.  An error in the energy scale
of 1.5\% leads to an error of 4\% in the yield at $\pt=2$~GeV/$c$ and
of 8\% at $\pt=10$~GeV/$c$.
The error of the high-$\pt$ trigger efficiency in p+p is different
from in d+Au: it amounts to 7.5$\%$ at $p_{T}$ = 3.5 GeV/$c$ and becomes
negligible at $p_{T}$ = 5 GeV/$c$ (see Section~\ref{sec:hipt_trigger}).\\


\subsubsection{$\eta\rightarrow\gaga$~ analysis (Au+Au)}

The sources of systematic errors in the Au+Au analysis are listed in
Table~\ref{tab:sys_errors}. The main sources of systematic errors in the
$\eta$ measurement are the uncertainties in the yield extraction (10--30\%),
the yield correction ($\sim$10\%), and the energy scale (a maximum of $\sim$8\%).
The energy scale uncertainty is basically the same as discussed before for the p+p and d+Au analyses.
The uncertainty on the raw yield extraction was studied by varying the normalization
region of the mixed event background and by comparing yields extracted from 2$\sigma$
and 3$\sigma$ integration windows.  The yields were found to vary within 10\% of the expectation
for all centralities. The final results obtained with different PID cut combinations
are found to be consistent within $\sim$8\%, and this was the assigned
systematic uncertainty for the photon identification procedure.
The final combined systematic errors on the spectra are at
the level of $\sim$10-15\% (type-A, point-to-point) and $\sim$10-15\% (type-B, $\pt$-correlated).


\subsubsection{$\eta\rightarrow\pi^{0}\pi^{+}\pi^{-}$ analysis (p+p and d+Au)}

Systematic errors for the $\pi^{0}\pi^{+}\pi^{-}$ channel
are summarized in Table~\ref{tab:errtable_3pi}. The p+p and d+Au data samples have
different systematic errors which are usually larger in d+Au. This is due to the
larger high-$\pt$ trigger threshold set during d+Au data taking.
The PC3-EMCal matching uncertainty is used to evaluate peak extraction uncertainty.
The dominant systematic uncertainties in the p+p (d+Au) measurement are in the yield
extraction and the phase-space corrections, with uncertainties of 10-30\% (10-30\%)
and $\sim$10\% ($\sim$25\%), respectively.
The final combined systematic errors on the spectra are at the level of
$\sim$30\% (p+p) and $\sim$40\% (d+Au).

s

\section{RESULTS AND DISCUSSION}
\label{sec:results}

In this section, the fully corrected spectra for $\eta$ production differential in $\pt$
in p+p, d+Au and Au+Au are presented, as well as the nuclear modification 
factors for d+Au and Au+Au collisions. The measured $\eta/\pi^0$ ratio as a function of $\pt$ 
for the three colliding systems is presented and discussed in comparison with a compilation of 
world data for hadron-hadron, hadron-nucleus, nucleus-nucleus and $e^+e^-$ collisions 
and to phenomenological (PYTHIA and ``$\mt$-scaling'') expectations.



\subsection{Transverse Momentum Spectra (p+p, d+Au, Au+Au)}
\label{sec:pt_spectra}

\begin{figure}[htb]
\begin{center}
\includegraphics[width=1.0\linewidth]{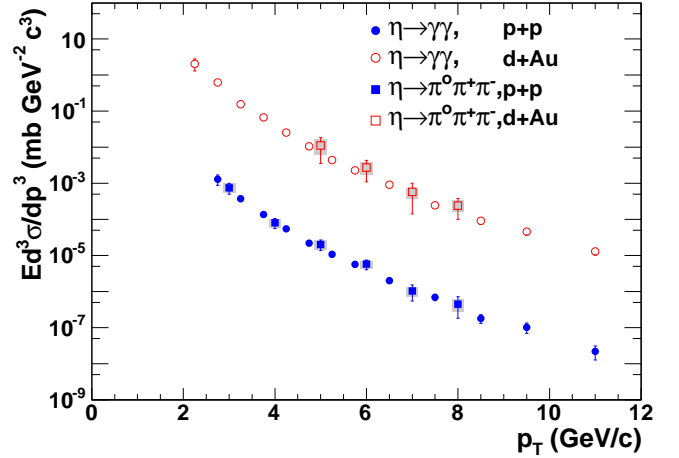}
\caption{Invariant $\eta$ cross-section as a function of transverse momentum
in p+p and d+Au collisions at $\sqrt{s_{_{NN}}}$= 200~GeV measured in the 
$\eta\rightarrow\gaga$ (circles) and $\eta\rightarrow\pi^{0}\pi^{+}\pi^{-}$ 
(squares) decay channels. The error bars of the $\eta\rightarrow\gaga$ are the 
quadratic sum of statistical and systematic uncertainties. The error bars (bands) 
of the $\pi^{0}\pi^{+}\pi^{-}$ spectra represent the statistical (systematic) 
uncertainties of the measurement.}
\label{fig:pp_dAu_sigma_eta}
\end{center}
\end{figure}

The fully corrected spectra for the $\eta$ meson are shown
in Fig.~\ref{fig:pp_dAu_sigma_eta} for MB events in
proton-proton and deuteron-gold collisions at $\sqrt{s_{NN}}$ = 200 GeV. 
The figure shows the spectra obtained in both the $\eta\rightarrow\gaga$ 
and $\eta\rightarrow\pi^{0}\pi^{+}\pi^{-}$ decay channels. 
For the $\gaga$ result, the error bars represent the total error, given by the 
quadratic sum of the statistical and the systematic uncertainties.
For the pion-triplet spectra, the error bars (bands) represent the 
statistical (systematic) uncertainties. These results agree well in 
spite of very different analysis approaches and sources
of systematic uncertainties. Due to higher acceptance and lower trigger
threshold (see Figs.~\ref{fig:acceptance}, \ref{fig:3pi-acceptance}, \ref{fig:trigger}),
the $\gaga$ channel has superior statistics and therefore these results 
alone are used henceforth.


\begin{figure}[htb]
\begin{center}
\includegraphics[width=1.0\linewidth]{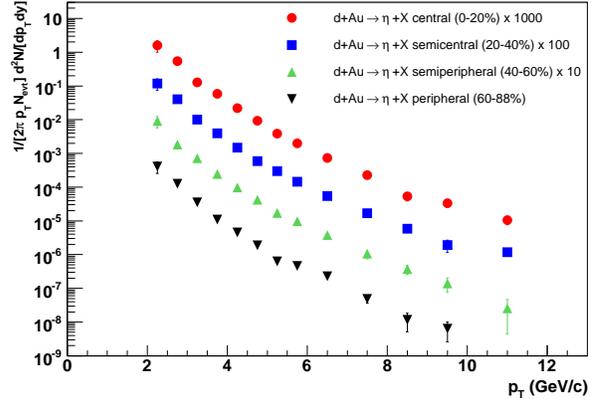}
\caption{Invariant $\eta$ yields as a function of transverse momentum
in d+Au collisions at $\sqrt{s_{NN}}$~=~200~GeV in four different
centralities (0--20\%, 20--40\%, 40--60\%, 60--88\%). The error bars are the 
quadratic sum of statistical and all systematic uncertainties. For clarity, 
the data points are scaled vertically as noted in the figure.}
\label{fig:dAu_eta}
\end{center}
\end{figure} 

\begin{figure}[htb]
\begin{center}
\includegraphics[width=1.0\linewidth]{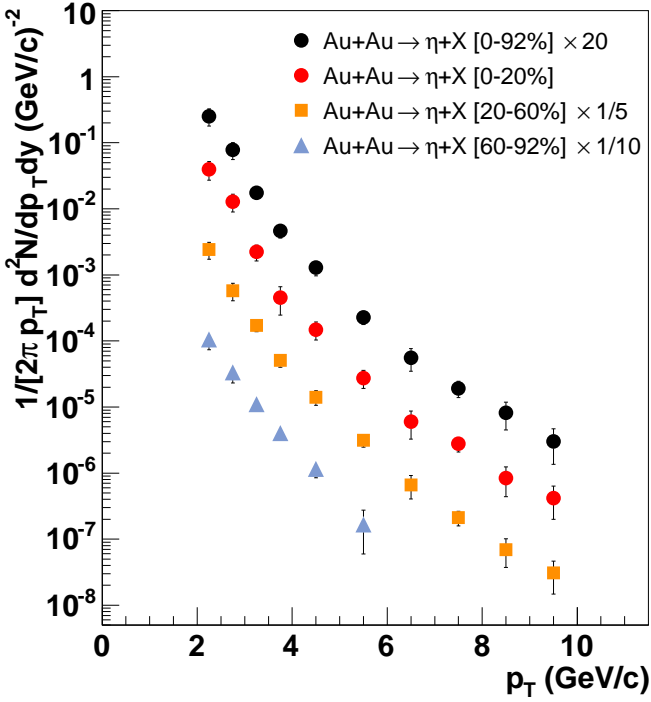}
\caption{Invariant $\eta$ yields as a function of transverse momentum
in Au+Au collisions at $\sqrt{s_{NN}}$~=~200~GeV for MB and three 
centralities (0--20\%, 20--60\%, 60--92\%). The error bars are the 
quadratic sum of statistical and point-to-point systematic uncertainties. 
For clarity, the data points are scaled vertically as noted in the figure.}
\label{fig:AuAu_eta}
\end{center}
\end{figure}

The invariant yields measured in four different centrality classes
in d+Au collisions at $\sqrt{s_{NN}}$ = 200~GeV are shown in Fig.~\ref{fig:dAu_eta}.
In Fig.~\ref{fig:AuAu_eta} the fully corrected invariant spectra 
for MB and three different centrality classes in Au+Au collisions 
at $\sqrt{s_{NN}}$ = 200 GeV are shown. The error bars represent the quadratic sum of 
the statistical and the point-to-point systematic uncertainties.


\subsection{Nuclear Modification Factor in d+Au, $R_{dAu}(\pt$)}
\label{sec:R_dA}

\begin{figure}[htb]
\begin{center}
\includegraphics[width=1.0\linewidth]{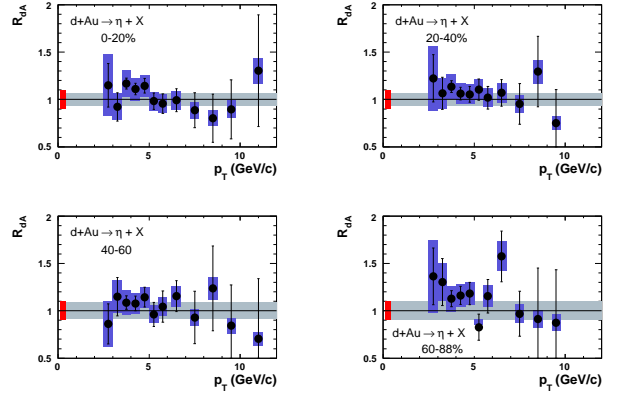}
\caption{Nuclear modification factors for $\eta$ production for four d+Au 
centralities: 0--20\%, 20--40\%, 40--60\%, 60--88\%. 
The error bars (bands) around each point are the statistical
(type-B systematic) uncertainties. The error band at $R_{dA}$ = 1 indicates 
the uncertainty in $\mean{T_{dA}}$ for each centrality. The error box at 
$R_{dA}$ = 1 indicates the p+p cross-section uncertainty of 9.7\%.}
\label{fig:RdA_diff_cent}
\end{center}
\end{figure}

Medium effects in d+A collisions are quantitatively determined using
the \emph{nuclear modification factor} given as the ratio of the measured 
d+A invariant yields, $d^2N_{dA}/d\pt dy$, over the measured p+p cross-sections,
$d^2\sigma_{pp}/d\pt dy$, scaled by the nuclear thickness function
$\langle T_{dA} \rangle$
in the centrality bin under consideration:
\begin{equation}
R_{dA}(\pt)\,=\,\frac{d^2N_{dA}/dy d\pt}{\langle T_{dA}\rangle \,\cdot\,d^2\sigma_{pp}/dy d\pt}\;.
\label{eq:R_dA}
\end{equation}
Deviations from $R_{dA}(\pt)$ = 1 quantify the degree of departure 
of the hard d+A yields from an incoherent superposition of $NN$ collisions.
The values of the nuclear thickness function for different centralities
are obtained in a Glauber MC calculation and tabulated in Table~\ref{tab:T_AA}.
The resulting $R_{dA}(\pt)$ for $\eta$ mesons in d+Au collisions is plotted
for different centralities in Fig.~\ref{fig:RdA_diff_cent}.


The data points at lower transverse momenta have large statistical errors. 
This is caused by the poor signal-to-background ratio of the $\eta$ peak in 
the sample that is not triggered with the Gamma trigger. 
The systematic uncertainties shown in the plot are computed propagating 
the experimental uncertainties in the p+p and d+Au measurements 
described in Section \ref{sec:sys_error}. Some of these uncertainties cancel out 
when calculating the nuclear modification factor~(Eq.~\ref{eq:R_dA}). 
The error due to the $\eta$ reconstruction efficiency as well as the error due 
to uncertainties in the energy scale are very similar for the measurement of 
$\eta$ mesons in p+p and d+Au collisions as the measured data have been taken 
in the same experimental run, and they cancel almost completely in the ratio.\\

\begin{figure}
 \includegraphics[width=1.0\linewidth]{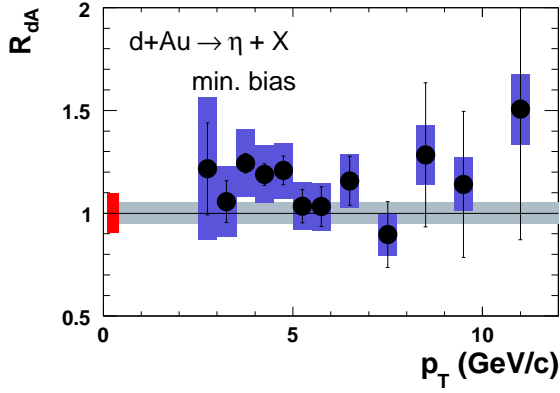}
\caption{Nuclear modification factor $R_{dA}$ for $\eta$ mesons as
  a function of $\pt$ for minimum-bias $\sqrt{s_{NN}} = 200$ GeV d+Au collisions.
The uncertainties are the same described in Fig.~\protect\ref{fig:RdA_diff_cent}.}
\label{fig:RdA_minB}
\end{figure}

In the case of {\it minimum bias} d+Au collisions, the nuclear modification
factor, shown in Fig.~\ref{fig:RdA_minB}, is more simply defined as the ratio of 
d+Au over p+p {\it cross sections} normalized by the total number of nucleons 
($2\cdot A$ for a d+A collision) with $A=197$ for a gold nucleus: 
\begin{eqnarray}
\label{eq:rda2}
R_{dA} (\pt) = \frac{d\sigma_{dA}}{2\cdot A\cdot d\sigma_{pp}}.
\end{eqnarray}

\begin{figure}
 \includegraphics[width=1.0\linewidth]{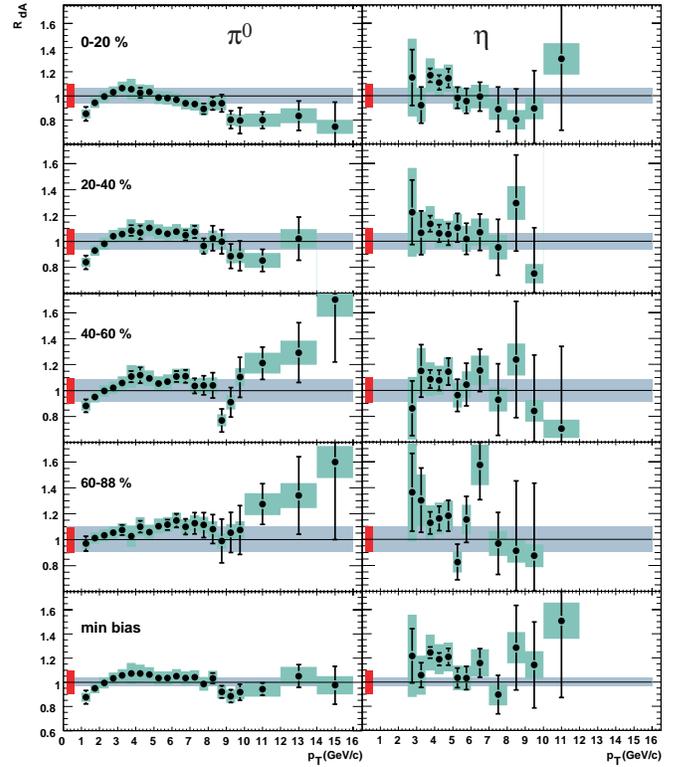}
\caption{Nuclear modification factor $R_{\rm{dA}}(\pt)$ for $\pi^0$ (left) 
and $\eta$ (right) production in different centrality selections and MB d+Au data. 
The bands around the data points show systematic uncertainties which can vary 
with $\pt$ (type-B errors). The shaded band around unity indicates the 
$\langle T_{dA}\rangle$ uncertainty and the small box on the left side of the 
data points indicates the normalization uncertainty of the p+p total inelastic
cross-section.}
\label{fig:RdA_pi0_eta}
\end{figure}

All the d+Au nuclear modification factors shown in Figs.~\ref{fig:RdA_diff_cent} 
and~\ref{fig:RdA_minB} are approximately 1 and show a very weak $\pt$ and/or 
centrality dependence.
Similar trends have been observed for $\pi^0$ production~\cite{ppg044}. 
As shown in the comparison plot of Fig.~\ref{fig:RdA_pi0_eta}, the $\pi^0$ 
nuclear modification factors indicate small shape modifications with centrality, 
with a possible Cronin enhancement on the level of $10\%$ around 4~GeV/$c$ 
disappearing for $p_\mathrm{T} > 10$~GeV/$c$. At high $\pt$ the 
$\pi^0$ MB result can be described well by next-to-leading-order 
pQCD calculations~\cite{deFlorian03,guzey04} without implementation of the 
Cronin effect. The contribution of (anti-)shadowing effects~\cite{eks98,deFlorian03} 
in the $\eta$ or $\pi^0$ production is very small, as expected for this 
kinematical region with $x_T = 2\pt/\sqrt{s}\approx$ 0.02-0.2.\\

The small role of initial-state cold nuclear effects observed in the mid-rapidity 
spectra of neutral mesons at high $\pt$ is also consistent with other similar  
observations in d+Au reactions at $\sqrtsnn$ = 200 GeV such as: 
(i) the absence of significant nuclear modifications in the yields of $J/\Psi$ 
compared to p+p collisions~\cite{ppg038}, and 
(ii) the very similar characteristics of near-side and away-side jet-like 
correlations in p+p and d+Au~\cite{ppg039}. Those results indicate that the 
nuclear medium has little influence on the hard processes in d+Au collisions
at top RHIC energies and $y$ = 0.



\subsection{Nuclear Modification Factor in Au+Au, $R_{AA}(\pt$)}
\label{sec:R_AA}

\begin{figure}[htbp]
\begin{center}
\includegraphics[width=1.0\linewidth]{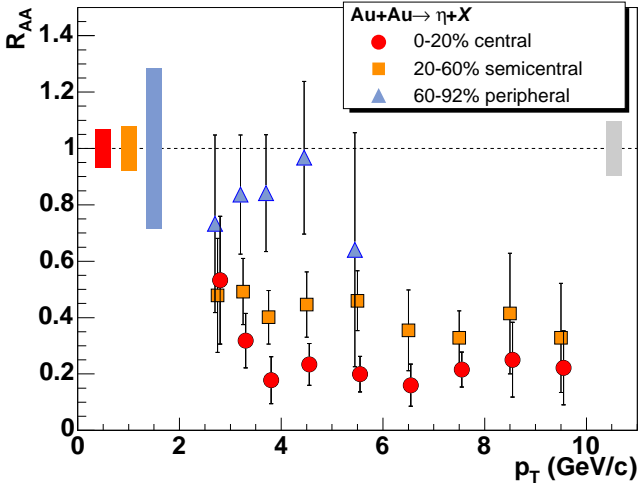}
\end{center}
\caption{Nuclear modification factors for $\eta$ in three Au+Au centralities
(0--20\%, 20--60\%, 60--92\%). The errors bars are point-to-point uncertainties.
The absolute normalization error bands at $R_{AA}$ = 1  are: 
(i) the uncertainties in $\mean{T_{AA}}$ for each centrality (left side),
and (ii) the p+p cross-section normalization uncertainty of 9.7\% (right side).
The $R_{AA}(\pt)$ for peripheral/central Au+Au have been slightly displaced 
to the left/right ($\pm$ 50 MeV/$c$) along the $\pt$ axis to improve the clarity 
of the plot.}
\label{fig:R_AA_eta}
\end{figure}

The nuclear modification factor, $R_{AA}(\pt)$, for $\eta$ production in
each centrality class in Au+Au collisions is computed using the standard formula:
\begin{equation}
R_{AA}(p_{T})\,=\,\frac{d^2N_{AA}/dy dp_{T}}{\langle T_{AA}
\rangle\,\cdot\, d^2\sigma_{pp}/dy dp_{T}},
\label{eq:R_AA}
\end{equation}
where (i) the Au+Au spectra $d^2N/dyd\pt$ are used in the numerator (Fig.~\ref{fig:AuAu_eta}), 
(ii) the p+p invariant spectrum $d^2\sigma/dyd\pt$ (Fig.~\ref{fig:pp_dAu_sigma_eta})
is used in the denominator, and 
(iii) $\langle T_{AA}\rangle$ are the values of the average Glauber overlap 
function for each Au+Au centrality (Table~\ref{tab:T_AA}). 
The $R_{AA}(\pt)$ is computed taking the bin-to-bin ratio of Au+Au and p+p spectra 
and propagating the corresponding uncertainties. Only the acceptance 
uncertainty ($\sim$5\%) cancels in the Au+Au/p+p ratio of spectra.
Figure~\ref{fig:R_AA_eta} compares the nuclear modification
factor for $\eta$ measured in central (0--20\%), semicentral (20--60\%) and
peripheral (60--92\%) Au+Au collisions. The error bars are the total point-to-point errors 
(including type-{\bf A} systematic and statistical uncertainties) of the
Au+Au and p+p measurements.
The error bands on the left are the uncertainties in $\langle T_{AuAu}\rangle$ for each 
centrality class. The error box on the right is the Run-3 p+p cross-section 
uncertainty of 9.7\%. As observed for high-$\pt$ 
$\pi^0$~\cite{ppg014,ppg054}, the Au+Au $\eta$ yields are consistent 
with the expectation of independent $NN$ scatterings in peripheral 
reactions ($R_{AA}\approx$1) but they are increasingly depleted with respect to this expectation 
for more central collisions. There is no $\pt$ dependence of $R_{AA}$, as seen also
for neutral pions.\\


\begin{figure}
\includegraphics[width=1.0\linewidth]{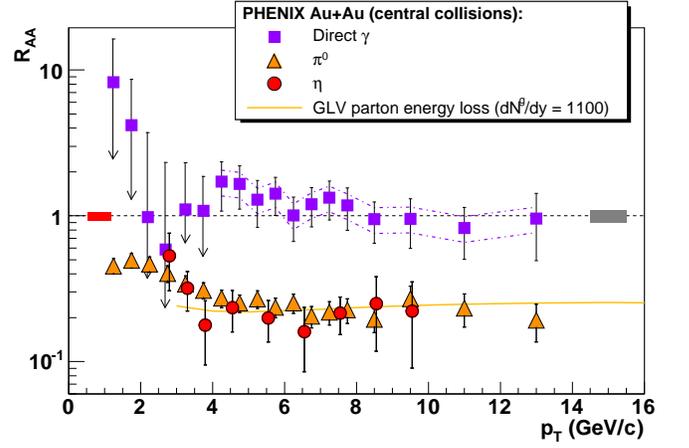}
\caption{(color online) $R_{AA}(\pt)$ measured in central Au+Au at $\sqrtsnn$~=~200 GeV for 
$\eta$, $\pi^0$~\protect\cite{ppg014,ppg054} and for direct $\gamma$~\protect\cite{ppg042}. 
The error bars include all point-to-point 
uncertainties. The error bands at $R_{AA}$ = 1 
have the same meaning as in Fig.~\ref{fig:R_AA_eta}.
The baseline p+p $\rightarrow\gamma+X$ reference used is a NLO pQCD calculation~\protect\cite{ppg042,vogelsang}
that reproduces our own data well~\protect\cite{pp_gamma}, with theoretical (scale)
uncertainties indicated by the dash-dotted lines around the points. 
The solid yellow curve is a parton energy loss prediction for the suppression
factor of leading pions in a medium with initial gluon density
$dN^g/dy=$ 1100~\protect\cite{vitev_gyulassy}.}
\label{fig:R_AA_eta_pi0_gamma} 
\end{figure}

Figure~\ref{fig:R_AA_eta_pi0_gamma} contrasts the nuclear modification factors measured
 in central Au+Au at $\sqrtsnn$ = 200 GeV for $\eta$, $\pi^0$~\cite{ppg014,ppg054} and
$\gamma$~\cite{ppg042}. Whereas direct photons are unsuppressed compared to the 
scaled reference given by a NLO pQCD calculation~\cite{ppg042,vogelsang} that
reproduces the PHENIX p+p $\gamma$ results well~\cite{pp_gamma}, neutral pions and
$\eta$ are  suppressed by a similar factor of $\sim$5 compared to the corresponding 
cross sections measured in p+p. 
Within the current uncertainties, light-quark neutral mesons at RHIC show a flat 
suppression in the range $\pt\approx$ 4 -- 15 GeV/$c$, independent of
their mass (note that the $\eta$ is $\sim$4 times heavier than the $\pi^0$). 
Those results are in agreement with parton energy loss calculations in a 
system with initial effective gluon densities of the order $dN^g/dy\sim$ 1000
(solid curve in the figure)~\cite{vitev_gyulassy}.
The equal suppression of $\eta$ and $\pi^0$ mesons and the agreement
with parton energy loss calculations suggest that
the final fragmentation of the quenched parton into a leading meson 
occurs in the vacuum according to the same probabilities (fragmentation functions) 
that govern high-$\pt$ hadroproduction in more elementary systems (p+p, $e^+e^-$).
This conclusion is examined in more detail in the next two sections.\\

\subsection{Ratio of $\eta$ to $\pi^0$ (p+p, d+Au, Au+Au)}
\label{sec:eta_pi0_ratios}

A useful way to determine possible differences in the suppression pattern of
$\pi^0$ and $\eta$ is to study the centrality dependence of the $\eta/\pi^0$
ratio, $R_{\eta/\pi^{0}}(\pt)$, in d+Au and Au+Au reactions and compare it
with the values measured in more elementary systems (p+p, $e^+e^-$).
The ``world'' $\eta/\pi^{0}$ ratio in hadronic and proton-nucleus collisions
increases rapidly with $\pt$ and flattens out above $\pt\approx$ 2.5 GeV/$c$ 
at values $R_{\eta/\pi^{0}}\sim$0.40 -- 0.50 (see Section~\ref{sec:eta_pi0_syst_hh_hA_AA}). 
Likewise, in electron-positron annihilations at the $Z$ pole ($\sqrt{s}=91.2$~GeV),  
$R_{\eta/\pi^{0}}\sim$0.5 for {\it energetic} $\eta$ and $\pi^0$ 
(with $x_{p} = p_{\ensuremath{\it hadron}}/p_{\ensuremath{\it beam}}>$ 0.4, 
consistent with the range of scaled momenta $\mean{z} = p_{hadron}/p_{jet}$ considered here),
as discussed in Section~\ref{sec:eta_pi0_ee}.
It is interesting to test if this ratio is modified in any way
by initial- and/or final-state effects in d+Au and Au+Au collisions
at RHIC energies.\\

\begin{figure}[htbp]
\includegraphics[width=1.0\linewidth]{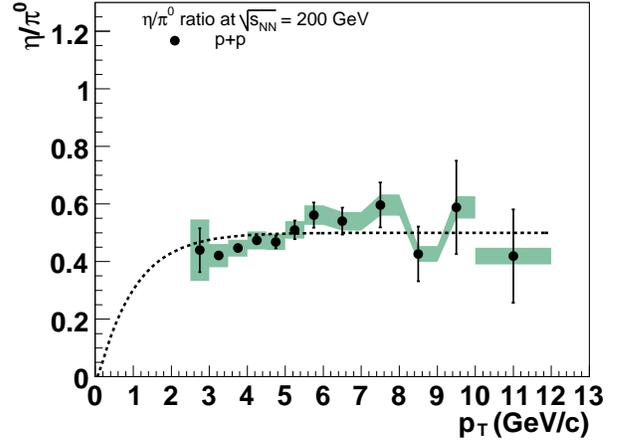}
\caption{Ratio $\eta/\pi^0$ measured in p+p collisions at $\sqrt{s}$ = 200 GeV. 
The error bars represent the point-to-point errors; the boxes represent 
the systematic uncertainties. The dashed line is the prediction of 
PYTHIA~\protect\cite{pythia} for p+p at this c.m.~energy.}
\label{fig:ratio_eta_pi_pp}
\end{figure}
\begin{figure}[htbp]
\includegraphics[width=1.0\linewidth]{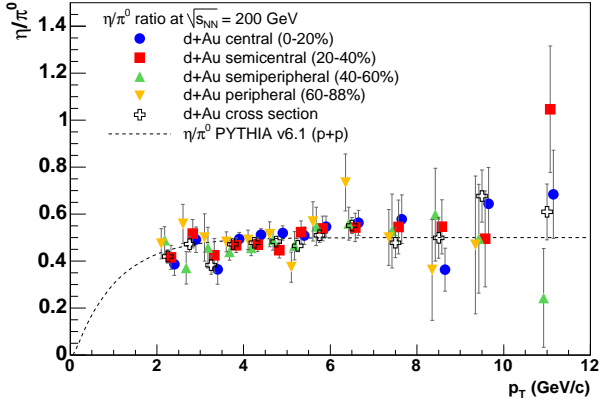}
\caption{Ratio $\eta/\pi^0$ measured in different centralities in d+Au collisions
at $\sqrtsnn$ = 200 GeV. The error bars represent all point-to-point uncertainties. 
The dashed line is the prediction of 
PYTHIA~\protect\cite{pythia} for p+p at this c.m.~energy. A few
$R_{\eta/\pi^0}(\pt)$ ratios have been slightly displaced to the left or right 
($\pm$ $<$ 150 MeV/$c$) along the $\pt$ axis to improve the clarity of the plot.}
\label{fig:ratio_eta_pi_dAu}
\end{figure}

The production ratio of $\eta$ and $\pi^0$ mesons is shown in 
Fig.~\ref{fig:ratio_eta_pi_pp} for p+p and in Fig.~\ref{fig:ratio_eta_pi_dAu}
for d+Au (MB and 4 centrality classes). The ratio is calculated point-by-point 
for the d+Au measurements, propagating the corresponding errors. 
In the p+p case, a fit to the $\pi^0$ spectrum~\cite{ppg054} was used. 
All the ratios are consistent with the PYTHIA~\cite{pythia} 
curve for p+p at $\sqrt{s}$ = 200 GeV
(dashed line, see discussion in Section~\ref{sec:eta_pi0_syst_hh_hA_AA})
with an asymptotic $R_{\eta/\pi^0}^{\infty}$ = 0.5 value.\\

\begin{figure}[htbp]
\includegraphics[width=1.0\linewidth]{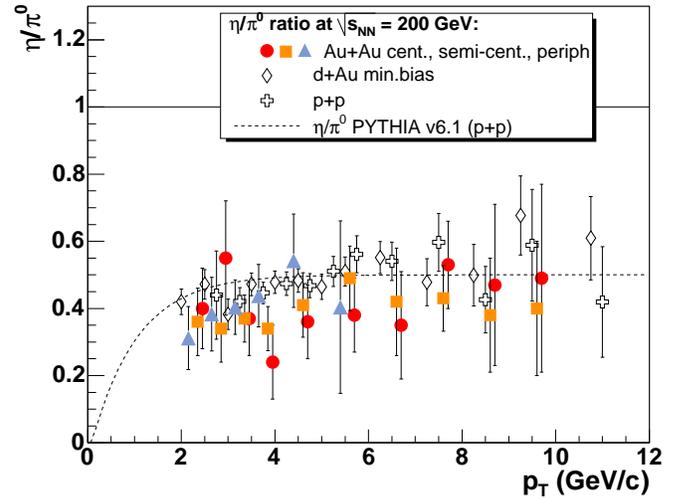}
\caption{Au+Au $R_{\eta/\pi^{0}}$ ratio in MB and three centrality classes
(0-20\%, 20-60\%, 60-92\%) as a function of $\pt$ compared to the ratio 
in d+Au and p+p collisions. The error bars include all point-to-point errors.
The dashed line is the prediction of PYTHIA~\protect\cite{pythia} for p+p at this c.m.~energy.
A few $R_{\eta/\pi^0}(\pt)$ ratios have been slightly displaced to the left or right 
($\pm$ 50 MeV/$c$) along the $\pt$ axis to improve the clarity of the plot.}
\label{fig:eta_pi0_ratio} 
\end{figure}

Figure~\ref{fig:eta_pi0_ratio} shows the $R_{\eta/\pi^{0}}(\pt)$ ratio for MB 
and three Au+Au centralities, obtained using the latest PHENIX $\pi^0$ spectra~\cite{ppg054}
and removing those systematic uncertainties which cancel in the ratio. 
The $R_{\eta/\pi^{0}}(\pt)$ data for Au+Au is compared to 
a PYTHIA~\cite{pythia} calculation that reproduces the hadronic 
collision data well 
(see next Section).
Within uncertainties, all the ratios are consistent with $R_{\eta/\pi^{0}}\approx$ 0.5 
(dashed line) and show no collision system, centrality, or $\pt$ dependence.
A simple fit to a constant above $\pt$ = 2 GeV/$c$ yields the following ratios:
\begin{itemize}
\item $R_{\eta/\pi^0}$ (Au+Au cent) = 0.40 $\pm$ 0.04 (stat) $\pm$ 0.02 (syst), $\chi^2$/ndf = 0.48
\item $R_{\eta/\pi^0}$ (Au+Au semicent) = 0.39 $\pm$ 0.03 (stat) $\pm$ 0.02 (syst), $\chi^2$/ndf = 0.26
\item $R_{\eta/\pi^0}$ (Au+Au periph) = 0.40 $\pm$ 0.04 (stat) $\pm$ 0.02 (syst), $\chi^2$/ndf = 0.42
\item $R_{\eta/\pi^0}$ (p+p)= 0.48 $\pm$ 0.02 (stat) $\pm$ 0.02 (syst), $\chi^2$/ndf = 0.89
\item $R_{\eta/\pi^0}$ (d+Au)= 0.47 $\pm$ 0.02 (stat) $\pm$ 0.02 (syst), $\chi^2$/ndf = 0.84 
\end{itemize}



\subsection{World Data on the $\eta/\pi^0$ Ratio in High-Energy Particle Collisions}
\label{sec:eta_pi0_syst}

In this last Section of the paper, we present a compilation of experimental 
$\eta/\pi^0$ ratios as a function of transverse momentum, 
$R_{\eta/\pi^0}(\pt)$, measured in different hadronic and nuclear colliding systems
in a wide range of center-of-mass energies ($\sqrtsnn\approx$ 3 -- 1800 GeV).
The collected world data on $\eta/\pi^0$ ratios includes:
\begin{description} 
\item (i) hadron-hadron collisions (26 p+p, p+$\bar{p}$, $\pi^\pm$+p data sets), 
\item (ii) hadron-nucleus collisions (17 p,$\pi^\pm$+A sets), and 
\item (iii) nucleus-nucleus collisions (7 A+A data sets).
\end{description} 
In addition, we present also the $R_{\eta/\pi^0}(x_{p})$ ratio obtained 
from inclusive $\pi^0$ and $\eta$ cross-sections in $e^+e^-$ 
as a function of scaled momentum $x_{p}=2\,p_{had}/\sqrt{s}$ measured by the 
four LEP experiments at the $Z$ pole ($\sqrt{s}=$ 91.2 GeV) .
In all cases, the ratio $R_{\eta/\pi^0}$ increases rapidly with $\pt$ 
(or $x_{p}$) and saturates at $R_{\eta/\pi^0}\approx$ 0.4 -- 0.5 
above $\pt\approx$ 3 GeV/$c$ ($x_p\approx$ 0.3). The experimental 
$R_{\eta/\pi^0}(\pt)$ ratios are also compared to PYTHIA and to $\mt$-scaling 
expectations. PHENIX p+p, d+Au and Au+Au $\eta/\pi^0$ 
ratios at $\sqrt{s}$ = 200 GeV are found to be consistent with the obtained
world data on $R_{\eta/\pi^0}$.


\subsubsection{$\eta/\pi^0$ ratio in hadron-hadron, hadron-nucleus and nuclear collisions 
($\sqrt{s}\approx$ 3 -- 1800 GeV)}
\label{sec:eta_pi0_syst_hh_hA_AA}

In Tables~\ref{tab:hh},~\ref{tab:hA} and \ref{tab:AA} we list all data sets with published 
$\eta$ and $\pi^0$ spectra and/or published $R_{\eta/\pi^0}(\pt)$ ratios in 
hadron-hadron, hadron-nucleus, and nucleus-nucleus collisions that we have found in the 
literature. Most of those measurements are performed around mid-rapidity.
Roughly half of the $R_{\eta/\pi^0}(\pt)$ listed 
have been directly taken from the original works whose references are 
provided in the data tables. A few others have been constructed by taking the ratio of the 
published $\pi^0$ and $\eta$ invariant cross-section spectra measured at the same $\sqrt{s}$. 
In the latter case, the error in the ratio has been computed by adding statistical and systematic
uncertainties quadratically. There were a few cases where the $p_{T}$ binning of the $\eta$ 
spectrum did not match that of the $\pi^0$. In these cases, the $\pi^{0}$ spectrum 
was fitted with a functional form that reproduced the data well
(usually a modified power law of the form discussed in~\cite{dde_hq04}) 
and the $\eta /\pi^{0}$ ratio was then obtained by dividing the $\eta$ spectrum data 
points by the values of the $\pi^{0}$ function at each point. In this case, the error 
was computed by dividing the quoted $\eta$ error by the function value at that point. 
The uncertainty arising from the $\pi^{0}$ spectrum fit was obtained by computing the 
minimum and maximum ratio values at each point. Both errors were then added in quadrature.\\

\begin{figure*}[htbp]
\centering
\includegraphics[width=1.0\linewidth]{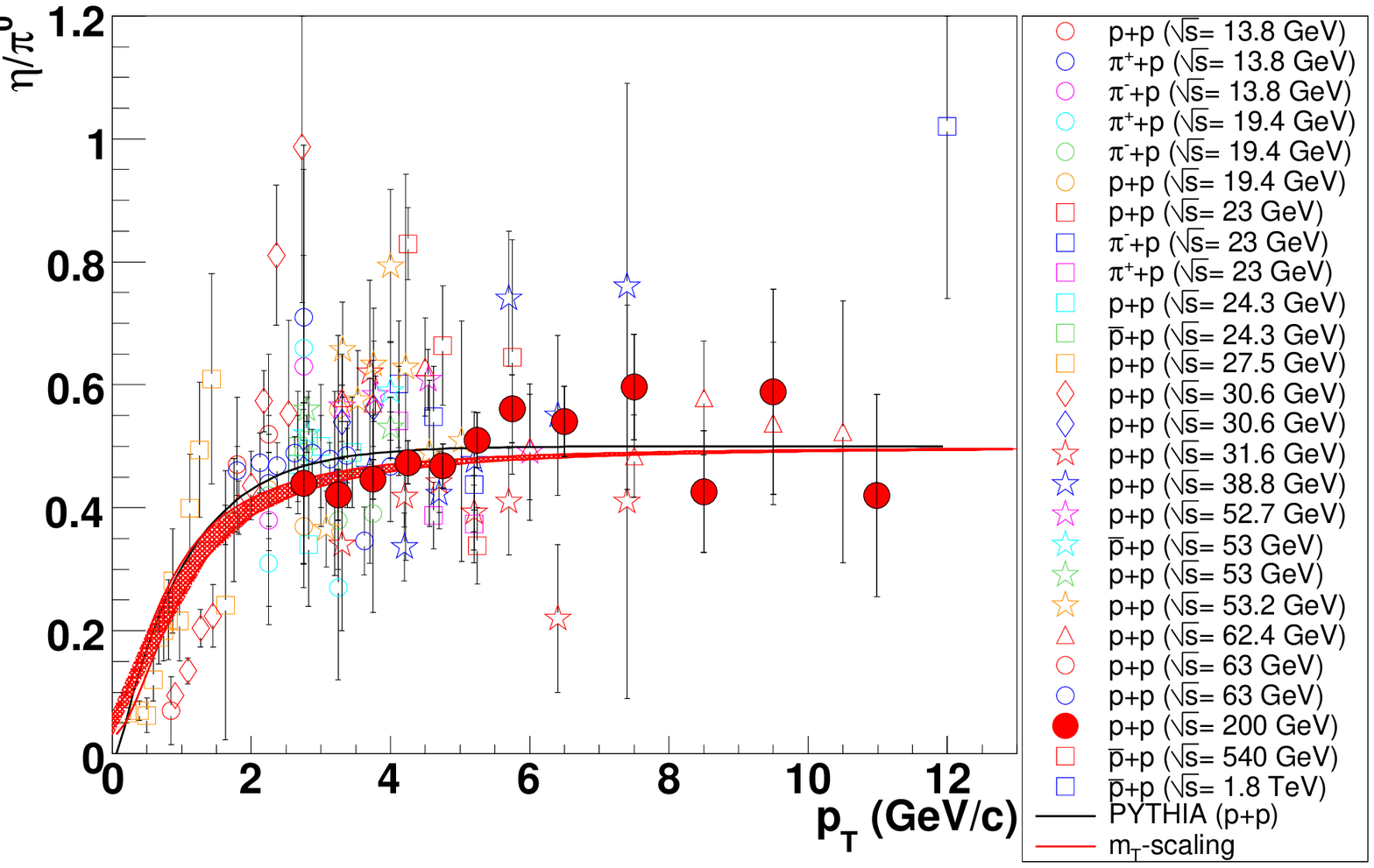}
\caption{(color online) Values of the $R_{\eta /\pi^{0}}$ ratios as a function of $p_{T}$
measured in the hadron-hadron collisions reported in  Table~\protect\ref{tab:hh}.
The black curve is the prediction of PYTHIA~\cite{pythia} for the ratio 
in p+p at $\sqrt{s}$ = 200 GeV, and the red shaded area indicates
the empirical $\mt$-scaling prescription Eq.~(\ref{eq:mt_pow}) with fixed $a$ = 1.2, 
power-law exponent $n$ = 10. -- 14., and an asymptotic $R_{\eta/\pi^0}^{\infty}$ = 0.5 ratio.}
\label{fig:hh_eta_pi0_ratios}
\end{figure*}

\begin{figure*}[htbp]
\centering
\includegraphics[width=1.0\linewidth]{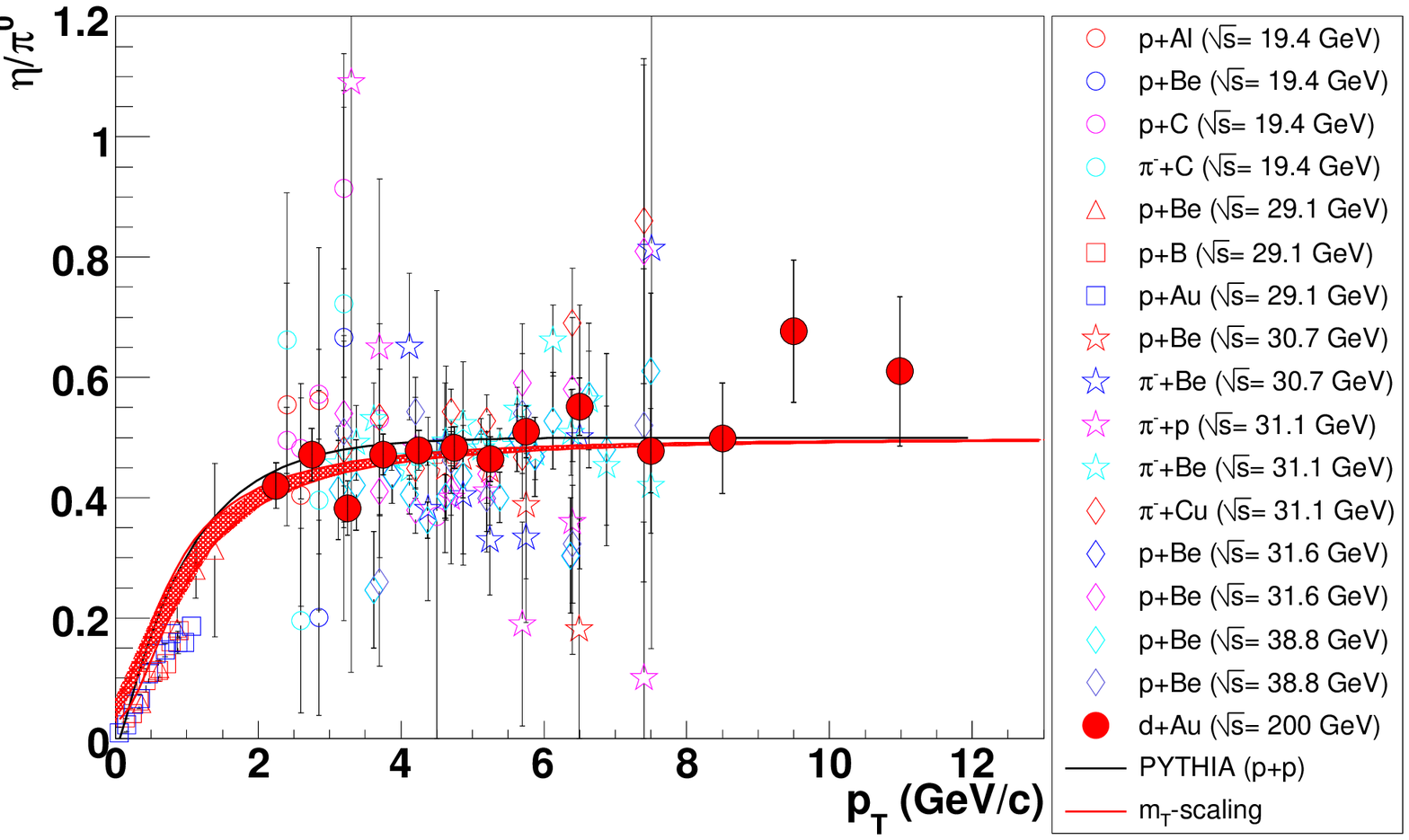}
\caption{(color online) Values of the $R_{\eta /\pi^{0}}$ ratios as a function of $p_{T}$
measured in the hadron-nucleus collisions reported in  Table~\protect\ref{tab:hA}.
The black curve is the prediction of PYTHIA~\cite{pythia} for the ratio 
in p+p at $\sqrt{s}$ = 200 GeV, and the red shaded area indicates
the empirical $\mt$-scaling prescription Eq.~(\ref{eq:mt_pow}) with fixed $a$ = 1.2, 
power-law exponent $n$ = 10. -- 14., and an asymptotic $R_{\eta/\pi^0}^{\infty}$ = 0.5 ratio.}
\label{fig:hA_eta_pi0_ratios}
\end{figure*}

\begin{figure*}[htbp]
\centering
\includegraphics[width=1.0\linewidth]{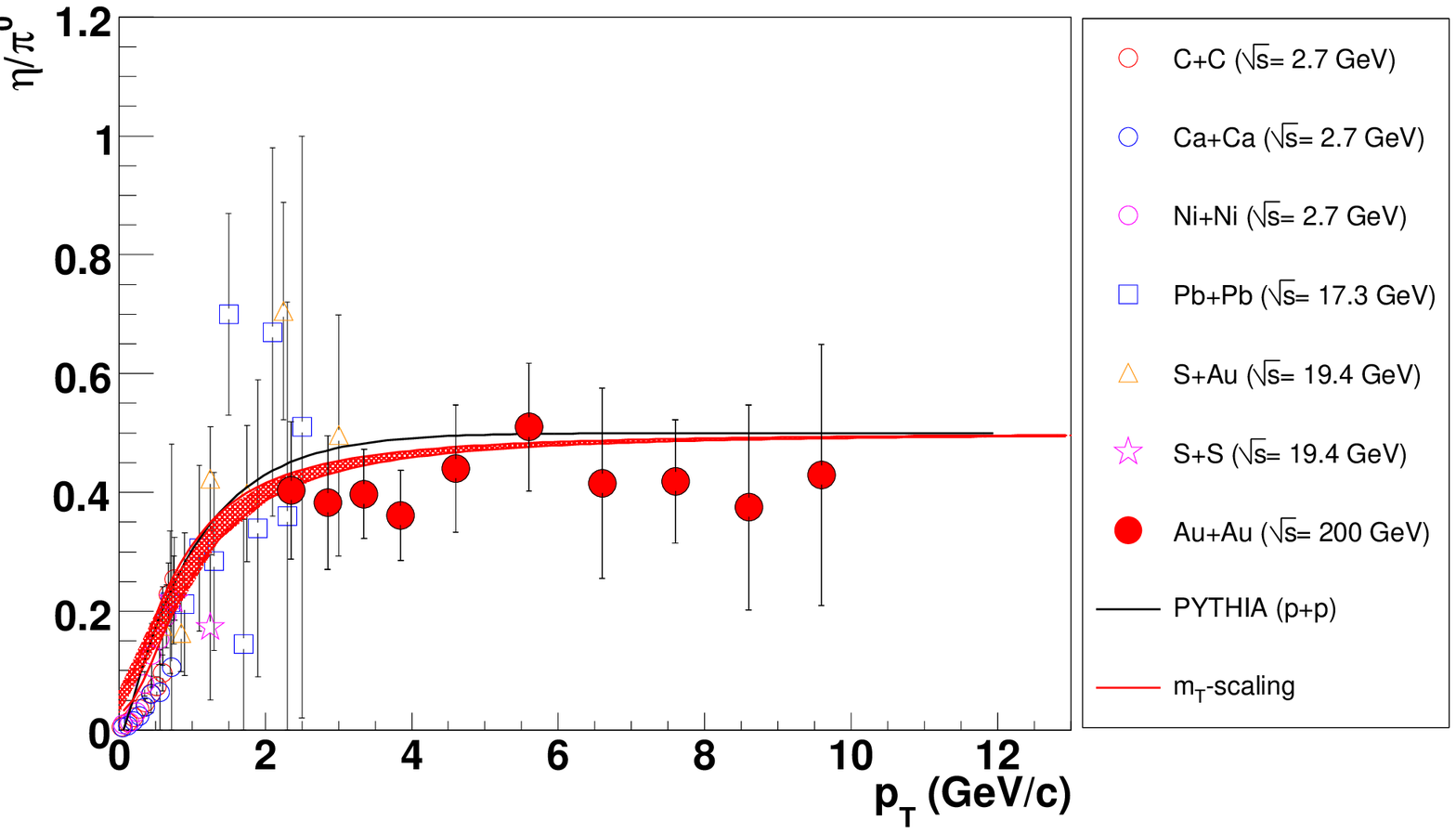}
\caption{(color online) Values of the $R_{\eta /\pi^{0}}$ ratios as a function of $p_{T}$
measured in the nucleus-nucleus collisions reported in  Table~\protect\ref{tab:AA}.
The black curve is the prediction of PYTHIA~\cite{pythia} for the ratio 
in p+p at $\sqrt{s}$ = 200 GeV, and the red shaded area indicates
the empirical $\mt$-scaling prescription Eq.~(\ref{eq:mt_pow}) with fixed $a$ = 1.2, 
power-law exponent $n$ = 10. -- 14., and an asymptotic $R_{\eta/\pi^0}^{\infty}$ = 0.5 ratio.}
\label{fig:AA_eta_pi0_ratios}
\end{figure*}

In Tables~\ref{tab:hh},~\ref{tab:hA} and~\ref{tab:AA}, together with the general info 
on the collected data sets, we indicate for each measurement the approximate $\pt$ and 
$x_T = 2\pt/\sqrt{s}$ ranges, as well as the average value of $R_{\eta/\pi^0}$ at 
high $\pt$, obtained by fitting the data to a constant above $\pt$ = 2 GeV/$c$. 
With the exception of the higher energy data ($\sqrtsnn\gtrsim$ 100 GeV), most of the 
experimental ratios have been measured in a fractional momentum range 
$x_T\approx$ 0.1 -- 0.3 where the parton distribution functions are dominated by 
valence quarks (rather than gluons) and, hence, the produced high-$\pt$ $\pi^0$ 
and $\eta$ mesons come largely from $q,\bar{q}$ fragmentation. 
Figures~\ref{fig:hh_eta_pi0_ratios}, \ref{fig:hA_eta_pi0_ratios}, and \ref{fig:AA_eta_pi0_ratios} 
show the corresponding $R_{\eta/\pi^0}(\pt)$ ratios for each type of colliding system. 
All the ratios show a rapid increase with $\pt$ and level off at 
$R_{\eta/\pi^0}\approx$ 0.4 -- 0.5 above $\pt\approx$ 3 GeV/$c$.
No difference is observed for different colliding systems. The PHENIX p+p, d+Au and Au+Au
high-$\pt$ data presented in the previous section are consistent with those ratios. 
A fit of the PHENIX ratios to a constant gives $R_{\eta/\pi^0}$ = 0.47 $\pm$ 0.03 
for both p+p and d+Au and, slightly lower but still consistent, 
$R_{\eta/\pi^0}$ = 0.40 $\pm$ 0.04 for Au+Au.
Together with the data points in Figs.~\ref{fig:hh_eta_pi0_ratios}--\ref{fig:AA_eta_pi0_ratios}, 
we also plot two phenomenological curves with PYTHIA 6.131~\cite{pythia} based on the
Lund fragmentation model~\cite{And83,And98}, and on $\mt$-scaling expectations 
for the $\eta/\pi^0$ ratio in p+p collisions at $\sqrt{s}$ = 200 GeV.\\

\paragraph{Lund string fragmentation:}

The fragmentation mechanism in PYTHIA is based on the phenomenological Lund string 
scheme~\cite{And83,And98} which considers the color field between the partons to be 
the fragmenting entity rather than the quarks and gluons themselves. The string is 
viewed as a color flux tube formed by gluon self-interaction between the partons.
As the partons move apart the potential energy stored in the string increases. 
At some point the string breaks via the production of new $q\bar{q}$ pairs according to the 
probability of a quantum-mechanical tunneling process, $\exp(-\pi\;m^2_{\mathrm{q,T}}/\kappa)$, 
which depends on the transverse-mass squared ($m^2_{\mathrm{q,T}} = m^2+\pt^2$) 
and the string tension $\kappa\approx$ 1 GeV/fm $\approx$ 0.2 GeV$^2$. 
The string break-up process proceeds repeatedly into color singlet systems as long 
as the invariant mass of the string pieces exceeds the on-shell mass of a hadron
(each hadron corresponding to a small piece of string with a quark at one end and 
an antiquark at the other). 
At each branching, probabilistic rules are given for the production of flavors
($u\bar{u}:d\bar{d}:s\bar{s}$ = 1 : 1 : 0.3 in the default settings), spin 
(e.g. a 3:1 mixture between the lowest lying vector and pseudoscalar multiplets
is used, suggested by spin counting arguments), and for the sharing of energy and momentum 
among the products. Regarding the latter, 
the probability that a hadron picks a fraction $z$ of $E+p_z$ out of the available 
$E+p$ ($p_z$ is the momentum of the formed hadron along the direction of the 
quark $q$) is given by the ``Lund symmetric fragmentation function'':
\begin{equation} 
f(z) \propto z^{-1}(1-z)^{a}\exp(-b\,\mt^2/z),
\label{eq:lund_ff}
\end{equation}
where $a$ and $b$ are free parameters adjusted to bring the fragmentation into 
accordance with measured LEP data, e.g. $a$ = 0.3 and $b$ = 0.58 GeV$^{-2}$ are
the current default values for PYTHIA 6.3~\cite{pythia6.3}.
In addition, for the flavor-diagonal meson states $u\bar{u}:d\bar{d}:s\bar{s}$, 
PYTHIA also includes mixing into the physical mesons. This is done according to a 
parametrization, based on the mixing angles given in the Review of Particle 
Properties~\cite{pdg}. In particular, the default choices correspond to 
$\eta = 1/2 (u\bar{u} + d\bar{d}) - 1/\sqrt{2}(s\bar{s})$ and 
$\eta^{'} = 1/2 (u\bar{u} + d\bar{d}) + 1/\sqrt{2}(s\bar{s})$.
Thus, in the  $\pi^0-\eta-\eta^{'}$ system, no account is taken of the difference 
in masses, an approximation which seems to lead to an overestimate of $\eta^{'}$ 
rates in $e^+e^-$ annihilation~\cite{aleph92_eta_Buskulic}. PYTHIA includes therefore parameters 
to allow an additional ``tunable'' suppression of $\eta$ and $\eta^{'}$ states.\\

The PYTHIA Monte Carlo simulations of $\pi^0$ and $\eta$ $\pt$-differential cross-sections 
were carried out with the default settings. In particular, no ad-hoc suppression of $\eta$ was
selected. Any uncertainty related to the choice of any (flavor-independent) settings 
should in principle cancel in the ratio of both $\pt$ spectra. 
As seen in Figs.~\ref{fig:hh_eta_pi0_ratios}--\ref{fig:AA_eta_pi0_ratios}, within the (relatively large in some cases) experimental uncertainties, good agreement between 
the $R_{\eta/\pi^0}(\pt)$ data and the model prediction is found
for all the colliding systems and $\pt$ ranges, despite being at very different
center-of-mass energies. We have also run PYTHIA at $\sqrt{s}$ = 30 GeV as a reference 
for lower energy results, but the resulting $\eta/\pi^0$ curve, though slightly lower 
at high-$\pt$ ($R_{\eta/\pi^0(\pt)}\approx$ 0.44) is still relatively close to the 
one obtained at $\sqrt{s}$ = 200 GeV. This is an indication that the $\pt$ dependence of 
the production mechanisms for both neutral mesons are very similar for all systems and 
c.m.~energies and, correspondingly, the ratio of $\pt$-differential cross-sections is basically 
independent of the characteristics of the initial collision process, but dominated by the 
ratio of $\eta$ and $\pi^0$ (vacuum) FF which is relatively constant in this kinematic 
range (see discussion in Section~\ref{sec:eta_pi0_ee}).\\


\begin{table*}[htbp]
\caption{\label{tab:hh}
Hadron-hadron collisions with a published $\eta/\pi^0$ ratio and/or 
$\eta$ and $\pi^0$ spectra. For each reaction we quote: the center-of-mass energy $\sqrt{s}$ 
(and $p_{lab}$ for fixed-target experiments), the $\pt$ and $x_T=2\pt/\sqrt{s}$ ranges of the measurement 
(the $x_T$ values are not quoted for ``soft'' spectra below $\pt$ = 1 GeV/$c$), and the average 
$\eta/\pi^{0}$ ratio above $\pt$ = 2 GeV/$c$ obtained by fitting $R_{\eta/\pi^0}(\pt>$ 2 GeV/$c$) to a constant.}
\begin{ruledtabular}\begin{tabular}{lcccccccc}
System  	& $\sqrt{s}$ (GeV) & $p_{lab}$ (GeV/$c$) & $p_{T}$ range (GeV/$c$)& $x_{T}$ range & 
$R_{\eta/\pi^0}(\pt>$ 2 GeV/$c$) & Authors & Collab./Exp. &  Ref. \\ \hline

p+p 	        & 13.8 & 100. & $1.6-2.4$     & $0.3-0.4$     & $0.52 \pm 0.13$ & Donaldson 78 & FNAL M2 & \cite{Donaldson78} \\
$\pi^+$+p 	& 13.8 & 100. & $1.6-3.$      & $0.3-0.4$     & $0.49 \pm 0.10$ & Donaldson 78 & FNAL M2 & \cite{Donaldson78} \\
$\pi^-$+p 	& 13.8 & 100. & $2.-3.$       & $0.3-0.4$     & $0.41 \pm 0.13$ & Donaldson 78 & FNAL M2 & \cite{Donaldson78} \\
$\pi^+$+p 	& 19.4 & 200. & $2.-3.5$      & $0.2-0.4$     & $0.40 \pm 0.07$ & Donaldson 78 & FNAL M2 & \cite{Donaldson78} \\
$\pi^-$+p 	& 19.4 & 200. & $1.5-4.$      & $0.2-0.4$     & $0.43 \pm 0.04$ & Donaldson 78 & FNAL M2 & \cite{Donaldson78} \\
p+p 	        & 19.4 & 200. & $2.-3.5$      & $0.2-0.4$     & $0.42 \pm 0.04$ & Donaldson 78 & FNAL M2 & \cite{Donaldson78} \\

p+p 		& 23.0 & 280. & $4.-5.5$      & $0.2-0.4$     & $0.60 \pm 0.04$ & Bonesini 88 & CERN WA70 & \cite{Bonesini88} \\
$\pi^+$+p 	& 23.0 & 280. & $4.-5.5$      & $0.2-0.4$     & $0.43 \pm 0.05$ & Bonesini 88 & CERN WA70 & \cite{Bonesini88} \\
$\pi^-$+p 	& 23.0 & 280. & $4.-5.5$      & $0.2-0.4$     & $0.57 \pm 0.06$ & Bonesini 88 & CERN WA70 & \cite{Bonesini88} \\

p+p 		& 24.3 &      & $2.5-4.$      & $0.2-0.3$     & $0.45 \pm 0.06$ & Antille 87 & CERN UA6 & \cite{Antille87} \\
$\bar{p}$+p 	& 24.3 &      & $2.5-4.$      & $0.2-0.3$     & $0.48 \pm 0.04$ & Antille 87 & CERN UA6 & \cite{Antille87} \\
p+p		& 27.5 & 400. & $0.2-1.6$     & $-$ 	      & $-$ &  Aguilar 91 & NA 27  & \cite{na27} \\

p+p 		& 30.6 &      & $0.8-3.$      & $\sim0.1-0.2$ & $0.55 \pm 0.04$  & Amaldi 79  &  ISR & \cite{Amaldi79} \\
p+p		& 30.6 &      & $3.-4.$       & $0.2-0.3$     & $0.54 \pm 0.05$  & Kourkoumelis 79 & ISR & \cite{Kourkoumelis79} \\
p+p 		& 31.6 & 530. & $3.-8.$       & $0.2-0.5$     & $0.41 \pm 0.03$  & Apanasevich 02 & FNAL 706 & \cite{Apanasevich02} \\
p+p 		& 38.8 & 800. & $3.-8.$       & $0.1-0.4$     & $0.44 \pm 0.03$  & Apanasevich 02 & FNAL 706 & \cite{Apanasevich02} \\
p+p		& 52.7 &      & $3.-6.$       & $0.1-0.3$     & $0.58 \pm 0.03$  & Kourkoumelis 79 & ISR & \cite{Kourkoumelis79} \\

$\bar{p}$+p	& 53.  &      & $2.5-4.$      & $0.1-0.2$     & $0.53 \pm 0.03$ & Akesson 85 & ISR AFS & \cite{Akesson85} \\
p+p     	& 53.  &      & $2.5-4.$      & $0.1-0.2$     & $0.55 \pm 0.02$ & Akesson 85 & ISR AFS & \cite{Akesson85} \\
p+p 		& 53.2 &      & $3.-6.$       & $0.1-0.2$     & $0.54 \pm 0.03$ & Amaldi 79  &  ISR & \cite{Amaldi79} \\
p+p 		& 62.4 &      & $3.-11.$      & $0.2-0.4$     & $0.55 \pm 0.03$ & Kourkoumelis 79 & ISR AFS & \cite{Kourkoumelis79} \\
p+p     	& 63.  &      & $0.2-1.5$     & $-$  &        $(0.07 \pm 0.055)$  & Akesson 86 & ISR AFS & \cite{Akesson86} \\
p+p     	& 63.  &      & $2.-4.$       & $0.06-0.13$   & $0.47 \pm 0.01$ & Akesson 83 & ISR AFS & \cite{Akesson83} \\
p+p 		& 200. &      & $2.-12.$      & $0.02-0.12$   & $0.48 \pm 0.03$ & S.S.Adler 06 & PHENIX & this work \\
$\bar{p}$+p	& 540. &      & $3.-6.$       & $0.01-0.02$   & $0.60 \pm 0.04(stat) \pm 0.15(syst)$ & Banner 85 & CERN UA2 & \cite{Banner85} \\
$\bar{p}$+p	& 1800.&      & $12.0$        & $0.01$        & $1.02 \pm 0.15(stat) \pm 0.23(syst)$ & Abe 93 & CDF & \cite{Abe93} \\
\end{tabular}\end{ruledtabular}
\end{table*}


\begin{table*}[htbp]
\caption{\label{tab:hA}
Hadron-nucleus collisions with a published $\eta/\pi^0$ ratio and/or 
$\eta$ and $\pi^0$ spectra. For each reaction we quote: the center-of-mass energy $\sqrtsnn$ 
(and $p_{lab}$ for fixed-target experiments), the $\pt$ and $x_T=2\pt/\sqrt{s}$ ranges of the measurement 
(the $x_T$ values are not quoted for ``soft'' spectra below $\pt$ = 1 GeV/$c$), and the average 
$\eta/\pi^{0}$ ratio above $\pt$ = 2 GeV/$c$ obtained by fitting $R_{\eta/\pi^0}(\pt>$ 2 GeV/$c$) to a constant.}
\begin{ruledtabular}\begin{tabular}{lcccccccc}
System  	& $\sqrtsnn$ (GeV) & $p_{lab}$ (GeV/$c$) & $p_{T}$ range (GeV/$c$)& $x_{T}$ range & 
$R_{\eta/\pi^0}(\pt>$ 2 GeV/$c$) & Authors & Collab./Exp. &  Ref. \\ \hline

p+Be		& 19.4 & 200. & $2.5-4.$      & $0.2-0.4$     & $0.28 \pm 0.15$  & Povlis 83 & FNAL E629 & \cite{Povlis83} \\
p+C		& 19.4 & 200. & $2.-5.$       & $0.2-0.5$     & $0.58 \pm 0.05$  & Povlis 83 & FNAL E629 & \cite{Povlis83} \\
p+Al		& 19.4 & 200. & $2.-3.$       & $0.2-0.3$     & $0.40 \pm 0.18$  & Povlis 83 & FNAL E629 & \cite{Povlis83} \\
$\pi^-$+C	& 19.4 & 200. & $2.-4.$       & $0.2-0.5$     & $0.32 \pm 0.11$  & Povlis 83 & FNAL E629 & \cite{Povlis83} \\
p+Be		& 23.8 & 300. & $2.5-5.$      & $0.2-0.4$     & $0.47 \pm 0.03$  & Deschamps 85 & FNAL E515 & \cite{Delchamps85} \\
p+Be		& 29.1 & 450. & $0.1-1.$      & $-$   	      & $-$ &  Agakichiev 98 & TAPS/CERES  & \cite{taps_ceres} \\
p+Au		& 29.1 & 450. & $0.1-1.2$     & $-$	      & $-$ &  Agakichiev 98 & TAPS/CERES  & \cite{taps_ceres} \\
p+Be		& 29.1 & 450. & $0.2-1.6$     & $-$	      & $-$ &  Tikhomirov 95 & HELIOS & \cite{helios95} \\
p+Be		& 30.7 & 500. & $4.-7.$       & $0.3-0.5$     & $0.40 \pm 0.06$ & Alverson 93 & FNAL E706 & \cite{Alverson93} \\
$\pi^-$+Be	& 30.7 & 500. & $4.-8.$       & $0.2-0.5$     & $0.43 \pm 0.05$ & Alverson 93 & FNAL E706 & \cite{Alverson93} \\
$\pi^-$+p 	& 31.1 & 515. & $3.-8.$       & $0.1-0.5$     & $0.41 \pm 0.05$ & Apanasevich 03 & FNAL 706 & \cite{Apanasevich03} \\
$\pi^-$+Be 	& 31.1 & 515. & $3.-8.$       & $0.1-0.5$     & $0.48 \pm 0.01$ & Apanasevich 03 & FNAL 706 & \cite{Apanasevich03} \\
$\pi^-$+Cu 	& 31.1 & 515. & $3.-8.$       & $0.1-0.5$     & $0.50 \pm 0.02$ & Apanasevich 03 & FNAL 706 & \cite{Apanasevich03} \\
p+Be 		& 31.6 & 530. & $3.-8.$       & $0.1-0.5$     & $0.42 \pm 0.01$ & Apanasevich 03 & FNAL 706 & \cite{Apanasevich03} \\
p+Cu 		& 31.6 & 530. & $3.-8.$       & $0.1-0.5$     & $0.42 \pm 0.02$ & Apanasevich 03 & FNAL 706 & \cite{Apanasevich03} \\
p+Be 		& 38.8 & 800. & $3.-8.$       & $0.1-0.4$     & $0.42 \pm 0.01$ & Apanasevich 03 & FNAL 706 & \cite{Apanasevich03} \\
p+Cu 		& 38.8 & 800. & $3.-8.$       & $0.1-0.4$     & $0.45 \pm 0.03$ & Apanasevich 03 & FNAL 706 & \cite{Apanasevich03} \\

d+Au 		& 200. &      & $2.-12.$      & $0.02-0.1$    & $0.47 \pm 0.03$ & S.S.Adler 06 & PHENIX & this work \\

\end{tabular}\end{ruledtabular}
\end{table*}


\begin{table*}[htbp]
\caption{\label{tab:AA}
Nucleus-nucleus collisions with a published $\eta/\pi^0$ ratio and/or 
$\eta$ and $\pi^0$ spectra. For each reaction we quote: the center-of-mass energy $\sqrtsnn$ 
(and $p_{lab}$ for fixed-target experiments), the $\pt$ and $x_T=2\pt/\sqrtsnn$ ranges of the measurement 
(the $x_T$ values are not quoted for ``soft'' spectra below $\pt$ = 1 GeV/$c$), and the average 
$\eta/\pi^{0}$ ratio above $\pt$ = 2 GeV/$c$ obtained by fitting $R_{\eta/\pi^0}(\pt>$ 2 GeV/$c$) to a constant.}
\begin{ruledtabular}\begin{tabular}{lcccccccc}
System  	& $\sqrtsnn$ (GeV) & $p_{lab}$ (GeV/$c$) & $p_{T}$ range (GeV/$c$)& $x_{T}$ range & 
$R_{\eta/\pi^0}(\pt>$ 2 GeV/$c$) & Authors & Collab./Exp. &  Ref. \\ \hline

C+C  		& 2.7 & 2.  &   $0.-0.8$     & $-$     & $-$ & Averbeck 97 & GSI TAPS & \cite{Averbeck97} \\
Ca+Ca  		& 2.7 & 2.  &   $0.-0.7$     & $-$     & $-$ & Averbeck 03 & GSI TAPS & \cite{Averbeck03} \\
Ni+Ni 		& 2.7 & 1.9 &   $0.-0.7$     & $-$     & $-$ & Averbeck 03 & GSI TAPS & \cite{Averbeck03} \\

Pb+Pb  		& 17.3 & 158. & $0.6-2.6$    & $\sim0.1-0.3$ & $0.53 \pm 0.21$ & Aggarwal 00 & CERN WA98  & \cite{Aggarwal00} \\

S+S  		& 19.4 & 200. & $0.5-1.5$    & $0.1-0.2$     & $0.21 \pm 0.06$ & Albrecht 95 & CERN WA80 & \cite{Albrecht95} \\
S+Au 		& 19.4 & 200. & $0.5-3.5$    & $0.1-0.3$     & $0.61 \pm0.14$  & Albrecht 95 & CERN WA80 & \cite{Albrecht95} \\

Au+Au		& 200. &      & $2.-10.$     & $0.02-0.1$    & $0.40 \pm 0.04$ & S.S.Adler 06 & PHENIX & this work \\

\end{tabular}\end{ruledtabular}
\end{table*}


\paragraph{{\it $\mt$ scaling}:}

The red shaded curve shown in Figs. \ref{fig:hh_eta_pi0_ratios}, \ref{fig:hA_eta_pi0_ratios}
and \ref{fig:AA_eta_pi0_ratios} corresponds to an empirical $\mt$-scaling 
observation~\cite{mt_scaling} which assumes that the hadron differential cross-sections, 
plotted as a function of the transverse mass of the produced particle $\mt = \sqrt{m_h^2+\pt^2}$, 
all have the same shape, $f(\mt)$, with an absolute normalization factor $C_h$ which 
can vary but which is found to be the same for many species:
\begin{equation}
E\frac{d^3\sigma_h}{d^3p} = C_h\cdot f(\mt)
\label{eq:mt}
\end{equation}
Assuming isospin symmetry for pion production, we have combined the measured PHENIX 
charged $(\pi^++\pi^-)/2$ (measured in the range $\pt$ = 0.2 -- 2.6 GeV/$c$)~\cite{ppg045} 
and neutral ($\pt$ = 1 -- 14 GeV/$c$)~\cite{ppg044} pion differential $\mt$ cross-sections 
in p+p collisions and fitted them with a modified power-law functional form\footnote{Note 
that the $a$ and $n$ parameters of Eq.~(\ref{eq:mt_pow}) are not independent but strongly 
correlated. They are actually related to the mean transverse mass of the spectrum via 
$\mean{\mt} = 2a/(n-3)$.} that reproduces the full spectra well in the range 
$\mt \approx$ 0.2 -- 14 GeV/$c^2$:
\begin{equation}
f(\mt) = (\mt+a)^{-n} \;,\mbox{  with $a$ = 1.2 and $n$ = 10.}
\label{eq:mt_pow}
\end{equation}
If one assumes that $\mt$-scaling holds for $\eta$, then its $\mt = \sqrt{m_{\eta}^2+\pt^2}$ 
spectrum can also be represented by Eqs.~(\ref{eq:mt}) and (\ref{eq:mt_pow}) 
(with, in principle, a different $C_{h}$) and therefore, the $\eta/\pi^0$ ratio 
as a function of $\pt$ should follow:
\begin{equation}
R_{\eta/\pi^0}(\pt) = R_{\eta/\pi^0}^{\infty}\cdot\left[\frac{a+\sqrt{m_{\eta}^2+\pt^2}}{a+\sqrt{m_{\pi^{0}}^2+\pt^2}}\right]^{n},
\label{eq:eta_pi0_ratio_mt}
\end{equation}
where $R_{\eta/\pi^0}^{\infty} = C_\eta/C_{\pi^0}$ is the asymptotic value 
of the ratio of $\eta$ over $\pi^0$ for large $\pt$. Note that since the assumption 
of $\mt$-scaling is that both $\mt$-differential cross-sections have the same shape, 
the same parameters $a$ = 1.2 and $n$ = 10 are valid for both spectra as well as 
for the ratio (Eq.~\ref{eq:eta_pi0_ratio_mt}). In all figures, the plotted $\mt$-scaling 
curve with an asymptotic value of $R_{\eta/\pi^0}^{\infty}$ = 0.5 is found to be 
in good agreement with both the data and the PYTHIA predictions. We note that the 
agreement between PYTHIA and $\mt$-scaling is not unexpected in as much as the Lund 
``fragmentation function,'' Eq.~(\ref{eq:lund_ff}), depends explicitly on the $\mt$ 
of the produced hadron. The upper red curve shown in all plots is that 
with the $a$ and $n$ parameters of Eq.~(\ref{eq:eta_pi0_ratio_mt}) that reproduce the 
power-law shape of the meson spectra at $\sqrt{s}$ = 200 GeV. At lower $\sqrt{s}$, 
the spectra get increasingly steeper and $a$ and $n$ change accordingly ($a$ and $n$ 
are correlated with $\mean{\mt}$ which itself is a logarithmically increasing function 
of $\sqrt{s}$, i.e. $\mean{\mt}=f(\sqrt{s})$). For illustrative purposes, we have 
(arbitrarily) fixed the parameter $a$ to the value $a$ = 1.2 
and refitted the $\pi^0$ spectra measured at different center-of-mass energies
with $n$ as a free exponent. 
With fixed $a$ the corresponding values of the power-law exponent increase with decreasing 
$\sqrt{s}$ as $n\approx$ 10., 11.5, 13.5 and 14.0 at $\sqrt{s}$ = 200, 63, 27 and 13 GeV, 
respectively. The shaded red area indicates the range of expected $\mt$-scaling ratios 
for power-law exponents $n$ = 10.--14. The differences are negligible at large $\pt$
-- where the $\eta$ and $\pi^0$ masses are much smaller than their $\pt$ and the ratio
(Eq.~\ref{eq:eta_pi0_ratio_mt}) $R_{\eta/\pi^0}(\pt) \approx R_{\eta/\pi^0}^{\infty}$ 
is independent of $n$ -- but become increasingly large at lower $\pt$ ($\pt\lesssim$ 3 GeV$c$).
Furthermore, it is worth noting that in the low-$\pt$ region below 1 GeV/$c$, the agreement 
between the data and the $\mt$-scaling curve is not always perfect for all data sets, even taking into account different 
power-law exponents. This is due to the fact that at very low $\mt\approx$ 0.--0.4 GeV/$c^2$, 
the pion yield rises due to contributions from multiple resonance decays and  
the formula~(Eq.~\ref{eq:mt_pow}) does not reproduce the spectral shape of the data anymore. 
Instead, an exponential behavior of the form $Ed^3\sigma/d^3p = B\cdot exp(-b\,\mt)$~\cite{hagedorn} 
extrapolates the spectra better in the soft regime all the way down to $\mt$ = 0 GeV/$c^2$. 
However, for all practical purposes in this analysis focused on high-$\pt$ production, 
we will consider Eq.~(\ref{eq:mt_pow}) (and correspondingly Eq.~\ref{eq:eta_pi0_ratio_mt}) 
to be a good enough approximation.\\

Lastly we want to mention that in the case of nucleus-nucleus collisions 
the existence of a strong collective radial flow ($\beta_{coll}\approx$ 0.6 at RHIC~\cite{phnx_whitepaper}),
absent in p+p collisions, changes the spectral shape of different hadrons produced at low 
transverse momenta ($\pt\lesssim$ 2 GeV/$c$) and should result in a violation of the 
$\mt$-scaling behavior~\cite{witt04}. Since hydrodynamical flow 
results in a larger boost for the (heavier) $\eta$ than for $\pi^0$, one expects a comparatively larger 
$R_{\eta/\pi^0}(\pt)$ ratio in Au+Au than in p+p collisions below $\pt\approx$ 2 GeV/$c$.
Unfortunately, we cannot test this assertion with RHIC data since our lowest $\pt$ value 
($\pt\approx$ 2 GeV/$c$) is just in the range where radial flow effects start to die out.
The same holds true also for the recent proposal~\cite{peitzmann05} to study 
the $\eta/\pi^0$ ratio as a tool to test different parton recombination scenarios 
in hadron production in nucleus-nucleus collisions. Lower-$\pt$ $\eta$ measurements, 
which are intrinsically more difficult due to the reduced PHENIX acceptance and the larger 
$\gaga$ combinatorial background, would be needed to better address the role of collective 
flow and/or parton recombination effects on the spectral shape and yields of light neutral 
mesons in Au+Au collisions at RHIC.


\subsubsection{$\eta/\pi^0$ ratio in $e^+e^-$ collisions at the $Z$ pole ($\sqrt{s}=$ 91.2 GeV)}
\label{sec:eta_pi0_ee}

In this last section 
we are interested in determining the $\eta/\pi^0$ ratio in an elementary colliding 
system such as $e^+e^-$ and comparing it to the corresponding ratios obtained in hadronic 
and nuclear collisions. In $e^+e^-$ the dominant high-momentum hadron production mechanism 
is $q,\bar{q}$ fragmentation since gluon production (and subsequent fragmentation) occurs 
with a probability which is suppressed by a factor $\alpha_{S}$, and therefore plays a comparatively 
less significant role than in the (highest energy) hadronic and nuclear collisions discussed 
in the previous section. Some of the experimental interest in the 
study of $\eta$ production in $e^+e^-$ collisions was in fact triggered by theoretical expectations 
that the isoscalar mesons contained a significant $g\,g$ component, and thus that gluon 
jets should exhibit an anomalously large tendency to fragment into $\eta$ and $\eta'(958)$
mesons~\cite{Peterson:1980ax,Ball:1995zv,Fritzsch:1997ps}. 
However, this hypothesis was not confirmed by a detailed analysis of the 
ALEPH $e^+e^-$ gluon fragmentation data~\cite{aleph99_eta_pi0_Barate}.
Table~\ref{tab:ee} lists all the existing measurements of inclusive $\pi^0$ and $\eta$ 
production in $e^+e^-$ collisions at LEP at energies around the $Z$ pole. At lower energies,
there are several results on inclusive $\pi^0$ production in $e^+e^-$ but few 
$\eta$ measurements exist ($\sqrt{s}=$ 29 and 35 GeV at SLAC 
PEP~\cite{Wormser:1988ru,Abachi:1987qd} and SLC~\cite{Bartel:1983pe,Bartel:1985wn,Pitzl:1989qy,Behrend:1989gn}, 
respectively), and we could not determine the corresponding ratios. 

\begin{table}[htbp]
\caption{Experimental measurements at LEP of $\eta$, $\pi^0$ spectra in $e^+e^-$ collisions
at $\sqrt{s}=$ 91.2 GeV.}
\label{tab:ee}
\begin{ruledtabular}\begin{tabular}{lcccc}
Collaboration - Year \hspace{0.5cm} & \hspace{0.5cm} Particle  \hspace{0.5cm} & \hspace{0.5cm} Authors [Ref.] \hspace{0.5cm} \\ \hline
ALEPH 92  &  $\eta$  &   Buskulic {\it et al.} \cite{aleph92_eta_Buskulic} \\
ALEPH 96a &  $\pi^0$  &   Barate {\it et al.} \cite{aleph96_pi0_Barate} \\
ALEPH 96b &  $\pi^0,\eta$  &   Barate {\it et al.} \cite{aleph96_eta_pi0_Barate} \\
ALEPH 99  &  $\pi^0,\eta$  &   Barate {\it et al.} \cite{aleph99_eta_pi0_Barate} \\
ALEPH 01  &  $\eta$  &   Heister {\it et al.} \cite{aleph01_eta_Heister} \\
DELPHI 95 &  $\pi^0$  &  Adam {\it et al.} \cite{delphi95_pi0_Adam} \\
L3 91     &  $\pi^0$  &  Adeva {\it et al.} \cite{l3_91_pi0_Adeva} \\
L3 92     &  $\eta$  &  Adriani {\it et al.} \cite{l3_92_eta_Adriani} \\
L3 94a    &  $\pi^0,\eta$  &  Acciarri{\it et al.} \cite{l3_94_eta_pi0_Acciarri} \\
OPAL 98   &  $\pi^0,\eta$  &  Ackerstaff {\it et al.} \cite{opal98_eta_pi0_Ackerstaff} \\
OPAL 00   &  $\pi^0,\eta$  &  Abbiendi {\it et al.} \cite{opal00_eta_pi0_Abbiendi} \\

\end{tabular}\end{ruledtabular}
\end{table}


\begin{figure}[htbp]
\centering
\includegraphics[width=1.0\linewidth]{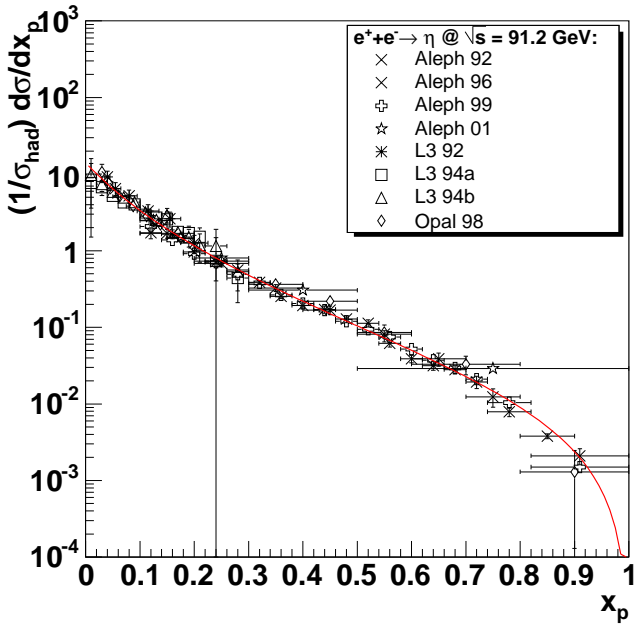}
\caption{Differential inclusive $\eta$ cross-section as a function of the 
scaled momentum $x_p=2\,p_{had}/\sqrt{s}$ measured at the $Z$ pole by 
the three LEP experiments listed in Table~\protect\ref{tab:ee}, fitted to  
Eq.~(\ref{eq:ff_eta_fit}) (solid curve).}
\label{fig:ee_eta}
\end{figure}

\begin{figure}[htbp]
\centering
\includegraphics[width=1.0\linewidth]{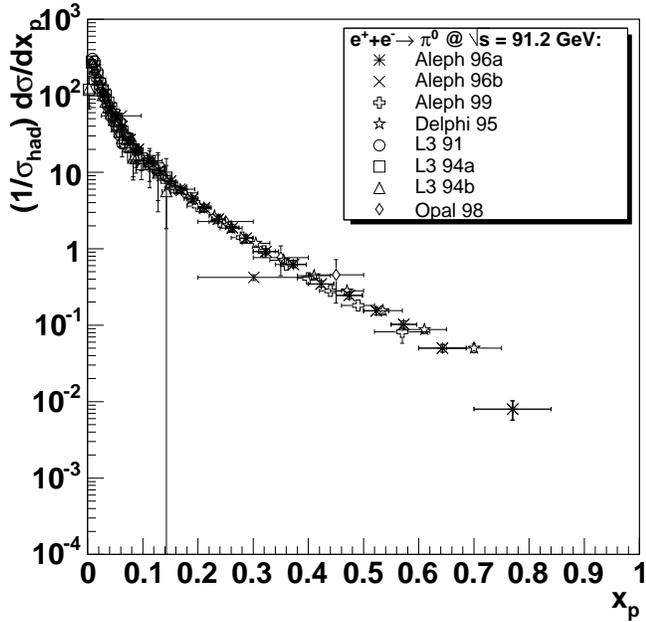}
\caption{Differential inclusive $\pi^0$ cross-section as a function of
the scaled momentum $x_p=2\,p_{had}/\sqrt{s}$ measured at the $Z$ pole 
by the four LEP experiments listed in Table~\protect\ref{tab:ee}.}
\label{fig:ee_pi0}
\end{figure}

Figures~\ref{fig:ee_eta} and \ref{fig:ee_pi0} show the combined inclusive $\eta$ and $\pi^0$ 
invariant cross-sections measured as a function of the scaled particle momentum $x_{p}=2\,p_{had}/\sqrt{s}$. 
Note that the overall $\eta$ and $\pi^0$ spectra have been measured in $x_p$ ranges which are not completely 
overlapping. There are more experimental measurements on inclusive $\eta$ ($\pi^0$) production 
at large (small) $x_p\gtrsim$0.7 ($x_p\lesssim$0.1). For this reason, in order 
to obtain the ratio of $\eta$ over $\pi^0$ cross-sections, we have parametrized the $\eta$ 
cross-section as 
\begin{equation}
\frac{1}{\sigma_{had}}\frac{d\sigma_{\eta}}{dx_{p}} = A \cdot (x_p+b)^n \cdot (1-x_p)^m\;\; 
\label{eq:ff_eta}
\end{equation}
and taken the ratio of the individual $\pi^0$ data points over the resulting fit. 
We note that there is currently no $\eta$ FF available 
in the standard FF sets at hand in the literature (BKK~\cite{bkk}, KKP~\cite{kkp}, Kretzer~\cite{kretzer}, 
BFGW~\cite{bfgw}). Namely, the LEP data compiled in Fig.~\ref{fig:ee_eta} 
have not been fitted and coded so far into any usable format that can be handled within
a QCD collinear factorization approach. We are aware of only two works (Rolli {\it et al.} at 
NLO~\cite{eta_FF_rolli}, and Indumathi and collaborators at LO~\cite{Indumathi:1998am}) 
which have tried to parametrize the $\eta$ FF from these data. 
An updated version of the $\eta$ FF would be useful as input to a NLO pQCD cross-section 
calculation for comparison to the results presented here and especially in the light of upcoming 
high-$\pt$ $\eta$ asymmetry results using polarized beams of relevance for the proton spin 
program at RHIC~\cite{phenix_eta_spin}. Fitting all the available $\eta$ data with Eq.~(\ref{eq:ff_eta}), 
we obtain the following empirical parametrization:
\begin{equation}
\frac{1}{\sigma_{had}}\frac{d\sigma_{\eta}}{dx_{p}} = 0.0975\cdot (x_p+0.186)^{-2.953}\cdot (1-x_p)^{1.507}\;\;,\;\mbox{  $\chi^2/$ndf = 0.37.}
\label{eq:ff_eta_fit}
\end{equation}


\begin{figure}[htbp]
\centering
\includegraphics[width=1.0\linewidth]{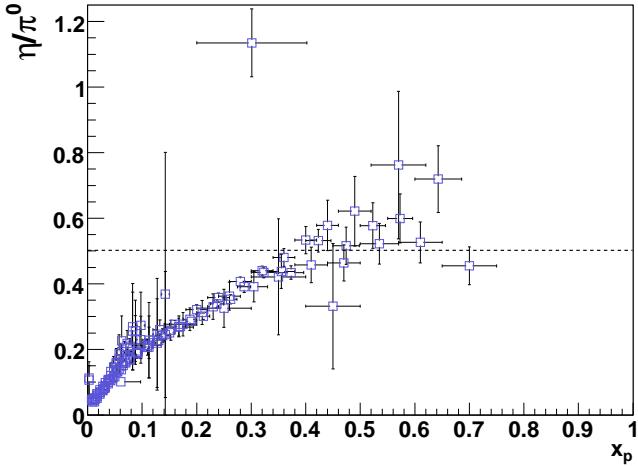}
\caption{Ratio $\eta/\pi^0$ versus scaled momentum $x_p=2\,p_{had}/\sqrt{s}$ 
measured in $e^+e^-$ collisions at LEP energies (Table~\protect\ref{tab:ee}), 
obtained from the $\pi^0$ results of Fig.~\protect\ref{fig:ee_pi0} and the $\eta$ fit, 
Eq.~(\protect\ref{eq:ff_eta_fit}). The dashed line is the asymptotic $R_{\eta/\pi^0}$ = 0.5 
measured in hadronic and nuclear collisions.}
\label{fig:ee_eta_pi0_ratio}
\end{figure}

Using the fit (Eq.~\ref{eq:ff_eta_fit}) and the $\pi^0$ data plotted in
Fig.~\ref{fig:ee_pi0} we have obtained the $R_{\eta/\pi^0}(x_p)$ ratio shown in 
Fig.~\ref{fig:ee_eta_pi0_ratio}. As seen for the corresponding $\eta/\pi^0$ ratios in hadronic 
and nuclear collisions, at low values of (scaled) momentum the $\pi^0$ production overwhelms 
that of $\eta$ (a significant fraction of low-energy pions issues from decay contributions of
heavier hadrons), but the ratio increases with $x_p$.  From $x_p\approx 0.35 - 0.7$, the ratio
is consistent with the asymptotic ratio of 0.5 found in hadron and nuclear collisions (dashed curve). 
This $x_p$ range corresponds to the values of fractional momenta 
$\mean{z}\gtrsim 0.3 - 0.7$ typically carried by the leading high-$\pt$ hadrons 
produced in high-energy h+p, h+A and A+A collisions~\cite{ppg029,kretzer2}.
New results on inclusive $\eta$ and $\pi^0$ production above $x_p$ = 0.6
in $e^+e^-$ collisions at the $B$-factories (BELLE and BaBar) would be useful to determine 
whether the value of the ratio indeed saturates at $R_{\eta/\pi^0}$ = 0.5 
or keeps increasing with $x_p$ as suggested by Fig.~\ref{fig:ee_eta_pi0_ratio}.


\vspace{4.cm}

\subsection{Summary of Experimental Results}
\label{sec:results_summary}

The studies presented here on high-$p_T$ $\pi^0$ and $\eta$ production in the three colliding 
systems (p+p, d+Au, Au+Au) provide interesting insights on initial- and final-state
QCD effects in cold nuclear matter (d+Au) and on the properties of the hot and dense medium
produced in central Au+Au collisions. The absence of any strong deviation from the point-like 
scaling expectations for the $p_T$-differential $\pi^0$ and $\eta$ yields measured in d+Au 
relative to p+p (Fig.~\ref{fig:RdA_pi0_eta}) indicates that the amount of nuclear shadowing 
and initial-state $p_T$ broadening is a small effect (at the 10\% level) at mid-rapidity at RHIC energies. 
This is in contrast with results at lower energies which showed a larger Cronin enhancement
for high-$p_T$ mesons than observed here. One reason for the difference is likely due to the
fact that hadron spectra at lower $\sqrtsnn$ have steeper slopes and thus initial-state $k_T$
``kicks'' produce a relatively larger net effect than on the harder spectra at RHIC energies.
The unsuppressed d+Au yields combined with the observation of strongly depleted yields of 
$\eta$ and $\pi^0$ in central Au+Au compared to binary-scaled p+p collisions (Fig.~\ref{fig:R_AA_eta}) indicate that the 
suppression is a final-state effect in the hot and dense matter produced in the central Au+Au reactions. 
The consistent values of the $\eta/\pi^0$ ratios measured at high $p_T$ 
in nuclear (Figs.~\ref{fig:hA_eta_pi0_ratios} and \ref{fig:AA_eta_pi0_ratios}) as well as 
in more elementary p+p  (Fig.~\ref{fig:hh_eta_pi0_ratios}) and $e^+e^-$  
(Fig.~\ref{fig:ee_eta_pi0_ratio}) collisions clearly supports the idea that the suppression
occurs at the parton level before the fragmentation of the parent quarks and gluons into 
a given leading meson. In particular, the overall agreement of the $\eta/\pi^0$ ratio measured 
in Au+Au and $e^+e^-$ collisions suggests that although the fast parent partons lose energy
while traversing the system produced in central Au+Au collisions, their relative probability to fragment 
into a given meson, given by universal fragmentation functions, is preserved as expected for 
final hadron formation in the vacuum.


\section{SUMMARY} 

\label{sec:summary}

In summary, the transverse momentum spectra of $\eta$ mesons in the range 
$\pt$ = 2--12~GeV$/c$ have been measured at mid-rapidity by 
the PHENIX experiment at RHIC in p+p, d+Au and Au+Au collisions at 
$\sqrt{s_{_{NN}}}$ = 200~GeV. The $\eta$ mesons are reconstructed through their
$\eta\rightarrow\gaga$ channel in the three colliding systems, as well as 
through the $\eta\rightarrow\pi^0\pi^+\pi^-$ decay mode in p+p and d+Au collisions. 
These data provide additional characterization of high-$\pt$ hadroproduction 
in hadronic and nuclear collisions at RHIC energies. 
The d+Au yields are largely consistent with the p+p differential 
cross-sections scaled by the number of incoherent nucleon-nucleon collisions
($R_{dA}\approx$ 1). No $\pt$ or centrality dependence is observed in the nuclear 
modification factor within uncertainties. Such an observation indicates a null or 
very weak $\pt$ broadening and, in general, a very modest influence of cold nuclear 
matter effects, such as shadowing of parton distribution functions, Cronin broadening, 
and/or hadronization by parton recombination, on high-$\pt$ meson production at 
mid-rapidity at top RHIC energies. 
In contrast, the invariant yields measured in Au+Au are increasingly depleted
with centrality compared to expectations from binary-scaled p+p collisions, 
up to a maximum factor of $\sim$5 suppression in central collisions. 
The magnitude, $\pt$ and centrality dependence of the Au+Au suppression is the same 
for $\eta$ mesons and neutral pions. The measured $\eta$/$\pi^0$ ratio in p+p, d+Au
and Au+Au is nearly flat over $\pt$ = 2--12 GeV/$c$ and is independent 
of the reaction centrality. A fit to a constant yields $R_{\eta/\pi^{0}}(\pt)$ = 0.4--0.5,
in agreement with the experimental world values at high $\pt$ collected 
here for hadron-hadron, hadron-nucleus and nucleus-nucleus collisions 
in a wide range of center-of-mass energies ($\sqrt{s}\approx$ 3--1800 GeV), 
as well as at high $x_p$ ($x_p\gtrsim$ 0.35) in electron-positron 
annihilations measured at $\sqrt{s}=$ 91.2 GeV at LEP.
These results indicate that any initial- and/or final-state nuclear effects 
influence the production of light neutral mesons at large $\pt$ in the same way.
The similar suppression pattern of $\eta$ and $\pi^0$ mesons is consistent with 
the expectations of final-state parton energy loss in the dense medium formed 
in Au+Au reactions. The approximately constant  $\eta$/$\pi^0 = $ 0.40 $\pm$ 0.04 ratio 
measured in central Au+Au collisions indicates that the attenuated parent 
partons fragment into leading mesons ($\eta, \pi^{0}$) in the vacuum according 
to the same probabilities that govern high-$\pt$ hadron production in more 
elementary ($e^+e^-$, p+p) collisions.


\section*{Acknowledgements}   

We thank the staff of the Collider-Accelerator and Physics
Departments at Brookhaven National Laboratory and the staff of
the other PHENIX participating institutions for their vital
contributions.  We acknowledge support from the Department of
Energy, Office of Science, Office of Nuclear Physics, the
National Science Foundation, Abilene Christian University
Research Council, Research Foundation of SUNY, and Dean of the
College of Arts and Sciences, Vanderbilt University (U.S.A),
Ministry of Education, Culture, Sports, Science, and Technology
and the Japan Society for the Promotion of Science (Japan),
Conselho Nacional de Desenvolvimento Cient\'{\i}fico e
Tecnol{\'o}gico and Funda\c c{\~a}o de Amparo {\`a} Pesquisa do
Estado de S{\~a}o Paulo (Brazil),
Natural Science Foundation of China (People's Republic of China),
Centre National de la Recherche Scientifique, Commissariat
{\`a} l'{\'E}nergie Atomique, and Institut National de Physique
Nucl{\'e}aire et de Physique des Particules (France),
Ministry of Industry, Science and Tekhnologies,
Bundesministerium f\"ur Bildung und Forschung, Deutscher
Akademischer Austausch Dienst, and Alexander von Humboldt Stiftung (Germany),
Hungarian National Science Fund, OTKA (Hungary), 
Department of Atomic Energy (India), 
Israel Science Foundation (Israel), 
Korea Research Foundation 
and Korea Science and Engineering Foundation (Korea),
Ministry of Education and Science, Rassia Academy of Sciences,
Federal Agency of Atomic Energy (Russia),
VR and the Wallenberg Foundation (Sweden), 
the U.S. Civilian Research and Development Foundation for the
Independent States of the Former Soviet Union, the US-Hungarian
NSF-OTKA-MTA, and the US-Israel Binational Science Foundation.


\begin{appendix}

\section*{APPENDIX: DATA TABLES}
\label{app:data_tables}

\subsection{Invariant $\eta$ cross-sections (p+p, d+Au) and yields (Au+Au)}
\label{app:spectra}

This Appendix collects the data tables of the $\pt$ spectra 
of $\eta$  mesons measured
at mid-rapidity in p+p, d+Au and Au+Au collisions at $\sqrtsnn$~=~200~GeV.
The invariant cross-sections for $\eta$ production in MB p+p
and d+Au collisions are tabulated in Table~\ref{tab:pp_eta} 
and Table~\ref{tab:dAu_eta_MB},
respectively. The invariant d+Au yields measured in 
centralities 0--20\%, 20--40\%, 40--60\%, 
and 60--88\% are tabulated in Table~\ref{tab:dAu_eta}.
Finally, the invariant yields in Au+Au reactions (MB and centralities 
0--20\%, 20--60\%, and 60--92\%) are presented in 
Tables~\ref{tab:AuAu_eta}. 
The quoted errors are categorized by type: 

({\bf A}) is a point-to-point error uncorrelated between $\pt$ bins, 

({\bf B}) is $\pt$ correlated, all points move in the same direction 
but not by the same factor, 

({\bf C}) is a normalization error in which all points 
move by the same factor independent of $\pt$.

\subsection{Nuclear modification factors (d+Au, Au+Au)}
\label{app:RAA}

We report below the $R_{AA}(\pt)$ $\eta$ data tables for various 
centralities in d+Au and Au+Au collisions.
The errors quoted are the point-to-point and absolute normalization ones. 
Note that there is an additional 9.7\% overall normalization uncertainty 
(Run-3 p+p BBC error, gray box in Fig.~\ref{fig:R_AA_eta}) not tabulated.

\subsection{$\eta/\pi^0$ ratios (p+p, d+Au, Au+Au)}
\label{app:eta_pi0}

The ratio of $\eta$ to $\pi^0$ invariant yields in p+p, d+Au and Au+Au 
collisions at $\sqrtsnn$~=~200~GeV at mid-rapidity are tabulated in 
Tables~\ref{tab:pp_eta_pi0}, \ref{tab:dAu_eta_pi0}, and 
\ref{tab:phenix_AuAu_200GeV_eta_pi0_ratio}. 
The data presented here are for minimum bias events
and various centrality bins in d+Au and Au+Au collisions. 



\begin{table*}[bh] 
\caption{Inelastic cross section measured in p+p at $\sqrt{s}$ = 200 GeV.}
\label{tab:pp_eta}
\begin{ruledtabular}\begin{tabular}{cccccccc} 
\footnotesize $\pt$ (GeV/$c$) & \footnotesize $\sigma$ (mb)& 
\footnotesize tot. err. & \footnotesize stat. err. & \footnotesize sys. err. & \footnotesize error
A & \footnotesize error B & \footnotesize error C\\ \hline
\multicolumn{8}{c}{$\eta\rightarrow\gamma\gamma$}\\\hline
2.75 & 1.30e-03 & 4.20e-04 & 2.25e-04 & 7.91e-06 & 0 & 3.34e-04 & 1.20e-04 \\
3.25 & 3.78e-04 & 6.81e-05 & 8.09e-06 & 1.37e-06 & 0 & 5.79e-05 & 3.48e-05 \\
3.75 & 1.37e-04 & 2.15e-05 & 3.53e-06 & 4.03e-07 & 0 & 1.70e-05 & 1.26e-05 \\
4.25 & 5.49e-05 & 8.20e-06 & 1.81e-06 & 1.47e-07 & 0 & 6.20e-06 & 5.05e-06 \\
4.75 & 2.22e-05 & 3.34e-06 & 1.03e-06 & 5.76e-08 & 0 & 2.43e-06 & 2.04e-06 \\
5.25 & 1.08e-05 & 1.70e-06 & 6.82e-07 & 2.83e-08 & 0 & 1.20e-06 & 9.90e-07 \\
5.75 & 5.66e-06 & 9.36e-07 & 4.39e-07 & 1.52e-08 & 0 & 6.42e-07 & 5.21e-07 \\
6.5 & 2.02e-06 & 3.47e-07 & 1.75e-07 & 5.58e-09 & 0 & 2.36e-07 & 1.86e-07 \\
7.5 & 6.99e-07 & 1.40e-07 & 9.14e-08 & 2.00e-09 & 0 & 8.45e-08 & 6.43e-08 \\
8.5 & 1.81e-07 & 4.92e-08 & 4.04e-08 & 5.35e-10 & 0 & 2.26e-08 & 1.67e-08 \\
9.5 & 1.02e-07 & 3.22e-08 & 2.80e-08 & 3.08e-10 & 0 & 1.30e-08 & 9.34e-09 \\
11 & 2.21e-08 & 9.24e-09 & 8.52e-09 & 1.35e-10 & 0 & 2.94e-09 & 2.03e-09 \\\hline
\multicolumn{8}{c}{$\eta\rightarrow\pi^{0}\pi^{+}\pi^{-}$}\\
\hline
3.0 & 7.5e-04 & 2.4e-04 & 1.3e-04 & 2.1e-04 & 1.1e-04 & 1.6e-04 & 8.5e-05 \\
4.0 & 8.1e-05 & 2.5e-05 & 1.4e-05 & 2.1e-05 & 1.3e-05 & 1.4e-05 & 9.1e-06 \\
5.0 & 2.0e-05 & 6.6e-06 & 3.3e-06 & 5.8e-06 & 4.5e-06 & 2.9e-06 & 2.3e-06 \\
6.0 & 5.8e-06 & 1.7e-06 & 1.1e-06 & 1.4e-06 & 9.5e-07 & 7.2e-07 & 6.5e-07 \\
7.0 & 1.0e-06 & 5.0e-07 & 3.9e-07 & 3.2e-07 & 2.6e-07 & 1.3e-07 & 1.2e-07 \\
8.0 & 4.5e-07 & 2.7e-07 & 2.1e-07 & 1.6e-07 & 1.5e-07 & 5.6e-08 & 5.1e-08 \\
\end{tabular}\end{ruledtabular}



\caption{Inelastic cross section measured in d+Au at $\sqrtsnn$ = 200 GeV.}
\label{tab:dAu_eta_MB}
\begin{ruledtabular}\begin{tabular}{cccccccc} 
\footnotesize $\pt$ (GeV/$c$) & \footnotesize $\sigma$ (mb) & \footnotesize tot. err. & 
\footnotesize stat. err. & \footnotesize sys. err. & \footnotesize error
A & \footnotesize error B & \footnotesize error C\\ \hline
\multicolumn{8}{c}{$\eta\rightarrow\gamma\gamma$}\\\hline
2.25 & 2.06e+00 & 7.49e-01 & 1.30e-01 & 3.37e-04 & 0 & 7.30e-01 & 1.08e-01 \\
2.75 & 6.25e-01 & 1.09e-01 & 4.06e-02 & 4.41e-05 & 0 & 9.55e-02 & 3.28e-02 \\
3.25 & 1.58e-01 & 2.40e-02 & 1.48e-02 & 7.84e-06 & 0 & 1.70e-02 & 8.27e-03 \\
3.75 & 6.72e-02 & 8.21e-03 & 1.86e-03 & 3.31e-06 & 0 & 7.18e-03 & 3.53e-03 \\
4.25 & 2.57e-02 & 3.18e-03 & 7.56e-04 & 1.28e-06 & 0 & 2.78e-03 & 1.35e-03 \\
4.75 & 1.06e-02 & 1.34e-03 & 3.64e-04 & 5.39e-07 & 0 & 1.17e-03 & 5.55e-04 \\
5.25 & 4.38e-03 & 5.65e-04 & 1.96e-04 & 2.20e-07 & 0 & 4.77e-04 & 2.30e-04 \\
5.75 & 2.30e-03 & 3.08e-04 & 1.22e-04 & 1.18e-07 & 0 & 2.56e-04 & 1.21e-04 \\
6.5 & 9.20e-04 & 1.27e-04 & 5.11e-05 & 4.86e-08 & 0 & 1.05e-04 & 4.83e-05 \\
7.5 & 2.47e-04 & 4.39e-05 & 3.00e-05 & 1.35e-08 & 0 & 2.93e-05 & 1.30e-05 \\
8.5 & 9.17e-05 & 1.89e-05 & 1.44e-05 & 5.18e-09 & 0 & 1.12e-05 & 4.82e-06 \\
9.5 & 4.56e-05 & 9.14e-06 & 6.69e-06 & 2.65e-09 & 0 & 5.75e-06 & 2.40e-06 \\
11 & 1.31e-05 & 2.88e-06 & 2.21e-06 & 7.92e-10 & 0 & 1.72e-06 & 6.88e-07 \\\hline
\multicolumn{8}{c}{$\eta\rightarrow\pi^{0}\pi^{+}\pi^{-}$}\\
\hline
5.0 & 1.1e-02  & 7.6e-03  & 5.7e-03  & 5.1e-03  & 3.7e-03 & 3.2e-03 & 9.9e-04 \\
6.0 & 2.7e-03  & 1.6e-03  & 1.2e-03  & 1.0e-03  & 6.9e-04 & 7.3e-04 & 2.4e-04 \\
7.0 & 5.8e-04  & 4.4e-04  & 3.9e-04  & 1.9e-04  & 1.3e-04 & 1.3e-04 & 5.1e-05 \\
8.0 & 2.4e-04  & 1.4e-04  & 1.1e-04  & 8.4e-05  & 6.1e-05 & 5.4e-05 & 2.1e-05 \\ 
 
\end{tabular}\end{ruledtabular}
\end{table*}




\begin{table*}[ht]
\caption{Invariant yields measured in d+Au for different centrality classes
from most central (0-20\%) to most peripheral (60-88\%).}
\label{tab:dAu_eta}
\begin{ruledtabular}\begin{tabular}{ccccccccc} 
\footnotesize Centrality & \footnotesize $\pt$ (GeV/$c$) & \footnotesize inv. yield & \footnotesize tot. err. & 
\footnotesize stat. err. & \footnotesize sys. err. & \footnotesize error
A & \footnotesize error B & \footnotesize error C\\ \hline 
	& 2.25 & 1.59e-03 & 5.96e-04 & 1.85e-04 & 5.67e-04 & 0 & 5.67e-04 & 0 \\	
	& 2.75 & 5.46e-04 & 1.02e-04 & 5.65e-05 & 8.49e-05 & 0 & 8.49e-05 & 0 \\
	& 3.25 & 1.27e-04 & 2.49e-05 & 2.05e-05 & 1.42e-05 & 0 & 1.42e-05 & 0 \\
	& 3.75 & 5.84e-05 & 6.89e-06 & 2.43e-06 & 6.45e-06 & 0 & 6.45e-06 & 0 \\
	& 4.25 & 2.22e-05 & 2.66e-06 & 9.62e-07 & 2.48e-06 & 0 & 2.48e-06 & 0 \\
	& 4.75 & 9.26e-06 & 1.15e-06 & 4.61e-07 & 1.06e-06 & 0 & 1.06e-06 & 0 \\
	& 5.25 & 3.85e-06 & 4.92e-07 & 2.36e-07 & 4.32e-07 & 0 & 4.32e-07 & 0 \\
0-20\%	& 5.75 & 1.97e-06 & 2.70e-07 & 1.47e-07 & 2.26e-07 & 0 & 2.26e-07 & 0 \\
	& 6.5 & 7.29e-07 & 1.06e-07 & 6.15e-08 & 8.60e-08 & 0 & 8.60e-08 & 0 \\
	& 7.5 & 2.26e-07 & 4.57e-08 & 3.64e-08 & 2.76e-08 & 0 & 2.76e-08 & 0 \\
	& 8.5 & 5.29e-08 & 1.36e-08 & 1.19e-08 & 6.65e-09 & 0 & 6.65e-09 & 0 \\
	& 9.5 & 3.31e-08 & 8.26e-09 & 7.06e-09 & 4.27e-09 & 0 & 4.27e-09 & 0 \\
	& 11 & 1.05e-08 & 2.84e-09 & 2.47e-09 & 1.41e-09 & 0 & 1.41e-09 & 0 \\ \hline
        & 2.25 & 1.18e-03 & 4.38e-04 & 1.37e-04 & 4.16e-04 & 0 & 4.16e-04 & 0 \\
        & 2.75 & 4.02e-04 & 7.52e-05 & 4.36e-05 & 6.13e-05 & 0 & 6.13e-05 & 0 \\
        & 3.25 & 1.02e-04 & 1.93e-05 & 1.59e-05 & 1.09e-05 & 0 & 1.09e-05 & 0 \\
        & 3.75 & 3.92e-05 & 4.60e-06 & 1.90e-06 & 4.18e-06 & 0 & 4.18e-06 & 0 \\
        & 4.25 & 1.47e-05 & 1.77e-06 & 7.97e-07 & 1.59e-06 & 0 & 1.59e-06 & 0 \\
        & 4.75 & 5.89e-06 & 7.53e-07 & 3.79e-07 & 6.50e-07 & 0 & 6.50e-07 & 0 \\
        & 5.25 & 2.99e-06 & 4.01e-07 & 2.35e-07 & 3.25e-07 & 0 & 3.25e-07 & 0 \\
20-40\% & 5.75 & 1.45e-06 & 2.06e-07 & 1.28e-07 & 1.61e-07 & 0 & 1.61e-07 & 0 \\
        & 6.5 & 5.44e-07 & 8.24e-08 & 5.41e-08 & 6.22e-08 & 0 & 6.22e-08 & 0 \\
        & 7.5 & 1.68e-07 & 3.69e-08 & 3.11e-08 & 1.99e-08 & 0 & 1.99e-08 & 0 \\
        & 8.5 & 5.91e-08 & 1.29e-08 & 1.07e-08 & 7.23e-09 & 0 & 7.23e-09 & 0 \\
        & 9.5 & 1.92e-08 & 7.74e-09 & 7.35e-09 & 2.42e-09 & 0 & 2.42e-09 & 0 \\
        & 11 & 1.19e-08 & 2.90e-09 & 2.45e-09 & 1.55e-09 & 0 & 1.55e-09 & 0 \\ \hline
        & 2.25 & 9.14e-04 & 3.40e-04 & 1.03e-04 & 3.24e-04 & 0 & 3.24e-04 & 0 \\
        & 2.75 & 1.85e-04 & 4.31e-05 & 3.25e-05 & 2.83e-05 & 0 & 2.83e-05 & 0 \\
        & 3.25 & 7.17e-05 & 1.47e-05 & 1.25e-05 & 7.73e-06 & 0 & 7.73e-06 & 0 \\
        & 3.75 & 2.46e-05 & 3.01e-06 & 1.47e-06 & 2.62e-06 & 0 & 2.62e-06 & 0 \\
        & 4.25 & 9.73e-06 & 1.22e-06 & 6.23e-07 & 1.05e-06 & 0 & 1.05e-06 & 0 \\
        & 4.75 & 4.19e-06 & 5.67e-07 & 3.27e-07 & 4.63e-07 & 0 & 4.63e-07 & 0 \\
        & 5.25 & 1.71e-06 & 2.69e-07 & 1.95e-07 & 1.85e-07 & 0 & 1.85e-07 & 0 \\
40-60\% & 5.75 & 9.73e-07 & 1.71e-07 & 1.33e-07 & 1.08e-07 & 0 & 1.08e-07 & 0 \\
        & 6.5 & 3.84e-07 & 6.14e-08 & 4.29e-08 & 4.39e-08 & 0 & 4.39e-08 & 0 \\
        & 7.5 & 1.07e-07 & 3.14e-08 & 2.87e-08 & 1.27e-08 & 0 & 1.27e-08 & 0 \\
        & 8.5 & 3.69e-08 & 1.15e-08 & 1.06e-08 & 4.52e-09 & 0 & 4.52e-09 & 0 \\
        & 9.5 & 1.41e-08 & 6.31e-09 & 6.06e-09 & 1.77e-09 & 0 & 1.77e-09 & 0 \\
        & 11 & 2.56e-09 & 2.11e-09 & 2.09e-09 & 3.35e-10 & 0 & 3.35e-10 & 0 \\ \hline
        & 2.25 & 4.07e-04 & 1.54e-04 & 5.17e-05 & 1.45e-04 & 0 & 1.45e-04 & 0 \\
        & 2.75 & 1.27e-04 & 2.64e-05 & 1.75e-05 & 1.97e-05 & 0 & 1.97e-05 & 0 \\
        & 3.25 & 3.52e-05 & 7.72e-06 & 6.66e-06 & 3.90e-06 & 0 & 3.90e-06 & 0 \\
        & 3.75 & 1.11e-05 & 1.45e-06 & 7.84e-07 & 1.22e-06 & 0 & 1.22e-06 & 0 \\
        & 4.25 & 4.55e-06 & 6.09e-07 & 3.40e-07 & 5.06e-07 & 0 & 5.06e-07 & 0 \\
        & 4.75 & 1.88e-06 & 2.71e-07 & 1.68e-07 & 2.13e-07 & 0 & 2.13e-07 & 0 \\
60-88\% & 5.25 & 6.34e-07 & 1.21e-07 & 9.76e-08 & 7.10e-08 & 0 & 7.10e-08 & 0 \\
        & 5.75 & 4.65e-07 & 8.14e-08 & 6.17e-08 & 5.31e-08 & 0 & 5.31e-08 & 0 \\
        & 6.5 & 2.27e-07 & 4.27e-08 & 3.34e-08 & 2.66e-08 & 0 & 2.66e-08 & 0 \\
        & 7.5 & 4.83e-08 & 1.17e-08 & 1.01e-08 & 5.87e-09 & 0 & 5.87e-09 & 0 \\
        & 8.5 & 1.18e-08 & 6.61e-09 & 6.44e-09 & 1.48e-09 & 0 & 1.48e-09 & 0 \\
        & 9.5 & 6.33e-09 & 3.76e-09 & 3.67e-09 & 8.15e-10 & 0 & 8.15e-10 & 0 \\
\end{tabular}\end{ruledtabular}
\end{table*}




\begin{table*}[ht] 
\caption{Invariant yields measured in Au+Au for different centrality classes, including 
minimum bias (0-92\%), most central (0-20\%), and most peripheral (60-92\%).}
\label{tab:AuAu_eta} 
\begin{ruledtabular}\begin{tabular}{cccccccc} 
Centrality & \footnotesize $\pt$ (GeV/$c$) & \footnotesize inv. yield & \footnotesize tot. err. & \footnotesize sys. err. & \footnotesize stat. err. + error A & \footnotesize error B & \footnotesize error C \\\hline 
       &      2.25  &  1.26e-02 & 3.66e-03 & 1.84e-03 & 3.16e-03 & 1.48e-03 & 0 \\ 
       &      2.75  &  3.90e-03 & 1.12e-03 & 5.78e-04 & 9.61e-04 & 4.66e-04 & 0 \\ 
       &      3.25  &  8.79e-04 & 1.70e-04 & 1.32e-04 & 1.08e-04 & 1.07e-04 & 0 \\ 
       &      3.75  &  2.33e-04 & 5.13e-05 & 3.53e-05 & 3.72e-05 & 2.88e-05 & 0 \\ 
0-92\% &  4.50  &  6.44e-05 & 1.60e-05 & 9.97e-06 & 1.25e-05 & 8.22e-06 & 0 \\ 
(MB)   &  5.50  &  1.14e-05 & 2.57e-06 & 1.82e-06 & 1.81e-06 & 1.52e-06 & 0 \\ 
       &      6.50  &  2.80e-06 & 1.05e-06 & 4.59e-07 & 9.48e-07 & 3.87e-07 & 0 \\ 
       &      7.50  &  9.60e-07 & 2.63e-07 & 1.62e-07 & 2.07e-07 & 1.38e-07 & 0 \\ 
       &      8.50  &  4.09e-07 & 1.84e-07 & 7.07e-08 & 1.70e-07 & 6.09e-08 & 0 \\ 
       &      9.50  &  1.51e-07 & 8.25e-08 & 2.67e-08 & 7.81e-08 & 2.32e-08 & 0 \\ \hline
       &      2.25  &  3.95e-02 & 1.24e-02 & 5.80e-03 & 1.09e-02 & 4.65e-03 & 0 \\ 
       &      2.75  &  1.29e-02 & 3.86e-03 & 1.91e-03 & 3.36e-03 & 1.54e-03 & 0 \\ 
       &      3.25  &  2.23e-03 & 5.99e-04 & 3.34e-04 & 4.97e-04 & 2.71e-04 & 0 \\ 
       &      3.75  &  4.53e-04 & 2.07e-04 & 6.88e-05 & 1.95e-04 & 5.61e-05 & 0 \\ 
       &      4.50  &  1.49e-04 & 4.55e-05 & 2.30e-05 & 3.93e-05 & 1.90e-05 & 0 \\ 
0-20\% &  5.50  &  2.74e-05 & 8.33e-06 & 4.37e-06 & 7.09e-06 & 3.65e-06 & 0 \\ 
       &      6.50  &  5.99e-06 & 2.73e-06 & 9.83e-07 & 2.55e-06 & 8.30e-07 & 0 \\ 
       &      7.50  &  2.79e-06 & 6.99e-07 & 4.71e-07 & 5.17e-07 & 4.02e-07 & 0 \\ 
       &      8.50  &  8.42e-07 & 4.02e-07 & 1.46e-07 & 3.75e-07 & 1.25e-07 & 0 \\ 
       &      9.50  &  4.18e-07 & 2.18e-07 & 7.40e-08 & 2.05e-07 & 6.43e-08 & 0 \\ \hline
       &      2.25  &  1.21e-02 & 3.51e-03 & 1.78e-03 & 3.03e-03 & 1.42e-03 & 0 \\ 
       &      2.75  &  2.88e-03 & 8.52e-04 & 4.26e-04 & 7.38e-04 & 3.43e-04 & 0 \\ 
       &      3.25  &  8.58e-04 & 1.66e-04 & 1.29e-04 & 1.06e-04 & 1.04e-04 & 0 \\ 
       &      3.75  &  2.54e-04 & 5.40e-05 & 3.85e-05 & 3.78e-05 & 3.14e-05 & 0 \\ 
       &      4.50  &  7.05e-05 & 1.74e-05 & 1.09e-05 & 1.36e-05 & 8.99e-06 & 0 \\ 
20-40\% &  5.50  &  1.57e-05 & 3.36e-06 & 2.51e-06 & 2.24e-06 & 2.09e-06 & 0 \\ 
       &      6.50  &  3.30e-06 & 1.28e-06 & 5.42e-07 & 1.16e-06 & 4.58e-07 & 0 \\ 
       &      7.50  &  1.06e-06 & 2.65e-07 & 1.79e-07 & 1.96e-07 & 1.53e-07 & 0 \\ 
       &      8.50  &  3.46e-07 & 1.60e-07 & 5.99e-08 & 1.48e-07 & 5.16e-08 & 0 \\ 
       &      9.50  &  1.54e-07 & 7.97e-08 & 2.72e-08 & 7.49e-08 & 2.36e-08 & 0 \\ \hline
       &      2.25  &  1.06e-03 & 3.27e-04 & 2.41e-04 & 2.21e-04 & 2.22e-04 & 0 \\ 
       &      2.75  &  3.34e-04 & 1.02e-04 & 7.61e-05 & 6.83e-05 & 7.03e-05 & 0 \\ 
       &      3.25  &  1.11e-04 & 2.34e-05 & 1.66e-05 & 1.65e-05 & 1.34e-05 & 0 \\ 
60-92\% &  3.75  &  4.04e-05 & 9.05e-06 & 6.13e-06 & 6.66e-06 & 5.00e-06 & 0 \\ 
       &      4.50  &  1.16e-05 & 3.11e-06 & 1.80e-06 & 2.54e-06 & 1.48e-06 & 0 \\ 
       &      5.50  &  1.67e-06 & 1.07e-06 & 2.65e-07 & 1.04e-06 & 2.22e-07 & 0 \\ 
\end{tabular}\end{ruledtabular} 
\end{table*}





\begin{table*}[ht]
\caption{\label{tab:R_dA} 
Nuclear modification factor $R_{dA}$ for $\eta$ in d+Au collisions
for different centrality classes, including
minimum bias (0-88\%), most central (0-20\%), and most peripheral (60-88\%).}
\begin{ruledtabular}\begin{tabular}{ccccccc}
Centrality & \footnotesize $\pt$ (GeV/$c$) & \footnotesize $R_{dA}$ & \footnotesize tot. err. & 
\footnotesize stat. err. + error A & \footnotesize error B & \footnotesize error C \\\hline
       &    2.75 & 1.22 & 0.435 & 0.224 & 0.340 & 0.134 \\
       &    3.25 & 1.06 & 0.233 & 0.102 & 0.162 & 0.117 \\
       &    3.75 & 1.24 & 0.220 & 0.047 & 0.148 & 0.137 \\
       &    4.25 & 1.19 & 0.201 & 0.052 & 0.125 & 0.131 \\
       &    4.75 & 1.21 & 0.205 & 0.070 & 0.119 & 0.133 \\
       &    5.25 & 1.03 & 0.183 & 0.080 & 0.102 & 0.114 \\
0-88\% & 5.75 & 1.03 & 0.191 & 0.097 & 0.102 & 0.114 \\
(MB)   & 6.5 & 1.16 & 0.220 & 0.119 & 0.115 & 0.128 \\
       &    7.5 & 0.896 & 0.215 & 0.160 & 0.089 & 0.099 \\
       &    8.5 & 1.28 & 0.406 & 0.350 & 0.128 & 0.142 \\
       &    9.5 & 1.14 & 0.400 & 0.356 & 0.114 & 0.126 \\
       &    11 & 1.51 & 0.680 & 0.635 & 0.151 & 0.166 \\ \hline
       &    2.75 & 1.15 & 0.421 & 0.231 & 0.322 & 0.130 \\
       &    3.25 & 0.922 & 0.236 & 0.150 & 0.141 & 0.104 \\
       &    3.75 & 1.17 & 0.210 & 0.057 & 0.140 & 0.132 \\
       &    4.25 & 1.11 & 0.191 & 0.060 & 0.116 & 0.126 \\
       &    4.75 & 1.15 & 0.198 & 0.078 & 0.113 & 0.130 \\
       &    5.25 & 0.982 & 0.179 & 0.087 & 0.097 & 0.111 \\
0-20\% & 5.75 & 0.956 & 0.184 & 0.103 & 0.094 & 0.108 \\
       &    6.5 & 0.992 & 0.199 & 0.120 & 0.098 & 0.112 \\
       &    7.5 & 0.887 & 0.232 & 0.184 & 0.088 & 0.101 \\
       &    8.5 & 0.802 & 0.285 & 0.254 & 0.080 & 0.091 \\
       &    9.5 & 0.894 & 0.343 & 0.312 & 0.089 & 0.101 \\
       &    11 & 1.30 & 0.626 & 0.590 & 0.131 & 0.148 \\ \hline
       &    2.75 & 1.22 & 0.450 & 0.249 & 0.342 & 0.139 \\
       &    3.25 & 1.07 & 0.270 & 0.169 & 0.163 & 0.121 \\
       &    3.75 & 1.14 & 0.206 & 0.062 & 0.136 & 0.129 \\
       &    4.25 & 1.06 & 0.186 & 0.067 & 0.111 & 0.120 \\
       &    4.75 & 1.05 & 0.188 & 0.084 & 0.104 & 0.120 \\
       &    5.25 & 1.10 & 0.208 & 0.112 & 0.109 & 0.125 \\
19-40\% & 5.75 & 1.02 & 0.201 & 0.120 & 0.100 & 0.116 \\
       &    6.5 & 1.07 & 0.222 & 0.141 & 0.106 & 0.121 \\
       &    7.5 & 0.953 & 0.264 & 0.216 & 0.095 & 0.108 \\
       &    8.5 & 1.30 & 0.425 & 0.371 & 0.129 & 0.147 \\
       &    9.5 & 0.750 & 0.374 & 0.354 & 0.075 & 0.085 \\
       &    11 & 2.13 & 0.995 & 0.934 & 0.214 & 0.242 \\ \hline
       &    2.75 & 0.86 & 0.338 & 0.212 & 0.241 & 0.102 \\
       &    3.25 & 1.15 & 0.305 & 0.202 & 0.176 & 0.136 \\
       &    3.75 & 1.09 & 0.201 & 0.071 & 0.130 & 0.128 \\
       &    4.25 & 1.08 & 0.192 & 0.078 & 0.113 & 0.127 \\
       &    4.75 & 1.14 & 0.210 & 0.104 & 0.113 & 0.135 \\
40-60\% & 5.25 & 0.962 & 0.198 & 0.126 & 0.095 & 0.114 \\
       &    5.75 & 1.04 & 0.234 & 0.164 & 0.103 & 0.123 \\
       &    6.5 & 1.15 & 0.246 & 0.163 & 0.114 & 0.136 \\
       &    7.5 & 0.929 & 0.314 & 0.277 & 0.092 & 0.110 \\
       &    8.5 & 1.24 & 0.490 & 0.449 & 0.123 & 0.146 \\
       &    9.5 & 0.842 & 0.451 & 0.430 & 0.084 & 0.099 \\
       &    11 & 0.704 & 0.645 & 0.636 & 0.071 & 0.083 \\ \hline
       &    2.75 & 1.36 & 0.515 & 0.301 & 0.381 & 0.165 \\
       &    3.25 & 1.30 & 0.358 & 0.248 & 0.200 & 0.158 \\
       &    3.75 & 1.13 & 0.213 & 0.085 & 0.135 & 0.137 \\
       &    4.25 & 1.16 & 0.212 & 0.095 & 0.122 & 0.141 \\
       &    4.75 & 1.18 & 0.223 & 0.119 & 0.117 & 0.143 \\
60-88\% & 5.25 & 0.826 & 0.190 & 0.138 & 0.081 & 0.010 \\
       &    5.75 & 1.15 & 0.255 & 0.177 & 0.114 & 0.140 \\
       &    6.5 & 1.57 & 0.368 & 0.269 & 0.156 & 0.191 \\
       &    7.5 & 0.969 & 0.285 & 0.239 & 0.096 & 0.117 \\
       &    8.5 & 0.913 & 0.558 & 0.538 & 0.091 & 0.111 \\
       &    9.5 & 0.874 & 0.578 & 0.561 & 0.087 & 0.106 \\ \hline
\end{tabular}\end{ruledtabular}
\end{table*}


\begin{table*}[ht] 
\caption{Nuclear modification factor $R_{AA}(\pt)$ for $\eta$ in Au+Au 
collisions for different centrality classes from most central (0-20\%) to most peripheral (60-92\%).
Note that there is an additional 9.7\% normalization uncertainty 
(Run-3 p+p BBC error, gray box in Fig.~\ref{fig:R_AA_eta}) not quoted.}
\label{tab:R_AA_0_20} 
\begin{ruledtabular}\begin{tabular}{ccccc} 
Centrality & \hspace{5mm} $\pt$ (GeV/$c$) \hspace{5mm} & \hspace{5mm} $R_{AA}$ \hspace{5mm}  
& \hspace{5mm} tot. err. \hspace{5mm} & \hspace{5mm} error C \hspace{5mm} \\\hline  
       &      2.75  &  0.532 & 0.227 (42.6\%) & 0.036 (6.8\%) \\ 
       &      3.25  &  0.318 & 0.096 (30.2\%) & 0.022 (6.8\%) \\ 
       &      3.75  &  0.178 & 0.084 (46.9\%) & 0.012 (6.8\%) \\ 
       &      4.50  &  0.234 & 0.074 (31.7\%) & 0.016 (6.8\%) \\ 
       &      5.50  &  0.199 & 0.063 (31.6\%) & 0.014 (6.8\%) \\ 
0-20\%  &   6.50  &  0.160 & 0.075 (46.9\%) & 0.011 (6.8\%) \\ 
       &      7.50  &  0.215 & 0.062 (28.9\%) & 0.015 (6.8\%) \\ 
       &      8.50  &  0.250 & 0.133 (53.0\%) & 0.017 (6.8\%) \\ 
       &      9.50  &  0.222 & 0.131 (59.3\%) & 0.015 (6.8\%) \\ \hline
       &      2.75  &  0.479 & 0.202 (42.3\%) & 0.037  (7.8\%) \\ 
       &      3.25  &  0.492 & 0.117 (23.8\%) & 0.038  (7.8\%) \\ 
       &      3.75  &  0.401 & 0.095 (23.6\%) & 0.031  (7.8\%) \\ 
       &      4.50  &  0.446 & 0.116 (26.0\%) & 0.035  (7.8\%) \\ 
       &      5.50  &  0.460 & 0.106 (23.1\%) & 0.036  (7.8\%) \\ 
20-60\% &  6.50  &  0.355 & 0.143 (40.3\%) & 0.028  (7.8\%) \\ 
       &      7.50  &  0.329 & 0.095 (28.9\%) & 0.026  (7.8\%) \\ 
       &      8.50  &  0.414 & 0.214 (51.6\%) & 0.032  (7.8\%) \\ 
       &      9.50  &  0.328 & 0.193 (58.9\%) & 0.026  (7.8\%) \\ \hline
       &      2.75  &  0.733 & 0.315 (43.0\%) & 0.209  (28.6\%) \\ 
       &      3.25  &  0.837 & 0.211 (25.3\%) & 0.239  (28.6\%) \\ 
60-92\% &  3.75  &  0.841 & 0.208 (24.7\%) & 0.240  (28.6\%) \\ 
       &      4.50  &  0.967 & 0.271 (28.0\%) & 0.276  (28.6\%) \\ 
       &      5.50  &  0.641 & 0.415 (64.8\%) & 0.183  (28.6\%) \\ 
\end{tabular}\end{ruledtabular} 
\end{table*}




\begin{table*}[ht]
\caption{Ratio of $\eta$ and $\pi^0$ for p+p collisions at $\sqrt{s}$ = 200 GeV.}
\label{tab:pp_eta_pi0}
\begin{ruledtabular}\begin{tabular}{ccccccc} 
\footnotesize $\pt$ (GeV/$c$) & \footnotesize $\eta/\pi^0$ & \footnotesize tot. err. & \footnotesize
stat. err. & \footnotesize error A & \footnotesize error B & \footnotesize error C \\\hline 
2.75 & 0.440 & 0.131 & 0.076 & 0 & 0.107 & 0 \\
3.25 & 0.421 & 0.041 & 0.009 & 0 & 0.040 & 0 \\
3.75 & 0.446 & 0.032 & 0.012 & 0 & 0.030 & 0 \\
4.25 & 0.473 & 0.035 & 0.016 & 0 & 0.031 & 0 \\
4.75 & 0.468 & 0.036 & 0.022 & 0 & 0.029 & 0 \\
5.25 & 0.510 & 0.045 & 0.032 & 0 & 0.031 & 0 \\
5.75 & 0.561 & 0.055 & 0.044 & 0 & 0.034 & 0 \\
6.5 & 0.540 & 0.057 & 0.047 & 0 & 0.033 & 0 \\
7.5 & 0.596 & 0.086 & 0.078 & 0 & 0.037 & 0 \\
8.5 & 0.426 & 0.099 & 0.095 & 0 & 0.027 & 0 \\
9.5 & 0.588 & 0.166 & 0.162 & 0 & 0.039 & 0 \\
11 & 0.419 & 0.164 & 0.162 & 0 & 0.029 & 0 \\

\end{tabular}\end{ruledtabular}
\end{table*}

\begingroup \squeezetable


\begin{table*}[ht] 
  \caption{\label{tab:dAu_eta_pi0}
Ratio of $\eta$ and $\pi^0$ for d+Au collisions
for different centrality classes, including
minimum bias (0-88\%), most central (0-20\%), and most peripheral (60-88\%).}
\begin{ruledtabular}\begin{tabular}{cccccccc} 
Centrality & $\pt$ (GeV/$c$) & $\eta/\pi^0$ & tot. err. & stat. err. + error A & error B & error C \\\hline
       & 2.25 & 0.420 & 0.038 & 0.028 & 0.025 & 0 \\
       & 2.75 & 0.472 & 0.044 & 0.033 & 0.028 & 0 \\
       & 3.25 & 0.383 & 0.045 & 0.039 & 0.023 & 0 \\
       & 3.75 & 0.472 & 0.034 & 0.018 & 0.028 & 0 \\
       & 4.25 & 0.478 & 0.033 & 0.017 & 0.029 & 0 \\
       & 4.75 & 0.483 & 0.035 & 0.020 & 0.029 & 0 \\
0-88\% & 5.25 & 0.465 & 0.037 & 0.025 & 0.028 & 0 \\
 (MB)  & 5.75 & 0.510 & 0.043 & 0.030 & 0.031 & 0 \\
       & 6.5 & 0.552 & 0.048 & 0.034 & 0.033 & 0 \\
       & 7.5 & 0.478 & 0.070 & 0.064 & 0.029 & 0 \\
       & 8.5 & 0.499 & 0.092 & 0.087 & 0.030 & 0 \\
       & 9.5 & 0.677 & 0.118 & 0.111 & 0.041 & 0 \\
       & 11 & 0.609 & 0.124 & 0.119 & 0.037 & 0 \\ \hline
       & 2.25 & 0.386 & 0.052 & 0.047 & 0.023 & 0 \\
       & 2.75 & 0.491 & 0.062 & 0.054 & 0.029 & 0 \\
       & 3.25 & 0.364 & 0.066 & 0.063 & 0.022 & 0 \\
       & 3.75 & 0.494 & 0.041 & 0.028 & 0.030 & 0 \\
       & 4.25 & 0.512 & 0.040 & 0.026 & 0.031 & 0 \\
       & 4.75 & 0.520 & 0.043 & 0.030 & 0.031 & 0 \\
       & 5.25 & 0.508 & 0.048 & 0.037 & 0.030 & 0 \\
0-20\% & 5.75 & 0.547 & 0.056 & 0.045 & 0.033 & 0 \\
       & 6.5 & 0.563 & 0.063 & 0.053 & 0.034 & 0 \\
       & 7.5 & 0.579 & 0.109 & 0.104 & 0.035 & 0 \\
       & 8.5 & 0.363 & 0.094 & 0.091 & 0.022 & 0 \\
       & 9.5 & 0.644 & 0.159 & 0.154 & 0.039 & 0 \\
       & 11 & 0.684 & 0.193 & 0.188 & 0.041 & 0 \\ \hline
       & 2.25 & 0.416 & 0.057 & 0.051 & 0.025 & 0 \\
       & 2.75 & 0.517 & 0.068 & 0.060 & 0.031 & 0 \\
       & 3.25 & 0.425 & 0.077 & 0.072 & 0.025 & 0 \\
       & 3.75 & 0.467 & 0.042 & 0.031 & 0.028 & 0 \\
       & 4.25 & 0.470 & 0.041 & 0.029 & 0.028 & 0 \\
       & 4.75 & 0.447 & 0.043 & 0.033 & 0.027 & 0 \\
20-40\% & 5.25 & 0.524 & 0.057 & 0.048 & 0.031 & 0 \\
       & 5.75 & 0.539 & 0.062 & 0.053 & 0.032 & 0 \\
       & 6.5 & 0.543 & 0.069 & 0.060 & 0.033 & 0 \\
       & 7.5 & 0.545 & 0.119 & 0.114 & 0.033 & 0 \\
       & 8.5 & 0.546 & 0.119 & 0.115 & 0.033 & 0 \\
       & 9.5 & 0.495 & 0.206 & 0.204 & 0.030 & 0 \\
       & 11 & 1.046 & 0.277 & 0.269 & 0.063 & 0 \\ \hline
       & 2.25 & 0.489 & 0.065 & 0.058 & 0.029 & 0 \\
       & 2.75 & 0.372 & 0.072 & 0.069 & 0.022 & 0 \\
       & 3.25 & 0.457 & 0.091 & 0.086 & 0.027 & 0 \\
       & 3.75 & 0.439 & 0.044 & 0.035 & 0.026 & 0 \\
       & 4.25 & 0.456 & 0.043 & 0.033 & 0.027 & 0 \\
       & 4.75 & 0.490 & 0.053 & 0.044 & 0.029 & 0 \\
40-60\% & 5.25 & 0.466 & 0.066 & 0.060 & 0.028 & 0 \\
       & 5.75 & 0.547 & 0.088 & 0.082 & 0.033 & 0 \\
       & 6.5 & 0.560 & 0.078 & 0.070 & 0.034 & 0 \\
       & 7.5 & 0.528 & 0.160 & 0.157 & 0.032 & 0 \\
       & 8.5 & 0.598 & 0.200 & 0.197 & 0.036 & 0 \\
       & 9.5 & 0.495 & 0.234 & 0.232 & 0.030 & 0 \\
       & 11 & 0.243 & 0.211 & 0.210 & 0.015 & 0 \\ \hline
       & 2.25 & 0.474 & 0.070 & 0.064 & 0.028 & 0 \\
       & 2.75 & 0.558 & 0.090 & 0.083 & 0.033 & 0 \\
       & 3.25 & 0.500 & 0.105 & 0.101 & 0.030 & 0 \\
       & 3.75 & 0.482 & 0.051 & 0.042 & 0.029 & 0 \\
       & 4.25 & 0.490 & 0.051 & 0.042 & 0.029 & 0 \\
       & 4.75 & 0.514 & 0.062 & 0.053 & 0.031 & 0 \\
60-88 \% & 5.25 & 0.374 & 0.068 & 0.064 & 0.022 & 0 \\
       & 5.75 & 0.569 & 0.090 & 0.083 & 0.034 & 0 \\
       & 6.5 & 0.734 & 0.129 & 0.121 & 0.044 & 0 \\
       & 7.5 & 0.501 & 0.123 & 0.119 & 0.030 & 0 \\
       & 8.5 & 0.362 & 0.216 & 0.215 & 0.022 & 0 \\
       & 9.5 & 0.469 & 0.294 & 0.293 & 0.028 & 0 \\
\end{tabular}\end{ruledtabular}
\end{table*}

\endgroup


\begin{table*}[ht] 
\caption{\label{tab:phenix_AuAu_200GeV_eta_pi0_ratio} 
Ratio of $\eta$ and $\pi^0$ in Au+Au collisions
for different centrality classes, including
minimum bias (0-92\%), most central (0-20\%), and most peripheral (60-92\%).}
\begin{ruledtabular}\begin{tabular}{cccc} 
Centrality & \hspace{5mm} $\pt$ (GeV/$c$) \hspace{5mm} & \hspace{5mm} $\eta/\pi^0$ \hspace{5mm} & \hspace{9mm} tot. err. \hspace{9mm} \\\hline 
       &   2.25  &  0.320 & 0.090 (28.1\%) \\ 
       &   2.75  &  0.410 & 0.120 (29.3\%) \\ 
       &   3.25  &  0.340 & 0.060 (17.6\%) \\ 
       &   3.75  &  0.290 & 0.058 (20.0\%) \\ 
0-92\% &  4.50  &  0.350 & 0.083 (23.7\%) \\ 
 (MB)  &  5.50  &  0.350 & 0.072 (20.6\%) \\ 
       &   6.50  &  0.350 & 0.130 (37.1\%) \\ 
       &   7.50  &  0.560 & 0.100 (17.9\%) \\ 
       &   8.50  &  0.480 & 0.210 (43.8\%) \\ 
       &   9.50  &  0.490 & 0.250 (51.0\%) \\ \hline
       &   2.25  &  0.400 & 0.120 (30.0\%) \\ 
       &   2.75  &  0.550 & 0.170 (30.9\%) \\ 
       &   3.25  &  0.370 & 0.110 (29.7\%) \\ 
       &   3.75  &  0.240 & 0.110 (45.8\%) \\ 
       &   4.50  &  0.360 & 0.110 (30.6\%) \\ 
0-20\%  &  5.50  &  0.380 & 0.110 (28.9\%) \\ 
       &   6.50  &  0.350 & 0.160 (45.7\%) \\ 
       &   7.50  &  0.530 & 0.130 (24.5\%) \\ 
       &   8.50  &  0.470 & 0.240 (51.1\%) \\ 
       &   9.50  &  0.490 & 0.280 (57.1\%) \\ \hline
       &   2.25  &  0.360 & 0.100 (27.8\%) \\ 
       &   2.75  &  0.340 & 0.100 (29.4\%) \\ 
       &   3.25  &  0.370 & 0.070 (18.9\%) \\ 
       &   3.75  &  0.340 & 0.066 (19.4\%) \\ 
       &   4.50  &  0.410 & 0.096 (23.4\%) \\ 
20-60\%  & 5.50  &  0.490 & 0.096 (19.6\%) \\ 
       &   6.50  &  0.420 & 0.160 (38.1\%) \\ 
       &   7.50  &  0.430 & 0.098 (22.8\%) \\ 
       &   8.50  &  0.380 & 0.170 (44.7\%) \\ 
       &   9.50  &  0.400 & 0.200 (50.0\%) \\ \hline
       &   2.25  &  0.312 & 0.094 (30.1\%) \\ 
       &   2.75  &  0.383 & 0.110 (28.7\%) \\ 
       &   3.25  &  0.404 & 0.081 (20.0\%) \\ 
60-92\% &  3.75  &  0.438 & 0.093 (21.2\%) \\ 
       &   4.50  &  0.542 & 0.139 (25.6\%) \\ 
       &   5.50  &  0.404 & 0.257 (63.6\%) \\ 
\end{tabular}\end{ruledtabular} 
\end{table*} 

\end{appendix}

\clearpage


\def\Journal#1#2#3#4{#1 {\bf #2},#3 (#4)}
\def\JournalAPPEAR#1#2{#1 {\bf #2}}
\def\ANR{Ann.\ Rev.\ Nucl.\ Part.\ Sci.~}
\def\RMP{Rev.\ Mod.\ Phys.~}
\def\IJMPA{Int. J. Mod. Phys.~A~}
\def\JPG{J.~Phys.~G~}
\def\NCA{Nuovo Cimento~}
\def\NIM{Nucl. Instrum. Methods~}
\def\NIMA{Nucl. Instrum. Methods~A~}
\def\NPA{Nucl. Phys.~A~}
\def\NPB{Nucl. Phys.~B~}
\def\PLB{Phys. Lett.~B~}
\def\PLC{Phys. Rept.~}
\def\PRL{Phys. Rev. Lett.~}
\def\PRD{Phys. Rev.~D~}
\def\PRC{Phys. Rev.~C~}
\def\ZPC{Z. Phys.~C~}
\def\EPJC{Eur.~Phys.~J.~C~}
\def\APPB{Acta Phys. Polon.~B~}


\begin{thebibliography}{2}
\bibitem{Indumathi:1998am} D.~Indumathi, H.~S.~Mani and A.~Rastogi, 
Phys.\ Rev.\ D {\bf 58}, 094014 (1998). 
Note that in the second case, the $\eta$ FF is not explicitly given but, within the framework, 
could be derived from the FF parameters for other octet mesons.
\end{thebibliography}
\end{document}